\documentclass[11pt,nofootinbib,preprint]{revtex4-1}
\pdfoutput=1
\usepackage[T1]{fontenc}
\usepackage[latin9]{inputenc}
\usepackage{times}
\usepackage{graphicx}
\usepackage{tabularx}
\usepackage{graphicx}
\usepackage{dcolumn}
\usepackage{bm}
\usepackage{booktabs}
\usepackage{makecell}
\usepackage{makebox}
\usepackage{multirow}
\usepackage[figuresright]{rotating}
\usepackage{amsmath}
\usepackage{amstext}
\usepackage{amssymb}
\usepackage{xfrac}
\usepackage[colorlinks=true,citecolor={blue},linktocpage=true]{hyperref}\usepackage{graphicx}

\usepackage{amsmath}
\usepackage{amstext}
\usepackage{amssymb}
\usepackage{amsfonts}
\usepackage{longtable,booktabs}
\usepackage{hyperref}
\usepackage{url}
\usepackage{subfigure}
\usepackage{dsfont}
\usepackage{booktabs}
\usepackage{amsbsy}
\usepackage{amsthm}
\usepackage{bm}
\usepackage{esint}
\usepackage{multirow}
\usepackage{hyperref}
\usepackage{cleveref}
\usepackage{mathrsfs}
\usepackage{amsfonts}
\usepackage{amsbsy}
\usepackage{dcolumn}
\usepackage{bm}
\usepackage{multirow}
\usepackage{color}
\usepackage{extarrows}
\usepackage{datetime}
\usepackage{comment}
\usepackage[super]{nth}
\usepackage{tikz-cd}
\usepackage{float}

\makeatletter

\def\Z{\mathbb{Z}}

\makeatother

\begin{document}
\title{Fusion rules and shrinking rules of topological orders in five dimensions}
\author{Yizhou Huang}
\thanks{These authors contributed equally to this work.}
 \author{Zhi-Feng Zhang}
\thanks{These authors contributed equally to this work.}
 \author{Peng Ye}
\email{yepeng5@mail.sysu.edu.cn}
\affiliation{School of Physics, State Key Laboratory of Optoelectronic Materials and Technologies, and Guangdong Provincial Key Laboratory of Magnetoelectric Physics and Devices, Sun Yat-sen University, Guangzhou 510275, China}
\date{\today}
\begin{abstract}
As a series of work about 5D (spacetime) topological orders, here we employ the path-integral formalism of 5D  topological quantum field theory  (TQFT) established in \href{https://doi.org/10.1007/JHEP04(2022)138}{Zhang and Ye, JHEP 04 (2022) 138}  to explore non-Abelian fusion rules,  hierarchical shrinking rules and quantum dimensions of particle-like, loop-like and membrane-like topological excitations in 5D topological orders.  To illustrate, we focus on a prototypical example of twisted $BF$ theories that comprise the twisted topological terms of the $BBA$ type. First, we classify topological excitations by establishing equivalence classes among all gauge-invariant Wilson operators. Then, we compute fusion rules from the path-integral and find that fusion rules may be non-Abelian; that is, the fusion outcome can be a direct sum of distinct excitations. We further compute shrinking rules. Especially, we discover exotic hierarchical structures hidden in shrinking processes of 5D or higher:  a membrane is shrunk into particles and loops, and the loops are subsequently shrunk into a direct sum of particles. We obtain the algebraic structure of shrinking coefficients and fusion coefficients.  We compute the quantum dimensions of all excitations and find that sphere-like membranes and torus-like membranes differ not only by their shapes but also by their quantum dimensions. We further study the algebraic structure that determines anomaly-free conditions on fusion coefficients and shrinking coefficients. Besides $BBA$, we explore general properties of all twisted terms in $5$D. Together with braiding statistics reported before, the theoretical progress here paves the way toward characterizing and classifying topological orders in higher dimensions where topological excitations consist of both particles and spatially extended objects. 
\end{abstract}
\maketitle
\newpage
\tableofcontents{}

\section{Introduction\label{s1}}
Originally discovered in the context of strongly correlated electron systems, such as the fractional quantum Hall effect and chiral spin liquids, \textit{topological order} has been widely explored in the literature of quantum many-body physics, high energy physics, mathematical physics, and quantum information~\cite{wenZootopoRMP}. While it is well-known that Ginzburg-Landau order parameters cannot characterize topological orders and their phase transitions, for the past decades, topological data, such as the number of topological excitations (i.e., anyons), ground state degeneracy, edge chiral central charge, braiding statistics, and fusion rules of anyons, have been extensively studied and utilized to characterize  3D (spacetime) topological orders.  The practical goal of characterizing and classifying topological orders can be regarded as the establishment of a set of such topological data or physical observables for every type of topological order, assuming that the ground states are realized by Hamiltonians with short-range interactions and a bulk gap. To achieve this goal, various theoretical approaches have been developed, including but not limited to, the construction of exactly solvable models (e.g., string-net models~\cite{string1}, Dijkgraaf-Witten models~\cite{dwitten}, and stabilizer codes~\cite{Kitaev2003faulttolerant}), quantum-informative perspectives (e.g.,  long-range entanglement nature of topological order~\cite{CGW2010LRE},   the correspondence between zero modes of entanglement spectrum and the physical edge state~\cite{Li_Haldane}, and topological entanglement entropy~\cite{KitaevTopoEE,Levin_Wen_TEE}), general algebraic structure based on unitary modular category~\cite{Kitaev2006,string1,Kong2020}, and topological quantum  field theoretical (TQFT) analysis~\cite{Witten1989,Turaev2016}.

As one of the most prominent   TQFTs, the Chern-Simons theory governs the low-energy physics of $3$D topological orders and provides a quantitative tool for computing topological properties of anyons which are a key ingredient of topological quantum computation~\cite{sarma_08_TQC}. Meanwhile, by placing the Chern-Simons theory on an open disc, the 2D boundary theory of 3D topological orders is shown to be dominated by gapless modes and mathematically described by conformal field theory (CFT)~\cite{moore1989classical} with  gravitational anomaly~\cite{Wen_grav_prd2013,string8}.  Furthermore, the Gutzwiller projective construction of both  bulk ground state wave functions and the associated edge CFT algebraic construction can be performed in a systematic manner~\cite{Wen99,Barkeshli2010,barkeshli_wen}.  It should be noticed that the Chern-Simons theory mentioned in the present paper belongs to  hydrodynamical field-theoretical approach, which is different from the flux-attachment   approach of composite bosons and fermions~\cite{composite_Jain_2,composite_Fradkin,PhysRevLett.62.82}. The Chern-Simons theory of 3D topological orders with global symmetry can be applied to characterizing and classifying 3D symmetry-enriched topological phases (SET) in which anyons may carry a fractional quantum number of global symmetry group~\cite{PhysRevB.93.155121,PhysRevB.87.195103}. Recently, the Chern-Simons theory at the trivial level~\cite{PhysRevB.86.125119,YW12,PhysRevB.93.115136,Ye14b}  has been   applied to describe the bulk effective field theory of bosonic symmetry-protected ``topological'' phases (SPT) as long as global symmetry is imposed nontrivially. Furthermore, the Chern-Simons theory at the trivial level, after weakly gauging, leads to   topological response theory of bosonic SPTs~\cite{PhysRevLett.112.141602,Hung_Wen_gauge}.

While the notion of topological order was born in $3$D,  lifting spatial dimensions of topological orders to $4$D and higher is an active branch of frontier research. In higher dimensions, the self-statistics of particle excitations is no longer anyonic~\cite{wu84} but the presence of spatially extended excitations (e.g., loops in $4$D and higher and membranes in $5$D and higher) may significantly enrich the topological data. Specifically, in $4$D topological orders, topological excitations include both  point-like particle excitations and loop-like excitations.  Consider a discrete Abelian gauge group $G=\prod_{i=1}^n{\mathbb{Z} _{N_i}}$. 
The underlying TQFTs are called ``twisted $BF$ theories'' where the topological action consists of  a multi-component $BF$ term \cite{Horowitz1990,hansson2004superconductors}  and some ``twisted (topological) terms''. The $BF$ term in $4$D is written as a wedge product of $B$ and $dA$: $BdA$  ($B$ and $A$ are respectively $2$- and $1$-form gauge fields), which can be physically understood as a result of exotic boson condensation and duality  transformations~\cite{hansson2004superconductors,YW13a,ye16a,Moy_Fradkin2023}. Twisted terms (or ``twists'' for short) consist of  $AAdA$, $AAAA$~\cite{2016arXiv161209298P,PhysRevB.99.235137,YeGu2015,ypdw}, and $AAB$~\cite{yp18prl}.  From these topological terms, braiding statistics among topological excitations  can be systematically computed, such as 
   particle-loop braiding~\cite{hansson2004superconductors,abeffect,PRESKILL199050,PhysRevLett.62.1071,PhysRevLett.62.1221,ALFORD1992251}, multi-loop braiding~\cite{wang_levin1}, and particle-loop-loop braiding (i.e., Borromean-Rings braiding) \cite{yp18prl}, which are tightly connected to link invariants, e.g., Hopf linking number of Hopf links and Milner's triple linking number of general Brunnian links~\cite{kauffman2006formal,mellor2003geometric}.  By excluding all actions that are not gauge-invariant, ref.~\cite{zhang2021compatible} exhausted all legitimate combinations of these topological terms to obtain a complete list of topological orders in $4$D.  
   
   Recently, fusion rules, shrinking rules, and quantum dimensions of all topological excitations in $4$D  twisted $BF$ theories are computed systematically~\cite{PhysRevB.107.165117}, where shrinking rules of loops are studied concretely and  non-Abelian fusion rules are found despite the Abelian nature of the gauge group.   By \textit{shrinking}, we mean that spatially extended topological excitations, i.e., loops, are shrunk into particles, which is a unique topological property for topological orders in $4$D and higher.  
   In addition, twisted $BF$ theories  of $4$D topological orders with various types of  symmetry can be applied to $4$D SET phases, leading to a field-theoretical classification and characterization of a large class of $4$D SET phases, see, e.g., refs.~\cite{ye16_set,2016arXiv161008645Y,Ye:2017aa,ye16a} and ref.~\cite{Ning2018prb} on the general theory. Again, twisted $BF$ theories at the trivial level have been applied to describe the bulk effective field theory of $4$D SPT phases~\cite{YeGu2015,bti2}, which, by weakly gauging, leads to the topological response theory of $4$D SPT phases~\cite{PhysRevLett.114.031601,PhysRevB.99.205120,bti6}. In addition, $BB$ and $dAdA$ type terms were also discussed respectively for fermionic topological phases and topological response theory of bosonic topological insulators~\cite{Kapustin:2014gua,PhysRevB.99.235137,zhang2023continuum,bti2,YW13a,lapa17}.

   By lifting spatial dimensions further, topological orders in $5$D admit particle-like, loop-like, and membrane-like topological excitations, which exhibit highly unexplored features of both the formalism of TQFTs and braiding statistics~\cite{Zhang:2021ycl}.   If we still consider the gauge group $G=\prod_{i=1}^n\Z_{N_i}$,  TQFTs in $5$D may contain two types of (generalized) $BF$ terms, i.e., $CdA$ and $\tilde{B}dB$, where $C$ is $3$-form, $B$ and $\tilde{B}$ are two different $2$-form, $A$ is $1$-form.  Therefore, for each $\Z_{N_i}$ subgroup, there are two choices for the assignment of gauge charges and corresponding $BF$ terms, which significantly enriches topological orders. On top of two types of $BF$ terms, there are many different twisted topological terms, e.g., $AAAAA$, $AAAdA$, $AdAdA$, $AAC$, $AAAB$, $BBA$, $AdAB$, $AAdB$, which lead to many different twisted $BF$ theories after proper combinations in $5$D,  as long as gauge invariance is guaranteed~\cite{zhang2021compatible}.  In ref.~\cite{Zhang:2021ycl}, braiding statistics encoded in all topological excitations has been investigated  via computing correlation functions of Wilson operators whose   expectation values  rely on  counting intersections of sub-manifolds in $5$D. The physical theory presented in ref.~\cite{Zhang:2021ycl}  provides an alternative route to understanding the link theory of higher-dimensional compact manifolds.  

While braiding statistics have shown how diverse the topological data of $5$D topological orders are~\cite{Zhang:2021ycl}, the fusion rules and shrinking rules of $5$D topological orders are still yet to be uncovered. Since  there are  membrane topological excitations that are spatially extended excitations of two dimensions, it is natural to expect more exotic fusion and shrinking rules than that of $4$D studied in ref.~\cite{PhysRevB.107.165117}, which motivates the present paper. In this paper, we adopt the path-integral formalism of $5$D topological orders constructed in ref.~\cite{Zhang:2021ycl}. We focus   on two prototypical twisted topological terms, namely, $BBA$ and $AAAB$ both of which exhibit not only non-Abelian fusion rules but also   hierarchical structure in shrinking rules.  By \textit{hierarchical}, we mean that a membrane is  shrunk into  particles and loops, and  the loops are subsequently shrunk into a direct sum of particles. Such phenomena can only exist in $5$D and higher.  We  realize that membrane excitations potentially possess very different topological properties, e.g., quantum dimensions if the membranes have different topologies, e.g., sphere and torus. We concretely compute all fusion rules and verify the consistency between fusion rules and shrinking rules, which leads to the algebraic structure hidden in the fusion coefficients and shrinking coefficients. Any violation of the algebraic structure may be associated with a quantum anomaly of some kind in the topological order. 
   
     This paper is organized as follows. In section~\ref{s2}, we first review the twisted $BF$ theory with the $BBA$ twisted term introduced in ref.~\cite{Zhang:2021ycl}. Then we construct Wilson operators for all topological excitations (see table~\ref{tab_excitation}) and classify them into different equivalence classes(see table~\ref{tab_equivalence_classes}). In section~\ref{s3}, we calculate some typical fusion rules (see table~\ref{table_fusionBBA}) for the $BBA$ twisted term explicitly from the path integral. We also compute quantum dimensions (see table~\ref{table_quantumdBBA}) for all topological excitations. In section~\ref{s4}, we calculate the shrinking rules (see table~\ref{table_shrinking}) for the $BBA$ twisted term. We concretely discuss the hierarchical structure in the shrinking rules. In section~\ref{section_algebraic}, we study the general algebraic structure that determines anomaly-free topological orders. In section~\ref{section_general_discuss_twisted5D}, we present several general properties (see table~\ref{table_comparison_5dtwisted}) of all twisted terms in $5$D. 
A brief summary is given in section~\ref{section_summary}. Several appendices are provided.

\section{Topological excitations and  Wilson operators for the $BBA$ twisted term}\label{s2}

\subsection{Twisted $BF$ theory with the $BBA$ twisted term}\label{ss21}
Any gauge group $G=\prod_{i=1}^n{\mathbb{Z} _{N_i}}$ with $n\geqslant3$ can support $BBA$ term in TQFT action. For simplicity, we consider $G=\left(\mathbb{Z} _{2}\right)^{3}$, that is, $\mathbb{Z}_{N_1}=\mathbb{Z}_{N_2}=\mathbb{Z}_{N_3}=\mathbb{Z} _{2}$, and take the TQFT action to be~\cite{Zhang:2021ycl}
\begin{equation}
	S=\int{\frac{N_1}{2\pi}\tilde{B}^1dB^1+}\frac{N_2}{2\pi}\tilde{B}^2dB^2+\frac{N_3}{2\pi}C^3dA^3+qB^1B^2A^3\,,
	\label{eq_action_BBA}
\end{equation}
where $q$ is a periodic and quantized coefficient. $A^3$ is a 1-form gauge field. $B^1$, $B^2$, $\tilde{B}^1$ and $\tilde{B}^2$ are 2-form gauge fields.  $C^3$ is a 3-form gauge field. The superscript of a gauge field labels its corresponding gauge group. Specifically, $B^1$ and $\tilde{B}^1$ correspond to the $\mathbb{Z} _{N_1}$ gauge group, $B^2$ and $\tilde{B}^2$ correspond to the $\mathbb{Z} _{N_2}$ gauge group, $A^3$ and $C^3$ correspond to the $\mathbb{Z} _{N_3}$ gauge group. We note that there are two kinds of $BF$ terms in the action~(\ref{eq_action_BBA}), i.e., $\tilde{B}dB$ and $CdA$, so the action represents a mixed $BF$ theory following the definition in ref.~\cite{Zhang:2021ycl}.  The coefficient $q$ is quantized and periodically identified: $q=\frac{pN_1N_2N_3}{\left( 2\pi \right) ^2N_{123}}$, $p\in \mathbb{Z} _{N_{123}}$, where $N_{123}$ is the greatest common divisor of $N_{1}$, $N_{2}$ and $N_{3}$, and  $N_{1}=N_{2}=N_{3}=N_{123}=2$ for $G=\left(\mathbb{Z} _{2}\right)^{3}$ considered here.  The gauge fields $\tilde{B}^1$, $\tilde{B}^2$ and $C^3$ serve as the Lagrange multipliers in action~(\ref{eq_action_BBA}), which enforces the flat connection conditions: $dB^1=0$, $dB^2=0$ and $dA^3=0$. The gauge transformations are given by:
\begin{align}
	&B^1\rightarrow B^1+dV^1, \tilde{B}^1\rightarrow \tilde{B}^1+d\tilde{V}^1+\frac{2\pi q}{N_1}\left( V^2A^3+B^2\chi ^3+V^2d\chi ^3 \right)\,,   
	\\
	&B^2\rightarrow B^2+dV^2, \tilde{B}^2\rightarrow \tilde{B}^2+d\tilde{V}^2+\frac{2\pi q}{N_2}\left( V^1A^3+B^1\chi ^3+V^1d\chi ^3 \right)\,,
	\\
	&A^3\rightarrow A^3+d\chi ^3, C^3\rightarrow C^3+dT^3+\frac{2\pi q}{N_3}\left( -V^1B^2-B^1V^2-\frac{1}{2}V^1dV^2-\frac{1}{2}V^2dV^1 \right) \,,\label{eq_gauge_C3}
\end{align}
where $\chi^3$, $V^i$, $\tilde{V}^i$ and $T^3$ are $0$-form, $1$-form, $1$-form and $2$-form gauge parameters respectively. They satisfy the following compactness conditions: $\int d\chi^3\in2\pi\mathbb{Z}$, $\int dV^i\in2\pi\mathbb{Z}$, $\int d\tilde{V}^i\in2\pi\mathbb{Z}$ and $\int dT^3\in2\pi\mathbb{Z}$, where the integrals are performed over $1$d, $2$d, $2$d and $3$d compact submanifolds respectively.

\subsection{Wilson operators for topological excitations\label{ss22}}
\begin{figure}
	\centering
	\includegraphics[scale=0.9,keepaspectratio]{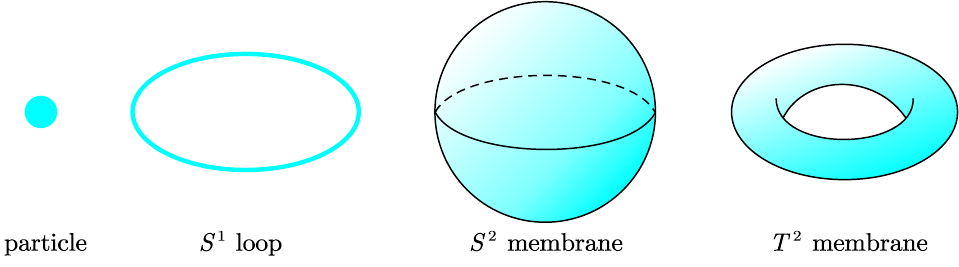}
	\caption{Illustration of particle excitations and spatially extended topological excitations.  We only consider excitations with these simple shapes in the $5$D twisted $BF$ theory with the $BBA$ twisted term.}
	\label{fig_excitation}
\end{figure}

Topological excitations in $5$D spacetime consist of point-like particle excitations, one-dimensional closed loop excitations and two-dimensional closed membrane excitations. We simply call them particles, loops, and membranes unless otherwise specified. Since loops and membranes are spatially extended excitations, they can have different shapes in space such as self-knotted loops, linking loops, or membranes in the shape of a torus with multiple handles. We only consider simple shapes, which are $S^1$ loop, $S^2$ sphere and $T^2$ torus (with genus$=1$), as shown in figure~\ref{fig_excitation}. 

In TQFTs, topological excitations  are represented by Wilson operators. In other words, by respectively using gauge fields of $1$-, $2$-, and $3$-form, we can construct Wilson loop operators for particle excitations, Wilson surface operators for loop excitations, and  Wilson volume operators for membrane excitations. Thus,   the equivalence classes of Wilson operators directly render the equivalence classes of topological excitations, which eventually leads to a finite set of topological excitations despite the infinite number of gauge-invariant Wilson operators that can be written.  Below, we demonstrate the details of some examples of  Wilson operators and leave all nonequivalent Wilson operators in table~\ref{tab_excitation}. In this table, there are $29$ topologically distinct excitations and one of them is the topologically trivial excitation denoted as $\mathsf{1}$.  \textit{It is important to point out that trivial particles, trivial loops, and trivial membranes are all identical to each other and are eventually denoted as $\mathsf{1}$.} We also list all equivalent excitations in different equivalence classes in table~\ref{tab_equivalence_classes}.
\begin{table}
	\caption{\label{tab_excitation}Operators for $29$ nonequivalent excitations in $5$D topological order with action $S=\int{\frac{N_1}{2\pi}\tilde{B}^1dB^1+}\frac{N_2}{2\pi}\tilde{B}^2dB^2+\frac{N_3}{2\pi}C^3dA^3+qB^1B^2A^3$. We define that $f_{1}=-\frac{1}{2}\frac{2\pi q}{N_1}\left( d^{-1}A^3B^2+d^{-1}B^2A^3 \right)$, $f_{2}=-\frac{1}{2}\frac{2\pi q}{N_2}\left( d^{-1}A^3B^1+d^{-1}B^1A^3 \right)$ and $f_{3}=\frac{1}{2}\frac{2\pi q}{N_3}\left( d^{-1}B^1B^2+d^{-1}B^2B^1 \right)$. The meanings of numbers in subscript and superscript are summarized at the end of section~\ref{ss23}. For simplicity, in the subscript of the loops, we rewrite ``$n_1n_2n_3,000$'' and ``$000,n_1n_2n_3$'' as ``$n_1n_2n_3,$'' and ``$,n_1n_2n_3$'' respectively.  In the superscript of membranes, we rewrite ``$n_1n_2n_3;000,000$'' and ``$000;000,n_1n_2n_3$'' as ``$n_1n_2n_3;$'' and ``$,n_1n_2n_3$'' respectively. }
	\centering
	\scriptsize
	\begin{tabular*}{\textwidth}{@{\extracolsep{\fill}}cc|cc}
		\hline
		\small excitation & \small operator for excitation & \small excitation & \small operator for excitation \\
		\hline
		$\mathsf{1}$ & $\exp \left( 0 \right)$ & $\mathsf{M^S}_{001}^{001;}$ & \makecell{\scriptsize$2\exp \left( i\int_{\gamma}{A^3}+i\int_{\omega}{C^3+f_{3}} \right)$ \\ \scriptsize$\times\delta \left( \int_{\sigma}{B^1} \right) \delta \left( \int_{\sigma}{B^2} \right)$}  \\
		\hline
		$\mathsf{P}_{001}$ & $\exp \left( i\int_{\gamma}{A^3} \right)$ & $\mathsf{M^S}_{001}^{,100}$ & \makecell{\scriptsize$4\exp \left( i\int_{\sigma}{\tilde{B}^1+f_{1}}+\int_{\omega}{C^3+f_{3}} \right)$ \\ \scriptsize$\times\delta \left( \int_{\sigma}{B^1} \right) \delta \left( \int_{\sigma}{B^2} \right)\delta \left( \int_{\gamma}{A^3} \right)$}   \\
		\hline
		$\mathsf{L}_{100,}$ & $\exp \left( i\int_{\sigma}{B^1} \right)$ & $\mathsf{M^S}_{001}^{,010}$ & \makecell{\scriptsize$4\exp \left( i\int_{\sigma}{\tilde{B}^2+f_{2}}+\int_{\omega}{C^3+f_{3}} \right)$ \\ \scriptsize$\times\delta \left( \int_{\sigma}{B^1} \right) \delta \left( \int_{\sigma}{B^2} \right)\delta \left( \int_{\gamma}{A^3} \right)$} \\
		\hline
		$\mathsf{L}_{010,}$ & $\exp \left( i\int_{\sigma}{B^2} \right)$ & $\mathsf{M^S}_{001}^{,110}$ & \makecell{\scriptsize$4\exp \left( i\int_{\sigma}{\tilde{B}^1+f_{1}}+i\int_{\sigma}{\tilde{B}^2+f_{2}}+\int_{\omega}{C^3+f_{3}} \right)$ \\ \scriptsize$\times\delta \left( \int_{\sigma}{B^1} \right) \delta \left( \int_{\sigma}{B^2} \right)\delta \left( \int_{\gamma}{A^3} \right)$} \\
		\hline
		$\mathsf{L}_{110,}$ & $\exp \left( i\int_{\sigma}{B^1}+i\int_{\sigma}{B^2} \right)$ & $\mathsf{M^T}_{001}$ & \makecell{\scriptsize$4\exp \left( i\int_{\tilde{\omega}}{C^3+f_{3}} \right)$ \\ \scriptsize$\times\delta \left( \int_{\sigma}{B^1} \right) \delta \left( \int_{\sigma}{B^2} \right)$}\\
		\hline
		$\mathsf{L}_{100,}^{001}$ & $\exp \left( i\int_{\sigma}{B^1}+i\int_{\gamma}{A^3} \right)$ & $\mathsf{M^T}_{001}^{001;}$ & \makecell{\scriptsize$4\exp \left( i\int_{\gamma}{A^3}+i\int_{\tilde{\omega}}{C^3+f_{3}} \right)$ \\ \scriptsize$\times\delta \left( \int_{\sigma}{B^1} \right) \delta \left( \int_{\sigma}{B^2} \right)$} \\
		\hline
		$\mathsf{L}_{010,}^{001}$ & $\exp \left( i\int_{\sigma}{B^2}+i\int_{\gamma}{A^3} \right)$ & $\mathsf{M^T}_{001}^{,100}$ & \makecell{\scriptsize$8\exp \left( i\int_{\sigma}{\tilde{B}^1+f_{1}}+i\int_{\tilde{\omega}}{C^3+f_{3}} \right)$ \\ \scriptsize$\times\delta \left( \int_{\sigma}{B^1} \right) \delta \left( \int_{\sigma}{B^2} \right)\delta \left( \int_{\gamma}{A^3} \right)$} \\
		\hline
		$\mathsf{L}_{110,}^{001}$ & $\exp \left( i\int_{\sigma}{B^1}+i\int_{\sigma}{B^2}+i\int_{\gamma}{A^3} \right)$ & $\mathsf{M^T}_{001}^{,010}$ & \makecell{\scriptsize$8\exp \left( i\int_{\sigma}{\tilde{B}^2+f_{2}}+i\int_{\tilde{\omega}}{C^3+f_{3}} \right)$ \\ \scriptsize$\times\delta \left( \int_{\sigma}{B^1} \right) \delta \left( \int_{\sigma}{B^2} \right)\delta \left( \int_{\gamma}{A^3} \right)$} \\
		\hline
		$\mathsf{L}_{,100}$ & \makecell{\scriptsize$2\exp \left( i\int_{\sigma}{\tilde{B}^1+f_{1}} \right)$\\ \scriptsize$\times\delta \left( \int_{\gamma}{A^3} \right)\delta \left( \int_{\sigma}{B^2} \right)$} & $\mathsf{M^T}_{001}^{,110}$ & \makecell{\scriptsize$8\exp \left( i\int_{\sigma}{\tilde{B}^1+f_{1}}+i\int_{\sigma}{\tilde{B}^2+f_{2}}+i\int_{\tilde{\omega}}{C^3+f_{3}} \right)$ \\ \scriptsize$\times\delta \left( \int_{\sigma}{B^1} \right) \delta \left( \int_{\sigma}{B^2} \right)\delta \left( \int_{\gamma}{A^3} \right)$} \\
		\hline
		$\mathsf{L}_{,010}$ &\makecell{\scriptsize$2\exp \left( i\int_{\sigma}{\tilde{B}^2+f_{2}} \right)$\\ \scriptsize$\times\delta \left( \int_{\gamma}{A^3} \right)\delta \left( \int_{\sigma}{B^1} \right)$} & $\mathsf{M^{ST}}$ & \makecell{\scriptsize$8\exp \left( i\int_{\omega}{C^3+f_{3}}+i\int_{\tilde{\omega}}{C^3+f_{3}} \right)$\\ \scriptsize$\times\delta \left( \int_{\sigma}{B^1} \right) \delta \left( \int_{\sigma}{B^2} \right)$}  \\
		\hline
		$\mathsf{L}_{,110}$ &\makecell{\scriptsize$2\exp \left( i\int_{\sigma}{\tilde{B}^1+f_{1}} + i\int_{\sigma}{\tilde{B}^2+f_{2}} \right)$\\ \scriptsize$\times\delta \left( \int_{\gamma}{A^3} \right)\delta \left( \int_{\sigma}{B^1-B^2} \right)$} & $\mathsf{M^{ST}}^{001;}$ & \makecell{\scriptsize$8\exp \left( i\int_{\gamma}{A^3}+i\int_{\omega}{C^3+f_{3}}+i\int_{\tilde{\omega}}{C^3+f_{3}} \right)$\\ \scriptsize$\times\delta \left( \int_{\sigma}{B^1} \right) \delta \left( \int_{\sigma}{B^2} \right)$} \\
		\hline
		$\mathsf{L}_{100,100}$ &\makecell{\scriptsize$2\exp \left( i\int_{\sigma}{B^1}+i\int_{\sigma}{\tilde{B}^1+f_{1}} \right)$\\ \scriptsize$\times\delta \left( \int_{\gamma}{A^3} \right)\delta \left( \int_{\sigma}{B^2} \right)$} & $\mathsf{M^{ST}}^{,100}$ & \makecell{\scriptsize$16\exp \left( i\int_{\sigma}{\tilde{B}^1+f_{1}}+i\int_{\omega}{C^3+f_{3}}+i\int_{\tilde{\omega}}{C^3+f_{3}} \right)$\\ \scriptsize$\times\delta \left( \int_{\sigma}{B^1} \right) \delta \left( \int_{\sigma}{B^2} \right)\delta \left( \int_{\gamma}{A^3} \right)$} \\
		\hline
		$\mathsf{L}_{010,010}$ &\makecell{\scriptsize$2\exp \left( i\int_{\sigma}{B^2}+i\int_{\sigma}{\tilde{B}^2+f_{2}} \right)$\\ \scriptsize$\times\delta \left( \int_{\gamma}{A^3} \right)\delta \left( \int_{\sigma}{B^1} \right)$} & $\mathsf{M^{ST}}^{,010}$ & \makecell{\scriptsize$16\exp \left( i\int_{\sigma}{\tilde{B}^2+f_{2}}+i\int_{\omega}{C^3+f_{3}}+i\int_{\tilde{\omega}}{C^3+f_{3}} \right)$\\ \scriptsize$\times\delta \left( \int_{\sigma}{B^1} \right) \delta \left( \int_{\sigma}{B^2} \right)\delta \left( \int_{\gamma}{A^3} \right)$} \\
		\hline
		$\mathsf{L}_{100,110}$ &\makecell{\scriptsize$2\exp \left( i\int_{\sigma}{B^1}+i\int_{\sigma}{\tilde{B}^1+f_{1}}+i\int_{\sigma}{\tilde{B}^2+f_{2}} \right)$\\ \scriptsize$\times\delta \left( \int_{\gamma}{A^3} \right)\delta \left( \int_{\sigma}{B^1-B^2} \right)$} & $\mathsf{M^{ST}}^{,110}$ & \makecell{\scriptsize$16\exp \left( i\int_{\sigma}{\tilde{B}^1+f_{1}}+i\int_{\sigma}{\tilde{B}^2+f_{2}}\right)$ \\ \scriptsize$\times\exp \left(i\int_{\omega}{C^3+f_{3}}+i\int_{\tilde{\omega}}{C^3+f_{3}} \right)$\\ \scriptsize$\times\delta \left( \int_{\sigma}{B^1} \right) \delta \left( \int_{\sigma}{B^2} \right)\delta \left( \int_{\gamma}{A^3} \right)$}  \\
		\hline
		$\mathsf{M^S}_{001}$ & \makecell{\scriptsize$2\exp \left( i\int_{\omega}{C^3+f_{3}} \right)$ \\ \scriptsize$\times\delta \left( \int_{\sigma}{B^1} \right)\delta \left( \int_{\sigma}{B^2} \right)$} & $-$ & $-$ \\
		\hline
	\end{tabular*}
\end{table}

\begin{table}
	\caption{\label{tab_equivalence_classes}Equivalence classes in $BBA$ twisted term. All excitations in the same  equivalence class are indistinguishable. For $BBA$ twisted term, we list all $29$ topologically distinct excitations and their equivalent excitations in this table. Any addition that appears in subscript and superscript is addition mod $2$ since all gauge subgroups are $\mathbb{Z}_{2}$. The meanings of numbers in subscript and superscript are summarized at the end of section~\ref{ss23}. For simplicity, in the subscript of the loops, we rewrite ``$n_1n_2n_3,000$'' and ``$000,n_1n_2n_3$'' as ``$n_1n_2n_3,$'' and ``$,n_1n_2n_3$'' respectively. In the superscript of membranes, we rewrite ``$n_1n_2n_3;000,000$'', ``$000;n_1n_2n_3,000$'', ``$000;000,n_1n_2n_3$'', ``$n_1n_2n_3;m_1m_2m_3,000$''' as ``$n_1n_2n_3;$'', ``$;n_1n_2n_3,$'', ``$,n_1n_2n_3$'', ``$n_1n_2n_3;m_1m_2m_3,$'' respectively.}
	\centering
	\begin{tabular*}{\textwidth}{@{\extracolsep{\fill}}cc|cc}
		\hline
		excitation & equivalent excitations & excitation & equivalent excitations \\
		\hline
		$\mathsf{1}$ & $-$ & $\mathsf{M^S}_{001}^{001;}$ &  $\mathsf{M^S}_{001}^{001;n_{1}n_{2}0,}\quad n_{1},n_{2}=0,1$ \\
		\hline
		$\mathsf{P}_{001}$ & $-$ & $\mathsf{M^S}_{001}^{,100}$ & $\mathsf{M^S}_{001}^{00n_{3};n_{1}n_{2}0,100}\quad n_{1},n_{2},n_{3}=0,1$ \\
		\hline
		$\mathsf{L}_{100,}$ & $-$ & $\mathsf{M^S}_{001}^{,010}$ & $\mathsf{M^S}_{001}^{00n_{3};n_{1}n_{2}0,010}\quad n_{1},n_{2},n_{3}=0,1$ \\
		\hline
		$\mathsf{L}_{010,}$ & $-$ & $\mathsf{M^S}_{001}^{,110}$ & $\mathsf{M^S}_{001}^{00n_{3};n_{1}n_{2}0,110}\quad n_{1},n_{2},n_{3}=0,1$ \\
		\hline
		$\mathsf{L}_{110,}$ & $-$ & $\mathsf{M^T}_{001}$ & $\mathsf{M^T}_{001}^{;n_{1}n_{2}0,}\quad n_{1},n_{2}=0,1$ \\
		\hline
		$\mathsf{L}_{100,}^{001}$ & $-$ & $\mathsf{M^T}_{001}^{001;}$ & $\mathsf{M^T}_{001}^{001;n_{1}n_{2}0,}\quad n_{1},n_{2}=0,1$ \\
		\hline
		$\mathsf{L}_{010,}^{001}$ & $-$ & $\mathsf{M^T}_{001}^{,100}$ & $\mathsf{M^T}_{001}^{00n_{3};n_{1}n_{2}0,100}\quad n_{1},n_{2},n_{3}=0,1$ \\
		\hline
		$\mathsf{L}_{110,}^{001}$ & $-$ & $\mathsf{M^T}_{001}^{,010}$ & $\mathsf{M^T}_{001}^{00n_{3};n_{1}n_{2}0,010}\quad n_{1},n_{2},n_{3}=0,1$  \\
		\hline
		$\mathsf{L}_{,100}$ & $\mathsf{L}_{0n_{2}0,100}^{00n_{3}}\quad n_{2},n_{3}=0,1$
		 & $\mathsf{M^T}_{001}^{,110}$ & $\mathsf{M^T}_{001}^{00n_{3};n_{1}n_{2}0,110}\quad n_{1},n_{2},n_{3}=0,1$ \\
		\hline
		$\mathsf{L}_{,010}$ & $\mathsf{L}_{n_{1}00,010}^{00n_{3}}\quad n_{1},n_{3}=0,1$ & $\mathsf{M^{ST}}$ & $\mathsf{M^{ST}}^{;n_{1}n_{2}0,}\quad n_{1},n_{2}=0,1$ \\
		\hline
		$\mathsf{L}_{,110}$ & $\mathsf{L}_{n_{1}n_{1}0,110}^{00n_{3}}\quad n_{1},n_{3}=0,1$ & $\mathsf{M^{ST}}^{001;}$ & $\mathsf{M^{ST}}^{001;n_{1}n_{2}0,}\quad n_{1},n_{2}=0,1$ \\
		\hline
		$\mathsf{L}_{100,100}$ & $\mathsf{L}_{1n_{2}0,100}^{00n_{3}}\quad n_{2},n_{3}=0,1$ & $\mathsf{M^{ST}}^{,100}$ & $\mathsf{M^{ST}}^{00n_{3};n_{1}n_{2}0,100}\quad n_{1},n_{2},n_{3}=0,1$ \\
		\hline
		$\mathsf{L}_{010,010}$ & $\mathsf{L}_{n_{1}10,010}^{00n_{3}}\quad n_{1},n_{3}=0,1$ & $\mathsf{M^{ST}}^{,010}$ & $\mathsf{M^{ST}}^{00n_{3};n_{1}n_{2}0,010}\quad n_{1},n_{2},n_{3}=0,1$ \\
		\hline
		$\mathsf{L}_{100,110}$ & $\mathsf{L}_{(1+n_{1})n_{1}0,110}^{00n_{3}}\quad n_{1},n_{3}=0,1$ & $\mathsf{M^{ST}}^{,110}$ & $\mathsf{M^{ST}}^{00n_{3};n_{1}n_{2}0,110}\quad n_{1},n_{2},n_{3}=0,1$  \\
		\hline
		$\mathsf{M^S}_{001}$ & $\mathsf{M^S}_{001}^{;n_{1}n_{2}0,}\quad n_{1},n_{2}=0,1$ & $-$ & $-$ \\
		\hline
	\end{tabular*}
\end{table}

\textbf{\textit{Wilson operators for particles.}} 
 A particle can only carry gauge charges of 1-form gauge fields $A$ and we generally use $\mathsf{P}_{n_{1}n_{2}\cdots n_{k}}$ to represent a particle simultaneously carrying $n_{1}$, $n_{2}$, $\cdots$ , $n_{k}$ units of $\mathbb{Z} _{N_1}$, $\mathbb{Z} _{N_2}$, $\cdots$, $\mathbb{Z} _{N_k}$ gauge charge. The Wilson operator for such excitation should be constructed by using $k$ gauge fields ${A^1, A^2, \cdots, A^k}$. However, given a specific $BF$ theory, some gauge charges of 1-form gauge fields $A$ may be absent because each gauge subgroup $\mathbb{Z} _{N_i}$ corresponds to one kind of $BF$ term, which is either $\tilde{B}^idB^i$ or $C^idA^i$. If the corresponding $BF$ term is $\tilde{B}^idB^i$, then $A^i$ and $C^i$ are absent, which means $n_i$ can only be 0 for $\mathsf{P}_{n_{1}n_{2}\cdots n_{k}}$.
 
 Consider $BBA$ twisted terms, a particle carrying unit gauge charge of $1$-form gauge field $A^3$ is called a \textit{$\mathbb{Z} _{N_3}$ particle}, which can be represented by the following gauge invariant Wilson operator:
\begin{align}
	\mathsf{P}_{001}=\exp \left( i\int_{\gamma}{A^3} \right) \,,
\end{align}
where $\gamma=S^1$ is the closed world-line of the particle. Since  the action~(\ref{eq_action_BBA}) contains only one $1$-form gauge field, i.e., $A^3$, only $n_3$ can be nonzero. The antiparticle of $\mathsf{P}_{n_{1}n_{2}n_{3}}$ is represented by $\bar{\mathsf{P}}_{n_{1}n_{2}n_{3}}$. For instance, the antiparticle of $\mathsf{P}_{001}$ is represented by $\bar{\mathsf{P}}_{001}$. 
In section~\ref{s3}, we will show that for $G=\left(\mathbb{Z} _{2}\right)^{3}$, the antiparticle of any excitation is itself because the fusion of two identical excitations can be written as $c\cdot\left(\mathsf{1}\oplus\cdots\right)$, where $c$ is just a constant. Also, $n_{i}$ only takes $0$ or $1$ mod $2$ due to the $\mathbb{Z} _{2}$ cyclic group structure, which largely simplifies our analysis.

\textbf{\textit{Wilson operators for loops.}}
Similarly, a loop can carry gauge charges of 2-form gauge fields $B$ and $\tilde{B}$ and we generally use $\mathsf{L}_{n_{1}n_{2}\cdots n_{k},\tilde{n}_{1}\tilde{n}_{2}\cdots\tilde{n}_{k}}$ to represent a loop. Here $n_{1}$, $n_{2}$, $\cdots$ , $n_{k}$ denote $n_{1}$, $n_{2}$, $\cdots$ , $n_{k}$ units of $\mathbb{Z} _{N_1}$, $\mathbb{Z} _{N_2}$, $\cdots$, $\mathbb{Z} _{N_k}$ gauge charge of $B^1$, $B^2$, $\cdots$, $B^k$ fields respectively. $\tilde{n}_{1}$, $\tilde{n}_{2}$, $\cdots$ , $\tilde{n}_{k}$ denote $\tilde{n}_{1}$, $\tilde{n}_{2}$, $\cdots$ , $\tilde{n}_{k}$ units of $\mathbb{Z} _{N_1}$, $\mathbb{Z} _{N_2}$, $\cdots$, $\mathbb{Z} _{N_k}$ gauge charge of $\tilde{B}^1$, $\tilde{B}^2$, $\cdots$, $\tilde{B}^k$ fields respectively. 

In $BBA$ twisted term, a loop carrying unit gauge charge of the $2$-form gauge field $B^1$ ($\tilde{B}^1$) is called a $\mathbb{Z} _{N_1}$ $B$-loop ($\tilde{B}$-loop). The gauge invariant Wilson operator for such a $\mathbb{Z} _{N_1}$ $B$-loop is
\begin{align}
	\mathsf{L}_{100,000}=\exp \left( i\int_{\sigma}{B^1} \right)\,,
\end{align}
where $\sigma=S^1\times S^1=T^2$ is the closed world sheet of the loop excitation.  Such a loop is a \textit{pure loop} that only carries gauge charges of  either one or more $2$-form gauge fields.  In contrast, we can  further  consider a \textit{decorated loop} that simultaneously carries gauge charges of both $1$-form gauge fields and $2$-form gauge fields. For instance, a $\left( \mathbb{Z} _{N_1},\mathbb{Z} _{N_2} \right)$ $B$-loop decorated by a $\mathbb{Z} _{N_3}$ particle is represented by
\begin{align}
	\mathsf{L}_{110,000}^{001}=\exp \left( i\int_{\sigma}{B^1}+i\int_{\sigma}{B^2}+i\int_{\gamma}{A^3} \right)\,.
\end{align}
Here, the superscript ``$001$'' denotes a $\mathbb{Z} _{N_3}$ particle decoration and the subscript ``$110,000$'' means that the loop simultaneously carries unit gauge charges of $B^1$ gauge field and  $B^2$ gauge field. $\gamma$ is an arbitrary closed line on $\sigma$, i.e., $\gamma\in \sigma$. If there is no decoration at all, $\mathsf{L}_{110,000}^{000}$ is a pure loop. For the notational simplicity, we omit 000 in the superscript and subscript, thus $\mathsf{L}_{100,000}$, $\mathsf{L}_{110,000}^{001}$ and $\mathsf{L}_{110,000}^{000}$ can be rewritten as $\mathsf{L}_{100,}$, $\mathsf{L}_{110,}^{001}$ and $\mathsf{L}_{110,}$ respectively. \textit{Note that the comma in the subscript should not be omitted. The numbers on the left side of the comma correspond to the $B$ fields while the numbers on the right side of the comma correspond to the $\tilde{B}$ fields}.

\textbf{\textit{Wilson operators for membranes.}}
As mentioned above, we only consider   membrane excitations of two different shapes: the membrane denoted by $\mathsf{M^S}$ is in the shape of $S^2$ and the membrane denoted by $\mathsf{M^T}$ is in the shape of $T^2$. Both of them can carry gauge charges of 3-form fields $C$ and we generally use $\mathsf{M^S}_{n_{1}n_{2}\cdots n_{k}}$ or $\mathsf{M^T}_{n_{1}n_{2}\cdots n_{k}}$ to represent them. For instance, one may write a Wilson (volume) operator as $
	\mathsf{M^S}_{001}=\exp \left( i\int_{\omega}{C^3} \right)$,  where the world volume $\omega$ is an $S^2\times S^1$ manifold in $5$D spacetime. However, the Wilson operator we construct above is not gauge invariant because $C^3$ is transformed with nontrivial shifts as shown in eq.~(\ref{eq_gauge_C3}). The correct form should be:
\begin{align}
	\mathsf{M^S}_{001}=N\exp \left[ i\int_{\omega}{C^3+\frac{1}{2}\frac{2\pi q}{N_3}\left( d^{-1}B^1B^2+d^{-1}B^2B^1 \right)} \right] \delta \left( \int_{\sigma}{B^1} \right) \delta \left( \int_{\sigma}{B^2} \right) \,,
	\label{eq_op_sphere_001}
\end{align}
where $N=2$ is a normalization factor required by several principles based on physical intuition. Details of deriving normalization factors are shown in appendix~\ref{ap_normalization_factors}. We also find that the normalization factor of an excitation equals its quantum dimension in section~\ref{section_quantum_dimension}. We define $d^{-1}B^1$ and $d^{-1}B^2$ as
$d^{-1}B^1=\int_{\mathcal{A} \in \omega}{B^1} \quad\text{and}\quad d^{-1}B^2=\int_{\mathcal{A} \in \omega}{B^2}$, where $\mathcal{A}$ is an open area on $\omega$. As 1-form fields, $d^{-1}B^1$ and $d^{-1}B^2$ are well defined on $\omega$ if and only if $B^1$ and $B^2$ are exact on $\omega$, i.e., $B^1$ and $B^2$ satisfy extra constraints:  $\int_{\sigma}{B^1}=0$ mod $2\pi$, $\int_{\sigma}{B^2}=0$ mod $2\pi$, where $\sigma$ is an arbitrary closed surface on $\omega$. To enforce these constraints, we introduce two  delta functionals in eq.~(\ref{eq_op_sphere_001}):
\begin{align}
	\delta \left( \int_{\sigma}{B^1} \right) =\begin{cases}
		1, \quad \int_{\sigma}{B^1}=0   \,\, \mathrm{mod}\,\, 2\pi\\
		0, \quad \mathrm{else}\\
	\end{cases}\,,\,\,
	\delta \left( \int_{\sigma}{B^2} \right) =\begin{cases}
		1, \quad \int_{\sigma}{B^2}=0 \,\,  \mathrm{mod} \,\, 2\pi\\
		0, \quad \mathrm{else}\\
	\end{cases}\,.
\end{align}
Similarly, the operator for  a $\mathbb{Z} _{N_3}$ membrane of $T^2$ shape can be written as:
\begin{equation}
	\mathsf{M^T}_{001}=\tilde{N}\exp \left[ i\int_{\tilde{\omega}}{C^3+\frac{1}{2}\frac{2\pi q}{N_3}\left( d^{-1}B^1B^2+d^{-1}B^2B^1 \right)} \right] \delta \left( \int_{\sigma}{B^1} \right) \delta \left( \int_{\sigma}{B^2} \right) \,,
\end{equation}
where $\tilde{\omega}$ is a $T^2\times S^1$ manifold in $5$D spacetime, and in appendix~\ref{ap_normalization_factors} we will show that  $\tilde{N}=4$ due to the hierarchical shrinking structure. Symbolically, once we write down the Wilson operator for an $S^2$ membrane, we can formally express  the Wilson operator for a $T^2$ membrane  simply through replacing $\mathsf{M^S}$, $N$ and $\omega=S^2\times S^1$ by $\mathsf{M^T}$, $\tilde{N}$ and $\tilde{\omega}=T^2\times S^1$ respectively. 

Just like loop excitations that can be decorated by particles, membranes can   be decorated by particles and loops simultaneously. For instance, a $\mathbb{Z} _{N_3}$ $S^2$ membrane decorated by a $\mathbb{Z} _{N_3}$ particle and a $\mathbb{Z} _{N_1}$ $B$-loop is
\begin{multline}
	\mathsf{M^S}_{001}^{001;100,000}=2\exp \left[ i\int_{\gamma}{A^3}+i\int_{\sigma}{B^1}+i\int_{\omega}{C^3+\frac{1}{2}\frac{2\pi q}{N_3}\left( d^{-1}B^1B^2+d^{-1}B^2B^1 \right)} \right]\\ \times\delta \left( \int_{\sigma}{B^1} \right) \delta \left( \int_{\sigma}{B^2} \right) \,.
	\label{eq2}
\end{multline}
\textit{In the superscript, the numbers on the left side of the semicolon, between the semicolon and the comma, on the right side of the comma denote particle decorations, $B$-loop decorations, $\tilde{B}$-loop decorations respectively}. In eq.~(\ref{eq2}), the number  $001$ and $100,000$ in the superscript respectively denote a $\mathbb{Z} _{N_3}$ particle  decoration and a $\mathbb{Z} _{N_1}$ $B$-loop decoration respectively. The subscript $001$ means that the membrane carries a unit gauge charge of $3$-form gauge field $C^3$. $\gamma$ and $\sigma$ are arbitrary closed line and closed surface on $\omega$, thus we can always choose $\gamma\in\sigma\in\omega$. $\delta \left( \int_{\sigma}{B^1} \right)$ in eq.~(\ref{eq2}) plays an important role because it induces an equivalence relation between $\mathsf{M^S}_{001}^{001;100,000}$ and $\mathsf{M^S}_{001}^{001;000,000}$, which will be  discussed in section~\ref{ss23}. If there is no particle decoration at all, $\mathsf{M^S}_{001}^{000;100,000}$ and $\mathsf{M^S}_{001}^{000;000,100}$ can be  rewritten as $\mathsf{M^S}_{001}^{;100,}$ and $\mathsf{M^S}_{001}^{,100}$ respectively for the notational simplicity.  If there is no loop decoration at all, $\mathsf{M^S}_{001}^{001;000,000}$ can be  rewritten as $\mathsf{M^S}_{001}^{001;}$ for the notational simplicity. If all decorations are absent, the pure membrane $\mathsf{M^S}_{001}^{000;000,000}$ is directly replaced by $\mathsf{M^S}_{001}$.

\begin{figure}
	\centering
	\includegraphics[scale=0.85,keepaspectratio]{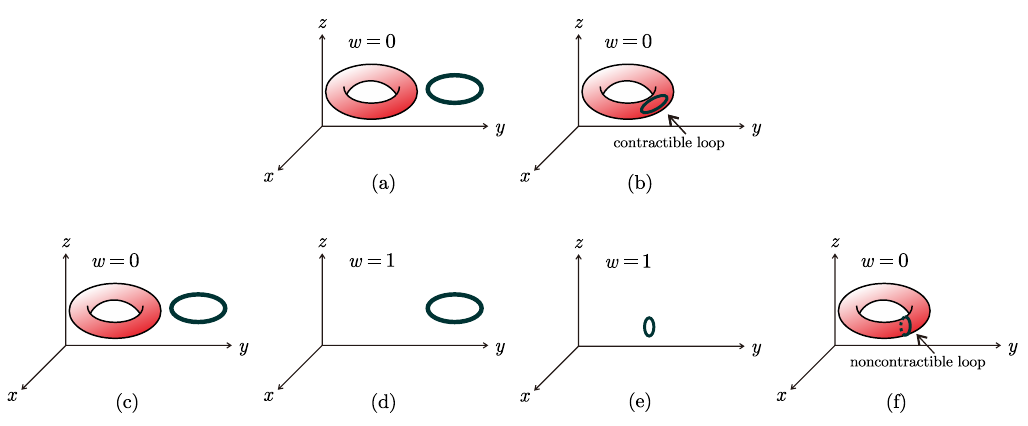}
	\caption{Illustration of constructing loop decorated $T^2$ membrane from fusion. (a) A $T^2$ membrane and a loop in $xyz$-space with $w=0$. The membrane is assumed to be static. (b) The loop moves to a contractible circle on the $T^2$ membrane in $xyz$-space with $w=0$. (c) The same initial state as shown in (a). (d) The loop moves to $xyz$-space with $w=1$ along $w$ axis. (e) The loop moves in $xyz$-space with $w=1$. (f) The loop moves back to $xyz$-space with $w=0$ along $w$ axis. Now we see that the loop decoration is placed on a non-contractible circle without cutting the $T^2$ membrane. Since (b) and (f) are both legitimate fusion results, the loop decoration can be adiabatically moved from one circle to another circle. Thus we conclude that the excitations shown in (b) and (f) are equivalent to each other. For loop decorations on other nonequivalent circles, we have similar arguments.}
	\label{fig_loop_decoration}
\end{figure}

Notice that in the above discussion, we do not have to consider how a loop decoration is placed on the $S^2$ membrane because all circles on an $S^2$ sphere can be smoothly deformed to each other. However, a $T^2$ torus has nonequivalent circles and these circles can not be smoothly deformed to each other. One may wonder whether loop decoration along different circles corresponds to nonequivalent excitations. We conclude that $T^2$ membranes with loop decoration along different circles are equivalent excitations. This can be understood from figure~\ref{fig_loop_decoration}. We consider fusing a pure $T^2$ membrane and a loop to create a loop decorated $T^2$ membrane (general discussion for fusion rules are shown in section~\ref{general_fusion}). Since the space is $4$d here, we assume that the membrane and the loop both live in $xyz$-space with $w=0$ at the beginning. When we consider fusing them, we can directly move the loop to a contractible circle on the membrane in $xyz$-space. However this is not the only legitimate way to fuse the $T^2$ membrane and the loop. Alternatively, we can first let the loop move a distance along the $w$ axis, then we move the loop in  $xyz$-space with $w=1$, and finally we move the loop back to $xyz$-space with $w=0$ along $w$ axis. The latter fusing process can lead to a $T^2$ membrane with loop decoration along a non-contractible circle. Since fusion is an adiabatic process, all legitimate fusion results of $2$ given excitations are equivalent to each other. Thus we conclude that placing loop decoration along different circles only leads to equivalent excitations. \textit{We only consider loop decoration along a contractible circle in the following discussion}.

\begin{figure}
	\centering
	\includegraphics[scale=0.8,keepaspectratio]{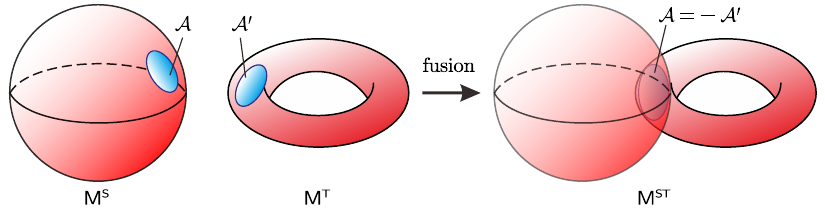}
	\caption{The shape of an $\mathsf{M^{ST}}$-excitation. An $\mathsf{M^{ST}}$-excitation can be seen as the fusion result of an $S^2$ membrane and a $T^2$ membrane. We choose open areas $\mathcal{A}$ and $\mathcal{A}^{\prime}$ on the $S^2$ and the $T^2$ membrane respectively. Then we let $\mathcal{A}$ overlaps $\mathcal{A}^{\prime}$ to glue the $S^2$ and the $T^2$ membrane together. This process defines the fusion between the $S^2$ and the $T^2$ membrane. On the resulting $\mathsf{M^{ST}}$-excitation we have $\mathcal{A}=-\mathcal{A}^{\prime}$ since $\mathcal{A}$ and $\mathcal{A}^{\prime}$ have opposite orientation.}
	\label{fig_MST}
\end{figure}

\textbf{\textit{Wilson operators for $\mathsf{M^{ST}}$-excitations.}}
We have already introduced two kinds of membranes: $S^2$ membranes and $T^2$ membranes. For the completeness of fusion rules to be discussed shortly, it is vital to incorporate another type of membrane formed by fusing an $S^2$ membrane and a $T^2$ membrane. The resulting object is no longer a manifold. As shown in figure~\ref{fig_MST}, geometrically this object can be understood as directly gluing an $S^2$ membrane and a $T^2$ membrane together. The corresponding Wilson operator is denoted as $\mathsf{M^{ST}}$:
\begin{align}
	\mathsf{M^{ST}}=&8\exp \left[ i\int_{\omega}{C^3+\frac{1}{2}\frac{2\pi q}{N_3}\left( d^{-1}B^1B^2+d^{-1}B^2B^1 \right)} \right] \nonumber
	\\
	&\times \exp \left[ i\int_{\tilde{\omega}}{C^3+\frac{1}{2}\frac{2\pi q}{N_3}\left( d^{-1}B^1B^2+d^{-1}B^2B^1 \right)} \right]  \delta \left( \int_{\sigma}{B^1} \right) \delta \left( \int_{\sigma}{B^2} \right)\,,
\end{align}
where $\sigma$ is an arbitrary closed surface on $\omega$ and $\tilde{\omega}$. An $\mathsf{M^{ST}}$-excitation can also have particle and loop decorations. We use the superscript introduced in $S^2$ membranes (see below eq.~(\ref{eq2})) to denote these decorations. All loop decorations are placed on a contractible circle on $\mathsf{M^{ST}}$-excitation.

\subsection{Equivalence classes of Wilson operators\label{ss23}}
In the above discussion, we mentioned that a   $\mathbb{Z} _{N_3}$ membrane that is decorated by a $\mathbb{Z} _{N_3}$ particle and a $\mathbb{Z} _{N_1}$ $B$-loop can be represented by eq.~(\ref{eq2}), and it is equivalent to a   $\mathbb{Z} _{N_3}$ membrane  that is only decorated by a $\mathbb{Z} _{N_3}$ particle, due to the constraint $\delta \left( \int_{\sigma}{B^1} \right)$: 
\begin{multline}
\mathsf{M^S}_{001}^{001;}\!\!=\!\!2\exp \left[ i\int_{\gamma}{A^3}+i\int_{\omega}{C^3+\frac{1}{2}\frac{2\pi q}{N_3}\!\left( d^{-1}B^1B^2\!+\!d^{-1}B^2B^1 \right)} \right] \!\!\delta \left( \int_{\sigma}{B^1} \right) \!\!\delta \left( \int_{\sigma}{B^2} \right) .\!\!
\end{multline}
To simplify, we omit $000,000$ and use $\mathsf{M^S}_{001}^{001;}$ to represent $\mathsf{M^S}_{001}^{001;000,000}$. We say $\mathsf{M^S}_{001}^{001;100,}$ and $\mathsf{M^S}_{001}^{001;}$ are equivalent to each other because they always give the same answers when we calculate any gauge-invariant correlation functions in the path integral. More concretely, the correlation function of $\mathsf{M^S}_{001}^{001;100,}$ and any operator $\mathcal{O}$ can be written as
\begin{align}
	\langle \mathcal{O} \mathsf{M^S}_{001}^{001;100,} \rangle =\frac{1}{\mathcal{Z}}\int\mathcal{D} \left[ ABC \right] \exp \left( iS \right) \times \mathcal{O} \times \mathsf{M^S}_{001}^{001;100,}\,,
\end{align}
where $\mathcal{Z} =\int\mathcal{D} \left[ABC\right] \exp \left( iS \right) $ is the partition function and $\mathcal{D}[ABC]$ denotes configurations of all gauge fields $A^3,B^1,\tilde{B}^1,B^2,\tilde{B}^2,C^3$. The constraint $\delta \left( \int_{\sigma}{B^1} \right) $ means that in the functional integration, the integration over $B^1$ is constrained within the subspace in which $B^1$ is flat, ensuring that $\exp \left( i\int_{\sigma}{{B}^1} \right)=1 $. Immediately, one can realize that $\langle \mathcal{O} \mathsf{M^S}_{001}^{001;100,} \rangle=\langle \mathcal{O} \mathsf{M^S}_{001}^{001;} \rangle$. 
Since $\mathcal{O}$ is an arbitrary operator, $\mathsf{M^S}_{001}^{001;100,}$ and $\mathsf{M^S}_{001}^{001;}$ are indistinguishable in all gauge-invariant observables. In this sense, we conclude that these two operators as well as the associated two topological excitations  are said to  belong to the same equivalence class. Similarly, noticing that $\delta \left( \int_{\sigma}{{B}^2} \right) $ enforces $\exp \left( i\int_{\sigma}{{B}^2} \right)=1$ in the functional integration, we can establish another equivalence class between  $\mathsf{M^S}_{001}^{001;010,}$ and $\mathsf{M^S}_{001}^{001;110,}$. In conclusion, we have
$	\mathsf{M^S}_{001}^{001;}=\mathsf{M^S}_{001}^{001;100,}=\mathsf{M^S}_{001}^{001;010,}=\mathsf{M^S}_{001}^{001;110,}\,$,
where ``$=$'' simply means that they are indistinguishable in all topological data and thus belong to the same equivalence class. 

Exhausting all these equivalence classes results in $29$ topologically distinct excitations, each belonging to its own unique equivalence classes. The details are listed in table~\ref{tab_equivalence_classes}. For loops, the superscript indicates gauge charges of $A$ fields (i.e., particle decorations), while the left and right sides of the comma in the subscript represent gauge charges of $B$ and $\tilde{B}$ fields, respectively. For membranes, numbers in the subscript denote gauge charges of $C$ fields, and the numbers on the left and right sides of the semicolon and comma represent particle and $\tilde{B}$-loop decorations. It's important to note that decorating ${B}$-loop decorations for membranes is considered trivial, as it leads to equivalent membranes, and thus, numbers between the semicolon and comma can be omitted for simplicity.

\section{Fusion rules for the $BBA$ twisted term\label{s3}}
\subsection{Fusion rules and the path-integral representation\label{general_fusion}}
\begin{figure}
	\centering
	\includegraphics[scale=0.9,keepaspectratio]{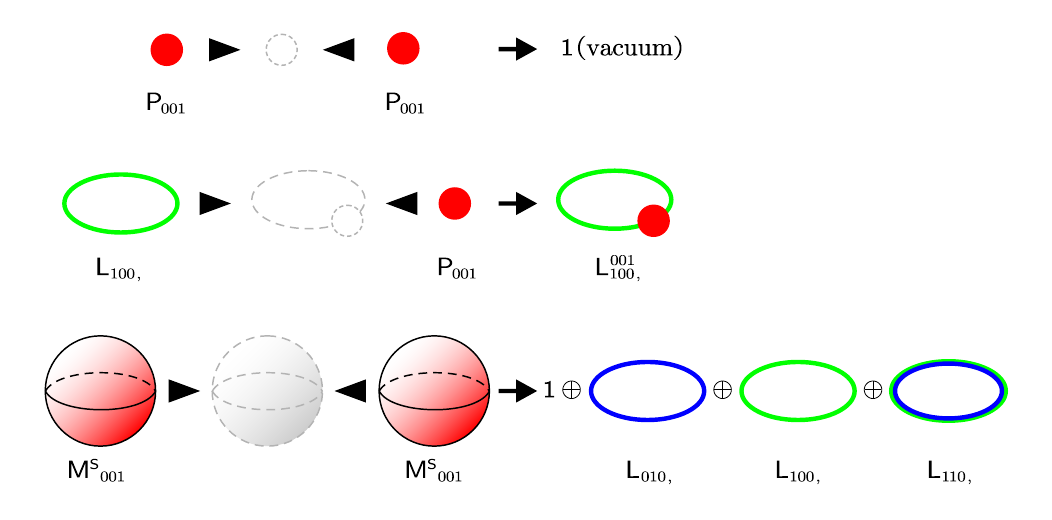}
	\caption{Typical fusion rules for the $5$D twisted $BF$ theory with the $BBA$ twisted term. Fusing two $\mathsf{P}_{001}$'s leads to a  vacuum (trivial excitation).  Fusing a $\mathsf{L}_{100,}$ and a $\mathsf{P}_{001}$ leads to a  decorated $B$-loop $\mathsf{L}_{100,}^{001}$. Fusing two $\mathsf{M^S}_{001}$'s  leads to a direct sum of a  vacuum and three different $B$-loops denoted by $\mathsf{L}_{100,}$, $\mathsf{L}_{010,}$ and $\mathsf{L}_{110,}$. The first two fusion rules shown in this figure are Abelian fusion rules since each of them only contains a single fusion channel. The  last fusion rule is non-Abelian fusion  since it has multiple fusion channels.}
	\label{fig_fusion}
\end{figure}

Fusion rules are crucial properties in topological orders. Figure~\ref{fig_fusion} illustrates several prototypical fusion processes, and detailed calculations of these fusion rules can be found in section~\ref{ss31}.
A general fusion process can be symbolically written as:
\begin{align}
	\mathsf{a}\otimes \mathsf{b}=\oplus _{\mathsf{c}}N_{\mathsf{c}}^{\mathsf{a}\mathsf{b}}\mathsf{c}
	\label{eq_general_fusion}\,,
\end{align}
where $\mathsf{a}$, $\mathsf{b}$ and $\mathsf{c}$ denote topological excitations. $N_{\mathsf{c}}^{\mathsf{a}\mathsf{b}}\in\Z$ is  fusion coefficient.  In the condensed matter literature, a fusion process is classified as Abelian if there is only one excitation $\mathsf{c}$ such that $N_{\mathsf{c}}^{\mathsf{a}\mathsf{b}}$ is nonzero, represented as $\mathsf{a}\otimes \mathsf{b}=\mathsf{c}$. In contrast, a fusion process is deemed non-Abelian if $\mathsf{a}\otimes \mathsf{b}=\mathsf{c}_1\oplus\mathsf{c}_2\oplus \cdots$, where there are multiple fusion channels. We adopt this terminology throughout the paper.  For an excitation $\mathsf{a}$, if $\mathsf{a}\otimes \mathsf{b}$ is always Abelian for any $\mathsf{b}$, we call $\mathsf{a}$ an Abelian excitation. Otherwise, we call $\mathsf{a}$ a non-Abelian excitation.
In this paper, we represent topological excitations using Wilson operators. In this context, the fusion process (\ref{eq_general_fusion}) can be understood as the equivalence of expectation values: 
\begin{align}
	\langle \mathcal{O}_\mathsf{a}\otimes \mathcal{O}_\mathsf{b} \rangle=\langle \oplus _{\mathsf{c}}N_{\mathsf{c}}^{\mathsf{a}\mathsf{b}}\mathcal{O}_\mathsf{c} \rangle\,.
\end{align}
In the path-integral representation, the above identity is represented by:   
\begin{align}
	 &\frac{1}{\mathcal{Z}}\int\mathcal{D} \left[ ABC\right] \exp \left( iS \right) \times \left( \mathcal{O} _{\mathsf{a}}\times \mathcal{O} _{\mathsf{b}} \right) =\frac{1}{\mathcal{Z}}\int \mathcal{D} \left[ ABC\right] \exp \left( iS \right) \times \left( \sum_\mathsf{c}{N_{\mathsf{c}}^{\mathsf{a}\mathsf{b}}}\mathcal{O} _{\mathsf{c}} \right)  \,.
	 	\label{eq_fusion_decomposition}
\end{align}
If the  decomposition (\ref{eq_fusion_decomposition}) is successful, we can obtain all fusion coefficients. This formula also verifies that: 
\begin{align}
	&\mathsf{a}\otimes \mathsf{b}=\mathsf{b}\otimes \mathsf{a}\,, \left(\mathsf{a}\otimes \mathsf{b}\right)\otimes \mathsf{c}=\mathsf{a}\otimes\left(\mathsf{b}\otimes \mathsf{c}\right)\,,\mathsf{a}\oplus \mathsf{b}=\mathsf{b}\oplus \mathsf{a}\,,\left(\mathsf{a}\oplus \mathsf{b}\right)\oplus \mathsf{c}=\mathsf{a}\oplus\left(\mathsf{b}\oplus \mathsf{c}\right)\,.
	\label{eq_fusion_algebra_rule}
\end{align} 
\subsection{Examples\label{ss31}}
In the following, through several examples, we  concretely compute the fusion rules of the twisted $BF$ theory with the $BBA$ twisted term whose action is given in eq.~(\ref{eq_action_BBA}).

\textbf{\textit{Fusing a $\mathbb{Z} _{N_3}$ particle and a $\mathbb{Z} _{N_3}$ particle.}}
Using eq.~(\ref{eq_fusion_decomposition}), we can calculate that
\begin{align}
	\langle \mathsf{P}_{001}\otimes \mathsf{P}_{001} \rangle 
	=&\frac{1}{\mathcal{Z}}\int{\mathcal{D} \left[ ABC   \right] \exp \left( iS \right) \exp \left( i\int_{\gamma}{A^3}+i\int_{\gamma}{A^3} \right) } \nonumber
	\\
	=&\frac{1}{\mathcal{Z}}\int{\mathcal{D} \left[ ABC   \right]  \exp \left( iS \right) \exp \left( i2\int_{\gamma}{A^3} \right) }  \,.
\end{align}
Integrating out Lagrange's multipliers, we obtain constraints for $A^3$, $B^1$, and $B^2$ respectively:
$\int{A^3}=\frac{2\pi m_3}{N_3}  \,,
\int{B^1}=\frac{2\pi m_1}{N_1}  \,,
\int{B^2}=\frac{2\pi m_2}{N_2} $, 
where $m_{1,2,3}\in\mathbb{Z}$. Since $N_1=N_2=N_3=2$, we have $\exp \left( i2\int_{\gamma}{A^3} \right) =1$, thus
$	\langle\mathsf{P}_{001}\otimes \mathsf{P}_{001} \rangle =1=\langle\mathsf{1} \rangle$,
i.e., 
$	\mathsf{P}_{001}\otimes \mathsf{P}_{001}=\mathsf{1}\,$.  This is an Abelian fusion process. The result indicates that fusing two $\mathsf{P}_{001}$'s leads to the trivial excitation. It also   means the antiparticle of $\mathsf{P}_{001}$ is itself, which is   consistent with the group structure of $\mathbb{Z}_2$ cyclic group.

\textbf{\textit{Fusing a $\mathbb{Z} _{N_3}$ particle and a $\mathbb{Z} _{N_1}$ $B$-loop.}}
In the path-integral representation, fusion of a $\mathbb{Z} _{N_3}$ particle and a $\mathbb{Z} _{N_1}$ $B$-loop is represented by:
\begin{align}
	&	\langle\mathsf{P}_{001}\otimes \mathsf{L}_{100,} \rangle  =\frac{1}{\mathcal{Z}}\int{\mathcal{D} \left[ ABC \right] \exp \left( iS \right) \exp \left( i\int_{\gamma}{A^3}+i\int_{\sigma}{B^1} \right)} =\langle \mathsf{L}_{100,}^{001} \rangle\,.
\end{align}
This is also an Abelian fusion process. This result tells us that a $\mathbb{Z} _{N_3}$ particle decorated $\mathbb{Z} _{N_1}$ $B$-loop can be constructed by fusing a $\mathbb{Z} _{N_3}$ particle and a $\mathbb{Z} _{N_1}$ $B$-loop.

\textbf{\textit{Fusing a $\mathbb{Z} _{N_1}$ $\tilde{B}$-loop and a $\mathbb{Z} _{N_1}$ $\tilde{B}$-loop.}}
Fusing a $\mathbb{Z} _{N_1}$ $\tilde{B}$-loop and a $\mathbb{Z} _{N_1}$ $\tilde{B}$-loop is 
\begin{align}
	&\langle \mathsf{L}_{,100}\otimes \mathsf{L}_{,100}\rangle \nonumber \\
	=&\frac{1}{\mathcal{Z}}\!\!\int{\!}\!\mathcal{D} \!\left[ ABC \right] \!\exp \left( iS \right) \nonumber
	\\
	&\times 2\exp \left[ i\int_{\sigma}{\tilde{B}^1-\frac{1}{2}\frac{2\pi q}{N_1}\left( d^{-1}A^3B^2+d^{-1}B^2A^3 \right)} \right] \delta \left( \int_{\sigma}{B^2} \right) \delta \left( \int_{\gamma}{A^3} \right) \nonumber
	\\
	&\times 2\exp \left[ i\!\int_{\sigma}{\tilde{B}^1-\frac{1}{2}\frac{2\pi q}{N_1}\left( d^{-1}A^3B^2+d^{-1}B^2A^3 \right)} \right] \delta \left( \int_{\sigma}{B^2} \right) \delta \left( \int_{\gamma}{A^3} \right) \,.
\end{align}
Integrating out Lagrange's multipliers, i.e., $\tilde{B}^1$, $\tilde{B}^2$, and $C^3$, we obtain $\int_{\sigma}{\hat{B}^1}=\frac{2\pi m_1}{N_1}$, $\int_{\sigma}{\hat{B}^2}=\frac{2\pi m_2}{N_2}$, $\int_{\omega}{\hat{A}^3}=\frac{2\pi m_3}{N_3}$ with $m_1,m_2,m_3\in\Z$, where we add a hat for each gauge field to specify the configurations after the Lagrange's multipliers are integrated. As a result, 
\begin{align}
	\!\!\!\!\!\!	\!\!\langle \mathsf{L}_{,100}\otimes \mathsf{L}_{,100}\rangle\!=&\frac{1}{\mathcal{Z}_{\rm eff}}   \!\!\sum_{\hat{B}^1,\hat{B}^2,\hat{A}^3}\!\! \bigg\{e^{iS_{\rm eff}}  \!\!
	\times \!\!4\exp\! \left[ -i2\int_{\sigma}{\frac{1}{2}\frac{2\pi}{N_1}\frac{pN_1N_2N_3}{\left( 2\pi \right) ^2N_{123}}\left( d^{-1}\hat{A}^3\hat{B}^2+d^{-1}\hat{B}^2\hat{A}^3 \!\right)} \!\right] \nonumber
	\\
	&\times \delta \left( \int_{\sigma}{\hat{B}^2} \right) \delta \left( \int_{\gamma}{\hat{A}^3} \right)     \bigg\}\,,
	\label{eq_illustrate_loop_nonAbelianfusion}
\end{align}
where two redundant delta functionals have been dropped as one delta functional  is sufficient for each constraint. The summation $\sum_{\hat{B}^1,\hat{B}^2,\hat{A}^3}$ denotes the functional integration over  the configurations after integrating Lagrange's multipliers. The effective partition function $\mathcal{Z}_{\rm eff}=\sum_{\hat{B}^1,\hat{B}^2,\hat{A}^3}\exp(iS_{\rm eff})$ with the effective action: 
\begin{align}
	S_{\rm eff}= \int{\frac{pN_1N_2N_3}{\left( 2\pi \right) ^2N_{123}}\hat{B}^1\hat{B}^2\hat{A}^3}\,.
\end{align}
In deriving eq.~(\ref{eq_illustrate_loop_nonAbelianfusion}), the exponential term ``$\exp\left(i2\times\int_{\sigma} \tilde{B}^1\right)$'' has been integrated. We define $d^{-1}\hat{B}^2$ as $d^{-1}\hat{B}^2=\int_{\mathcal{A} \in \sigma}{\hat{B}^2}$ where $\mathcal{A}$ is an open area on $\sigma$($\sigma$ should be replace by $\omega$ if we consider membranes here). Since $\int_{\sigma}{\hat{B}^2}=\frac{2\pi m_2}{N_2}$, $\int_{S^1}{d^{-1}\hat{B}^2}=\int_{S^1}{\int_{\mathcal{A} \in \sigma}{\hat{B}^2}}=\frac{2\pi k_2}{N_2}$ with $k_2$ being an integer and there exists another integer $k_{2}^{\prime}$ such that $k_2+k_{2}^{\prime}=m_2$, we have 
$	\int_{\sigma}{d^{-1}\hat{B}^2\hat{A}^3}=\int_{S^1}{d^{-1}\hat{B}^2}\times \int_{\gamma}{\hat{A}^3}=\frac{2\pi k_2}{N_2}\frac{2\pi m_3}{N_3}
$. Likewise, we define $d^{-1}\hat{A}^3$ as $d^{-1}\hat{A}^3=\int_{\left[a,b\right] \in \gamma}{\hat{A}^3}=\frac{2\pi k_3}{N_3}$ where $\left[a,b\right]$ is an open interval on $\gamma$, $k_3$ and $k_{3}^{\prime}$ are integers satisfying $k_3+k_{3}^{\prime}=m_3$. Thus, $\int_{\sigma}{d^{-1}\hat{A}^3\hat{B}^2}=d^{-1}\hat{A}^3\int_{\sigma}{\hat{B}^2}=\frac{2\pi k_3}{N_3}\frac{2\pi m_2}{N_2}$. According to the discussion in Appendix \ref{ap1},  $\exp \left[ -i2\times \int_{\sigma}{\frac{1}{2}\frac{2\pi q}{N_1}\left( d^{-1}\hat{A}^3\hat{B}^2+d^{-1}\hat{B}^2\hat{A}^3\! \right)} \right]=1$ can also be dropped.
By further noting that the delta functionals in eq.~(\ref{eq_illustrate_loop_nonAbelianfusion}) can be expanded via Fourier-like transformations:
\begin{gather}
	\delta \left( \int_{\sigma}{\hat{B}^2} \right) =\delta \left( \frac{2\pi m_2}{N_2} \right) =\frac{1}{2}\left[ 1+\exp \left( \frac{i2\pi m_2}{2} \right) \right] \,, \nonumber 
	\\
	\delta \left( \int_{\gamma}{\hat{A}^3} \right) =\delta \left( \frac{2\pi m_3}{N_3} \right) =\frac{1}{2}\left[ 1+\exp \left( \frac{i2\pi m_3}{2} \right) \right] \,,
\end{gather}
we finally    obtain the following relation between fusion input and outcome:
\begin{align}
	&\langle \mathsf{L}_{,100}\otimes \mathsf{L}_{,100}\rangle \,\, \nonumber
	\\
	=&\frac{1}{\mathcal{Z} _{\mathrm{eff}}}\!\sum_{\hat{B}^1,\hat{B}^2,\hat{A}^3}{\!}\exp \!\left( iS_{\mathrm{eff}} \right) \!\left[ \!1+\exp \left( i\int_{\sigma}{\hat{B}^2} \right) +\exp \left( i\int_{\gamma}{\hat{A}^3} \right) +\exp \left( i\int_{\sigma}{\hat{B}^2}+i\int_{\gamma}{\hat{A}^3} \right) \! \right] \nonumber
	\\
	\!\!=&\frac{1}{\mathcal{Z}}\!\!\int{\!}\!\mathcal{D} \!\left[ ABC \right] \!\exp \left( iS \right) \!\!\left[ 1\!+\!\exp \left( i\int_{\sigma}{\!}B^2 \right) \!+\!\exp \left( \!i\int_{\gamma}{\!}A^3\! \right) \!+\!\exp \left( i\int_{\sigma}{\!}B^2+i\int_{\gamma}{\!}A^3 \right) \right] \nonumber
	\\
	=&\langle 1\oplus \mathsf{P}_{001}\oplus \mathsf{L}_{010,}\oplus \mathsf{L}_{010,}^{001}\rangle \label{equation_L,100L,100}\,.
\end{align}
This is a non-Abelian fusion with multiple fusion outcomes. Fusing two identical $\mathbb{Z} _{N_1}$ $\tilde{B}$-loops renders   one  vacuum (i.e., trivial excitation), one $\mathbb{Z} _{N_3}$ particle, one $\mathbb{Z} _{N_2}$ $B$-loop, and one $\mathbb{Z} _{N_3}$ particle decorated $\mathbb{Z} _{N_2}$ $B$-loop.

\textbf{\textit{Fusing  a $\mathbb{Z} _{N_1}$ $B$-loop and a $\mathbb{Z} _{N_3}$ membrane.}}
Fusion of a $\mathbb{Z} _{N_1}$ $B$-loop and a $\mathbb{Z} _{N_3}$ membrane is
\begin{align}
	&	\langle \mathsf{L}_{100,}\otimes \mathsf{M^S}_{001} \rangle \nonumber\\
	=&\frac{1}{\mathcal{Z}}\int\!\! \mathcal{D} \!\left[ ABC \right]\! \exp \left( iS \right) \times2\exp \left( i\int_{\sigma}{B^1}+i\int_{\omega}{C^3+f} \right)\delta \left( \int_{\sigma}{B^1} \right) \delta \left( \int_{\sigma}{B^2} \right) \nonumber
	\\
	=&\frac{1}{\mathcal{Z}}\int\!\! \mathcal{D} \!\left[ ABC \right]\!   \exp \left( iS \right) \times 2\exp \left( i\int_{\omega}{C^3+f} \right)\delta \left( \int_{\sigma}{B^1} \right) \delta \left( \int_{\sigma}{B^2} \right) =\langle \mathsf{M^S}_{001} \rangle \,,
\end{align}
where $f=\frac{1}{2}\frac{2\pi q}{N_3}\left( d^{-1}B^1B^2+d^{-1}B^2B^1 \right) $.
This is reasonable because $\mathsf{M^S}_{001}$ is equivalent to $\mathsf{M^S}_{001}^{;100,}$ due to the delta functional $\delta \left( \int_{\sigma}{B^1} \right) $. Thus, $\mathsf{L}_{100,}$ behaves trivially on $\mathsf{M^S}_{001}$.

\textbf{\textit{Fusing a $\mathbb{Z} _{N_3}$ membrane and a $\mathbb{Z} _{N_3}$ membrane.}}
Fusion of a $\mathbb{Z} _{N_3}$ membrane and a $\mathbb{Z} _{N_3}$ membrane is
\begin{align}
	&\langle \mathsf{M^S}_{001}\otimes \mathsf{M^S}_{001} \rangle  \nonumber\\
	=& \frac{1}{\mathcal{Z}}\!\!\int\!\! \mathcal{D} \!\left[ ABC \right]\!   \exp \left( iS \right)\nonumber
	\\
	&\times 2\exp \left[ i\int_{\omega}{C^3+\frac{1}{2}\frac{2\pi q}{N_3}\left( d^{-1}B^1B^2+d^{-1}B^2B^1 \right)} \right] \delta \left( \int_{\sigma}{B^1} \right) \delta \left( \int_{\sigma}{B^2} \right) \nonumber
	\\
	&\times 2\exp \left[ i\!\int_{\omega}{C^3+\frac{1}{2}\frac{2\pi q}{N_3}\left( d^{-1}B^1B^2+d^{-1}B^2B^1 \right)} \right] \!\delta \left( \int_{\sigma}{B^1} \right)\! \delta \left( \int_{\sigma}{B^2} \right) \,.
\end{align}
Similar to the derivation eq.~(\ref{equation_L,100L,100}),  we can obtain:
\begin{align}
	&\langle \mathsf{M^S}_{001}\otimes \mathsf{M^S}_{001} \rangle \nonumber\\
	=& \frac{1}{\mathcal{Z}_{\rm eff}}  \!  \sum_{\hat{B}^1,\hat{B}^2,\hat{A}^3} \! \exp \! \left( iS_{\rm eff}\right) \! 
	\left[\!  1+\exp \left( i\int_{\sigma}{\hat{B}^1} \right) +\exp \left( i\int_{\sigma}{\hat{B}^2} \right) +\exp \left( i\int_{\sigma}{\hat{B}^1+\hat{B}^2} \right) \! \right] \nonumber
	\\
	\!\!=& \frac{1}{\mathcal{Z}}\!\!\int\!\! \mathcal{D} \!\left[ ABC \right]\!\exp \left( iS \right) \!\!
	\left[ 1\!+\!\exp \left( i\int_{\sigma}\!{ {B}^1} \right) \!+\!\exp \left(\! i\int_{\sigma}\!{ {B}^2} \!\right) \!+\!\exp \left(\! i\int_{\sigma}\!{ {B}^1\!+\! {B}^2} \!\right) \right] \nonumber
	\\
	=&\langle \mathsf{1}\oplus \mathsf{L}_{100,}\oplus \mathsf{L}_{010,}\oplus \mathsf{L}_{110,} \rangle	\label{equation_Ms001Ms001}\,.
\end{align}
This is also a non-Abelian fusion process. Fusing two identical $\mathbb{Z} _{N_3}$ membranes of $S^2$ shape renders   one  vacuum (i.e., trivial excitation), one $\mathbb{Z} _{N_1}$ $B$-loop, one $\mathbb{Z} _{N_2}$ $B$-loop, and one $ \left( \mathbb{Z} _{N_1},\mathbb{Z} _{N_2} \right) $ $B$-loop.

\textbf{\textit{Fusing an $\mathsf{M^{ST}}$-excitation and a $\mathbb{Z} _{N_3}$ membrane.}} 
Fusion of an $\mathsf{M^{ST}}$ and a $\mathbb{Z} _{N_3}$ membrane is
\begin{align}
	&\langle \mathsf{M^{ST}}\otimes \mathsf{M^S}_{001} \rangle \nonumber\\
	=&\frac{1}{\mathcal{Z}}\int{\mathcal{D} \left[ ABC\right]  \exp \left( iS \right)} 
	\times 8\exp \left[ i\int_{\omega}{C^3+\frac{1}{2}\frac{2\pi q}{N_3}\left( d^{-1}B^1B^2+d^{-1}B^2B^1 \right)} \right] \nonumber
	\\
	&\times \exp \left[ i\int_{\tilde{\omega}}{C^3+\frac{1}{2}\frac{2\pi q}{N_3}\left( d^{-1}B^1B^2+d^{-1}B^2B^1 \right)} \right] \delta \left( \int_{\sigma}{B^1} \right) \delta \left( \int_{\sigma}{B^2} \right) \nonumber
	\\
	&\times 2\exp \left[ i\int_{\omega}{C^3+\frac{1}{2}\frac{2\pi q}{N_3}\left( d^{-1}B^1B^2+d^{-1}B^2B^1 \right)} \right] \delta \left( \int_{\sigma}{B^1} \right) \delta \left( \int_{\sigma}{B^2} \right) \nonumber
	\\
	=&\langle 4\cdot \mathsf{M^T}_{001} \rangle \,.
\end{align}
As we mentioned above, $\mathsf{M^{ST}}$ is  constructed by fusing an $\mathsf{M^S}_{001}$ and an $\mathsf{M^T}_{001}$. So we can also calculate $\mathsf{M^{ST}}\otimes \mathsf{M^S}_{001}$ in an alternative way:
\begin{align}
	\mathsf{M^{ST}}\otimes \mathsf{M^S}_{001}=&\mathsf{M^S}_{001}\otimes \mathsf{M^T}_{001}\otimes \mathsf{M^S}_{001}=\mathsf{M^T}_{001}\otimes \mathsf{M^S}_{001}\otimes \mathsf{M^S}_{001} \nonumber
	\\
	=&\mathsf{M^T}_{001}\otimes \left(\mathsf{1}\oplus \mathsf{L}_{100,}\oplus \mathsf{L}_{010,}\oplus \mathsf{L}_{110,}\right) 
	=\mathsf{M^T}_{001}\oplus \mathsf{M^T}_{001}\oplus \mathsf{M^T}_{001}\oplus \mathsf{M^T}_{001} \nonumber
	\\
	=&4\cdot \mathsf{M^T}_{001}\,,
\end{align}
where we have applied the fusion rule in eq.~(\ref{equation_Ms001Ms001}) and the properties of fusion rules listed in eq.~(\ref{eq_fusion_algebra_rule}). 
 
\subsection{Fusion table and quantum dimensions\label{section_quantum_dimension}}
For the $BBA$ twisted term, we can exhaust all combinations of two topological excitations and calculate their fusion rules. In principle, we will get a $29\times29$ symmetric fusion table which includes all information about fusion rules in $BBA$ twisted term. We find that there are $8$ Abelian excitations and $21$ non-Abelian excitations. We list a part of the complete fusion table in table~\ref{table_fusionBBA}. We can see that this fusion sub-table is symmetric, which is consistent with the condition $\mathsf{a}\otimes \mathsf{b}=\mathsf{b}\otimes \mathsf{a}$. 
\begin{sidewaystable*}
	\caption{\label{table_fusionBBA}A $14\times14$ fusion sub-table for the TQFT action~(\ref{eq_action_BBA}). This closed fusion subtable is a part of the  complete $29\times29$ fusion table. We call an excitation an Abelian excitation if it always has a single fusion channel with any other excitations; otherwise, we call it a non-Abelian excitation. This subtable includes $8$ Abelian excitations and $6$ non-Abelian excitations. The meanings of numbers in subscript and superscript are summarized at the end of section~\ref{ss23}. The Wilson operators for excitations are shown in table~\ref{tab_excitation}.  }
	\centering
	\tiny
	\begin{tabular*}{\textwidth}{@{\extracolsep{\fill}}|c|c|c|c|c|c|c|c|c|c|c|c|c|c|c|}
		 \hline
		 & \multicolumn{8}{c|}{\textbf{ABELIAN EXCITATIONS}} & \multicolumn{6}{c|}{\textbf{NON-ABELIAN EXCITATIONS}} \\
		 \hline
		 $\otimes$ & $\mathsf{1}$ & $\mathsf{P}_{001}$ & $\mathsf{L}_{100,}$ & $\mathsf{L}_{010,}$ & $\mathsf{L}_{110,}$ & $\mathsf{L}_{100,}^{001}$ & $\mathsf{L}_{010,}^{001}$ & $\mathsf{L}_{110,}^{001}$ & $\mathsf{M^S}_{001}$ & $\mathsf{M^S}_{001}^{001;}$ & $\mathsf{M^T}_{001}$ & $\mathsf{M^T}_{001}^{001;}$ & $\mathsf{M^{ST}}$ & $\mathsf{M^{ST}}^{001;}$ \\
		 \hline
		 $\mathsf{1}$ & $\mathsf{1}$ & $\mathsf{P}_{001}$ & $\mathsf{L}_{100,}$ & $\mathsf{L}_{010,}$ & $\mathsf{L}_{110,}$ & $\mathsf{L}_{100,}^{001}$ & $\mathsf{L}_{010,}^{001}$ & $\mathsf{L}_{110,}^{001}$ & $\mathsf{M^S}_{001}$ & $\mathsf{M^S}_{001}^{001;}$ & $\mathsf{M^T}_{001}$ & $\mathsf{M^T}_{001}^{001;}$ &  $\mathsf{M^{ST}}$ & $\mathsf{M^{ST}}^{001;}$\\
		 \hline
		 $\mathsf{P}_{001}$ & $\mathsf{P}_{001}$ & $\mathsf{1}$ & $\mathsf{L}_{100,}^{001}$ & $\mathsf{L}_{010,}^{001}$ & $\mathsf{L}_{110,}^{001}$ &  $\mathsf{L}_{100,}$ & $\mathsf{L}_{010,}$ & $\mathsf{L}_{110,}$ & $\mathsf{M^S}_{001}^{001;}$ & $\mathsf{M^S}_{001}$ & $\mathsf{M^T}_{001}^{001;}$ & $\mathsf{M^T}_{001}$  & $\mathsf{M^{ST}}^{001;}$ & $\mathsf{M^{ST}}$ \\
		 \hline
		 $\mathsf{L}_{100,}$ & $\mathsf{L}_{100,}$ & $\mathsf{L}_{100,}^{001}$ & $\mathsf{1}$ & $\mathsf{L}_{110,}$ & $\mathsf{L}_{010,}$ & $\mathsf{P}_{001}$ & $\mathsf{L}_{110,}^{001}$ & $\mathsf{L}_{010,}^{001}$ & $\mathsf{M^S}_{001}$ & $\mathsf{M^S}_{001}^{001;}$ & $\mathsf{M^T}_{001}$ & $\mathsf{M^T}_{001}^{001;}$ & $\mathsf{M^{ST}}$ & $\mathsf{M^{ST}}^{001;}$\\
		 \hline
		 $\mathsf{L}_{010,}$ & $\mathsf{L}_{010,}$ & $\mathsf{L}_{010,}^{001}$ & $\mathsf{L}_{110,}$ & $\mathsf{1}$ & $\mathsf{L}_{100,}$ & $\mathsf{L}_{110,}^{001}$ & $\mathsf{P}_{001}$ & $\mathsf{L}_{100,}^{001}$ & $\mathsf{M^S}_{001}$ & $\mathsf{M^S}_{001}^{001;}$ & $\mathsf{M^T}_{001}$ & $\mathsf{M^T}_{001}^{001;}$ & $\mathsf{M^{ST}}$ & $\mathsf{M^{ST}}^{001;}$ \\
		 \hline
		 $\mathsf{L}_{110,}$ & $\mathsf{L}_{110,}$ & $\mathsf{L}_{110,}^{001}$ & $\mathsf{L}_{010,}$ & $\mathsf{L}_{100,}$ & $\mathsf{1}$ & $\mathsf{L}_{010,}^{001}$ & $\mathsf{L}_{100,}^{001}$ & $\mathsf{P}_{001}$ & $\mathsf{M^S}_{001}$ &  $\mathsf{M^S}_{001}^{001;}$& $\mathsf{M^T}_{001}$ & $\mathsf{M^T}_{001}^{001;}$ & $\mathsf{M^{ST}}$ & $\mathsf{M^{ST}}^{001;}$\\
		 \hline
		 $\mathsf{L}_{100,}^{001}$ & $\mathsf{L}_{100,}^{001}$ & $\mathsf{L}_{100,}$ & $\mathsf{P}_{001}$ & $\mathsf{L}_{110,}^{001}$ & $\mathsf{L}_{010,}^{001}$ & $\mathsf{1}$ & $\mathsf{L}_{110,}$ & $\mathsf{L}_{010,}$ & $\mathsf{M^S}_{001}^{001;}$ & $\mathsf{M^S}_{001}$ & $\mathsf{M^T}_{001}^{001;}$ & $\mathsf{M^T}_{001}$ & $\mathsf{M^{ST}}^{001;}$ & $\mathsf{M^{ST}}$ \\
		 \hline
		 $\mathsf{L}_{010,}^{001}$ & $\mathsf{L}_{010,}^{001}$ & $\mathsf{L}_{010,}$ & $\mathsf{L}_{110,}^{001}$ & $\mathsf{P}_{001}$ & $\mathsf{L}_{100,}^{001}$ & $\mathsf{L}_{110,}$ & $\mathsf{1}$ & $\mathsf{L}_{100,}$ & $\mathsf{M^S}_{001}^{001;}$ & $\mathsf{M^S}_{001}$ & $\mathsf{M^T}_{001}^{001;}$ & $\mathsf{M^T}_{001}$ & $\mathsf{M^{ST}}^{001;}$ & $\mathsf{M^{ST}}$ \\
		 \hline
		 $\mathsf{L}_{110,}^{001}$ & $\mathsf{L}_{110,}^{001}$ & $\mathsf{L}_{110,}$ & $\mathsf{L}_{010,}^{001}$ & $\mathsf{L}_{100,}^{001}$ & $\mathsf{P}_{001}$ & $\mathsf{L}_{010,}$ & $\mathsf{L}_{100,}$ & $\mathsf{1}$ & $\mathsf{M^S}_{001}^{001;}$ & $\mathsf{M^S}_{001}$ & $\mathsf{M^T}_{001}^{001;}$ & $\mathsf{M^T}_{001}$ & $\mathsf{M^{ST}}^{001;}$ & $\mathsf{M^{ST}}$ \\
		 \hline
		 $\mathsf{M^S}_{001}$ & $\mathsf{M^S}_{001}$ & $\mathsf{M^S}_{001}^{001;}$ & $\mathsf{M^S}_{001}$ & $\mathsf{M^S}_{001}$ & $\mathsf{M^S}_{001}$ & $\mathsf{M^S}_{001}^{001;}$ & $\mathsf{M^S}_{001}^{001;}$ & $\mathsf{M^S}_{001}^{001;}$ & \makecell{ $\mathsf{1}$\\  $\oplus \mathsf{L}_{100,}$\\  $\oplus \mathsf{L}_{010,}$\\  $\oplus \mathsf{L}_{110,}$} & \makecell{ $\mathsf{P}_{001}$\\  $\oplus \mathsf{L}_{100,}^{001}$\\  $\oplus \mathsf{L}_{010,}^{001}$\\  $\oplus \mathsf{L}_{110,}^{001}$} & $\mathsf{M^{ST}}$ & $\mathsf{M^{ST}}^{001;}$ & $4\cdot\mathsf{M^T}_{001}$ & $4\cdot\mathsf{M^T}_{001}^{001;}$ \\
		 \hline
		 $\mathsf{M^S}_{001}^{001;}$ & $\mathsf{M^S}_{001}^{001;}$ & $\mathsf{M^S}_{001}$ & $\mathsf{M^S}_{001}^{001;}$ & $\mathsf{M^S}_{001}^{001;}$ & $\mathsf{M^S}_{001}^{001;}$ & $\mathsf{M^S}_{001}$ & $\mathsf{M^S}_{001}$ & $\mathsf{M^S}_{001}$ & \makecell{ $\mathsf{P}_{001}$\\  $\oplus \mathsf{L}_{100,}^{001}$\\  $\oplus \mathsf{L}_{010,}^{001}$\\  $\oplus \mathsf{L}_{110,}^{001}$} & \makecell{ $\mathsf{1}$\\  $\oplus \mathsf{L}_{100,}$\\  $\oplus \mathsf{L}_{010,}$\\  $\oplus \mathsf{L}_{110,}$} & $\mathsf{M^{ST}}^{001;}$ & $\mathsf{M^{ST}}$ & 4$\cdot$$\mathsf{M^T}_{001}^{001;}$ & $4\cdot\mathsf{M^T}_{001}$ \\
		 \hline
		 $\mathsf{M^T}_{001}$ & $\mathsf{M^T}_{001}$ & $\mathsf{M^T}_{001}^{001;}$ & $\mathsf{M^T}_{001}$ & $\mathsf{M^T}_{001}$ & $\mathsf{M^T}_{001}$ & $\mathsf{M^T}_{001}^{001;}$ & $\mathsf{M^T}_{001}^{001;}$ & $\mathsf{M^T}_{001}^{001;}$ & $\mathsf{M^{ST}}$ & $\mathsf{M^{ST}}^{001;}$ & \makecell{ $4\cdot\mathsf{1}$\\  $\oplus4\cdot \mathsf{L}_{100,}$\\  $\oplus4\cdot \mathsf{L}_{010,}$\\  $\oplus4\cdot \mathsf{L}_{110,}$} & \makecell{ $4\cdot \mathsf{P}_{001}$\\  $\oplus4\cdot \mathsf{L}_{100,}^{001}$\\  $\oplus4\cdot \mathsf{L}_{010,}^{001}$\\  $\oplus4\cdot \mathsf{L}_{110,}^{001}$} & $16\cdot \mathsf{M^S}_{001}$ & $16\cdot \mathsf{M^S}_{001}^{001;}$ \\
		 \hline
		 $\mathsf{M^T}_{001}^{001;}$ & $\mathsf{M^T}_{001}^{001;}$ & $\mathsf{M^T}_{001}$ & $\mathsf{M^T}_{001}^{001;}$ & $\mathsf{M^T}_{001}^{001;}$ & $\mathsf{M^T}_{001}^{001;}$ & $\mathsf{M^T}_{001}$ & $\mathsf{M^T}_{001}$ & $\mathsf{M^T}_{001}$ & $\mathsf{M^{ST}}^{001;}$ & $\mathsf{M^{ST}}$ & \makecell{ $4\cdot \mathsf{P}_{001}$\\  $\oplus4\cdot \mathsf{L}_{100,}^{001}$\\  $\oplus4\cdot \mathsf{L}_{010,}^{001}$\\  $\oplus4\cdot \mathsf{L}_{110,}^{001}$} & \makecell{ $4\cdot\mathsf{1}$\\  $\oplus4\cdot \mathsf{L}_{100,}$\\  $\oplus4\cdot \mathsf{L}_{010,}$\\  $\oplus4\cdot \mathsf{L}_{110,}$} & $16\cdot \mathsf{M^S}_{001}^{001;}$ & $16\cdot \mathsf{M^S}_{001}$ \\
		 \hline
		 $\mathsf{M^{ST}}$ & $\mathsf{M^{ST}}$ & $\mathsf{M^{ST}}^{001;}$ & $\mathsf{M^{ST}}$ & $\mathsf{M^{ST}}$ & $\mathsf{M^{ST}}$ & $\mathsf{M^{ST}}^{001;}$ & $\mathsf{M^{ST}}^{001;}$ & $\mathsf{M^{ST}}^{001;}$ & $4\cdot\mathsf{M^T}_{001}$ & $4\cdot\mathsf{M^T}_{001}^{001;}$ & $16\cdot \mathsf{M^S}_{001}$ & $16\cdot \mathsf{M^S}_{001}^{001;}$ & \makecell{ $16\cdot\mathsf{1}$\\  $\oplus16\cdot \mathsf{L}_{100,}$\\  $\oplus16\cdot \mathsf{L}_{010,}$\\  $\oplus16\cdot \mathsf{L}_{110,}$} & \makecell{ $16\cdot \mathsf{P}_{001}$\\  $\oplus16\cdot \mathsf{L}_{100,}^{001}$\\  $\oplus16\cdot \mathsf{L}_{010,}^{001}$\\  $\oplus16\cdot \mathsf{L}_{110,}^{001}$} \\
		 \hline
		 $\mathsf{M^{ST}}^{001;}$ & $\mathsf{M^{ST}}^{001;}$ & $\mathsf{M^{ST}}$ & $\mathsf{M^{ST}}^{001;}$ & $\mathsf{M^{ST}}^{001;}$ & $\mathsf{M^{ST}}^{001;}$ & $\mathsf{M^{ST}}$ & $\mathsf{M^{ST}}$ & $\mathsf{M^{ST}}$ & $4\cdot\mathsf{M^T}_{001}^{001;}$ & $4\cdot\mathsf{M^T}_{001}$ & $16\cdot \mathsf{M^S}_{001}^{001;}$ & $16\cdot \mathsf{M^S}_{001}$ & \makecell{ $16\cdot \mathsf{P}_{001}$\\  $\oplus16\cdot \mathsf{L}_{100,}^{001}$\\  $\oplus16\cdot \mathsf{L}_{010,}^{001}$\\  $\oplus16\cdot \mathsf{L}_{110,}^{001}$} & \makecell{ $16\cdot\mathsf{1}$\\  $\oplus16\cdot \mathsf{L}_{100,}$\\  $\oplus16\cdot \mathsf{L}_{010,}$\\  $\oplus16\cdot \mathsf{L}_{110,}$} \\
		 \hline
	\end{tabular*}
\end{sidewaystable*}

For a topological excitation $\mathsf{a}$, we can define a fusion coefficient matrix $N_{\mathsf{a}}$, whose element is $ \left( N_{\mathsf{a}} \right) _{\mathsf{b}\mathsf{c}}=N_{\mathsf{c}}^{\mathsf{a}\mathsf{b}}$. Quantum dimension $d_{\mathsf{a}}$ is defined by the greatest eigenvalue of the matrix $N_{\mathsf{a}}$. We can obtain all matrices $N_{\mathsf{a}}$ from the complete fusion table and then calculate the quantum dimensions of all excitations, as shown in table~\ref{table_quantumdBBA}. We find that \textit{the quantum dimension of an excitation equals to the normalization factor of the excitation}.  For a non-Abelian excitation, the quantum dimension is larger than $1$. For an Abelian excitation, the quantum dimension is always $1$. 

After obtaining all quantum dimensions, we can also verify that  
\begin{align}
	d_{\mathsf{a}}d_{\mathsf{b}}=\sum_{\mathsf{c}}{N_{\mathsf{c}}^{\mathsf{ab}}d_{\mathsf{c}}}\,. \label{eq_dadb}
\end{align}
For example, from eq.~(\ref{equation_Ms001Ms001}) we have $\langle {\mathsf{M}^{\mathsf{S}}}_{001}\otimes {\mathsf{M}^{\mathsf{S}}}_{001}\rangle =\langle \mathsf{1}\oplus \mathsf{L}_{100,}\oplus \mathsf{L}_{010,}\oplus \mathsf{L}_{110,}\rangle $. We can calculate that quantum dimensions for ${\mathsf{M}^{\mathsf{S}}}_{001}$, $\mathsf{1}$, $\mathsf{L}_{100,}$, $\mathsf{L}_{010,}$ and $\mathsf{L}_{110,}$ are $2$, $1$, $1$, $1$ and $1$ respectively. Then the LHS of eq.~(\ref{eq_dadb}) is $2\times2=4$, the RHS of eq.~(\ref{eq_dadb}) is $1\times1+1\times1+1\times1+1\times1=4=\text{LHS}$.


\begin{table}
	\caption{\label{table_quantumdBBA}Quantum dimension of excitations in $BBA$ twisted term. For a non-Abelian excitation, the quantum dimension is larger than $1$ and equal to the normalization factor of the Wilson operator. For an Abelian excitation, the quantum dimension is always $1$.  }
	\centering
	\footnotesize
	\begin{tabular*}{\textwidth}{@{\extracolsep{\fill}}ccccccccccc}
		\toprule
		Excitation & $\mathsf{1}$ & $\mathsf{P}_{001}$ & $\mathsf{L}_{100,}$ & $\mathsf{L}_{010,}$ & $\mathsf{L}_{110,}$ & $\mathsf{L}_{100,}^{001}$ & $\mathsf{L}_{010,}^{001}$ & $\mathsf{L}_{110,}^{001}$ & $\mathsf{L}_{,100}$ & $\mathsf{L}_{,010}$ \\ 
		\hline
		Quantum dimension & $1$ & $1$ & $1$ & $1$ & $1$ & $1$ & $1$ & $1$ & $2$ & $2$ \\
		\toprule
		Excitation & $\mathsf{L}_{,110}$ & $\mathsf{L}_{100,100}$ & $\mathsf{L}_{010,010}$ & $\mathsf{L}_{100,110}$ & $\mathsf{M^S}_{001}$ & $\mathsf{M^S}_{001}^{001;}$ & $\mathsf{M^S}_{001}^{,100}$ & $\mathsf{M^S}_{001}^{,010}$ & $\mathsf{M^S}_{001}^{,110}$ & $\mathsf{M^T}_{001}$ \\
		\hline
		Quantum dimension &$2$&$2$&$2$&$2$&$2$&$2$&$4$&$4$&$4$&$4$ \\
		\toprule
		Excitation &$\mathsf{M^T}_{001}^{001;}$ & $\mathsf{M^T}_{001}^{,100}$ & $\mathsf{M^T}_{001}^{,010}$ & $\mathsf{M^T}_{001}^{,110}$ & $\mathsf{M^{ST}}$ & $\mathsf{M^{ST}}^{001;}$ & $\mathsf{M^{ST}}^{,100}$ & $\mathsf{M^{ST}}^{,010}$ & $\mathsf{M^{ST}}^{,110}$ & \\
		\hline
		Quantum dimension &$4$&$8$&$8$&$8$&$8$&$8$&$16$&$16$&$16$& \\
		\bottomrule
	\end{tabular*}
\end{table}

\section{Shrinking rules for the $BBA$ twisted term\label{s4}}

\subsection{General discussion on shrinking in $5$D and its hierarchical structure\label{general_shrinking}}
In $3$D topological orders, shrinking rules are absent because all topological excitations are  point-like particles, i.e., anyons. However, in $4$D and higher, spatially extended excitations exist, such that the inclusion of shrinking processes makes sense. While loop excitations in $4$D bring us  exotic braiding statistics, as reviewed in section~\ref{s1}, one may wonder  what are the resulting outcomes  by shrinking loop excitations. Indeed, a recent study in $4$D indicates that  the topological data given by shrinking rules   significantly enrich  the physics of  higher-dimensional topological orders~\cite{PhysRevB.107.165117}. 
More concretely, the world-sheet of a loop excitation is an $S^1\times S^1=T^2$ spacetime manifold. If we shrink the loop to a point, then the world-sheet is simultaneously shrunk  to a world-line, i.e., $T^2\rightarrow S^1$, which can be either the world-line of a particle or many particles. The former case is an Abelian shrinking process while the latter is a non-Abelian shrinking process.  
\begin{figure}
	\centering
	\includegraphics[scale=0.9,keepaspectratio]{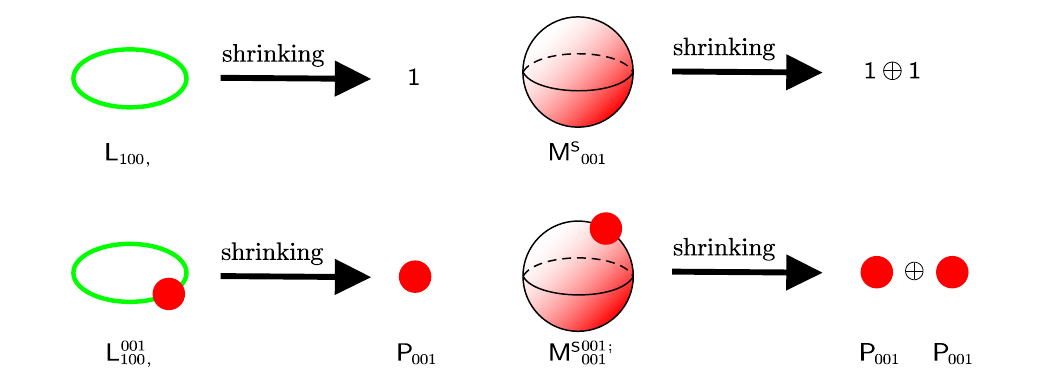}
	\caption{Typical shrinking rules for loops and $S^2$ membranes in the $5$D twisted $BF$ theory with the $BBA$ twisted term. Shrinking an $\mathsf{L}_{100,}$ and shrinking  an $\mathsf{L}_{100,}^{001}$ respectively lead to a vacuum and a $\mathsf{P}_{001}$. Shrinking an $\mathsf{M^S}_{001}$ and shrinking an $\mathsf{M^S}_{001}^{001;}$ respectively lead to a direct sum of $2$ vacua and a direct sum  of $2$ $\mathsf{P}_{001}$'s.}
	\label{fig_shrinking}
\end{figure}
\begin{figure}
	\centering
	\includegraphics[scale=0.9,keepaspectratio]{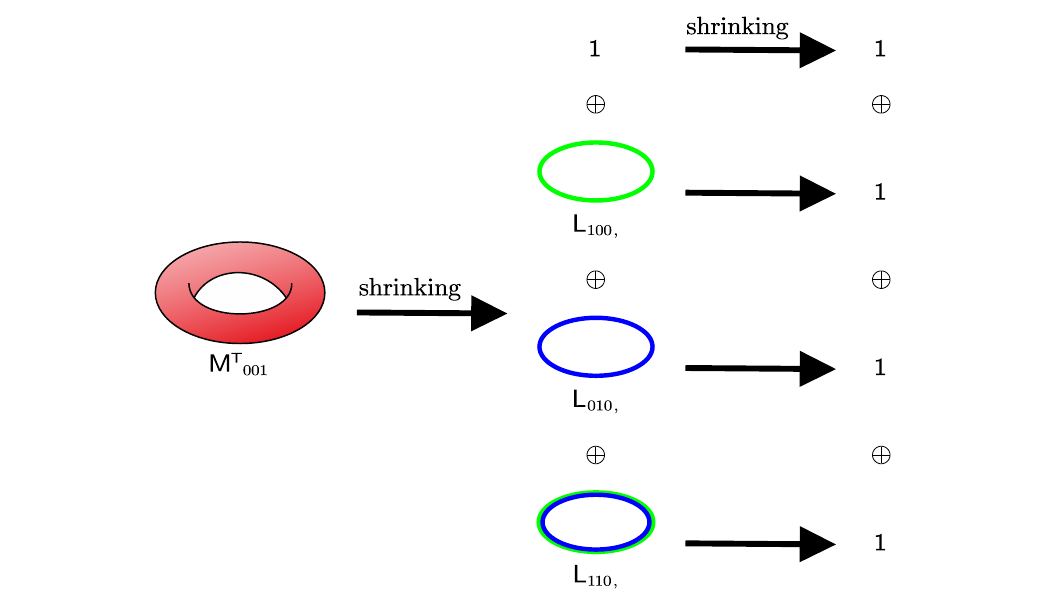}
	\caption{A typical example of hierarchical shrinking rules for $T^2$ membranes in the $5$D twisted $BF$ theory with the $BBA$ twisted term. We can shrink a $T^2$ membrane $\mathsf{M^T}_{001}$ to a superposition of a vacuum, a $\mathsf{L}_{100,}$, a $\mathsf{L}_{010,}$ and a $\mathsf{L}_{110,}$ first. Then we can continue to shrink them to $4$ vacua.}
	\label{fig_hierarchy_nontrivial}
\end{figure}
\begin{figure}
	\centering
	\includegraphics[scale=0.9,keepaspectratio]{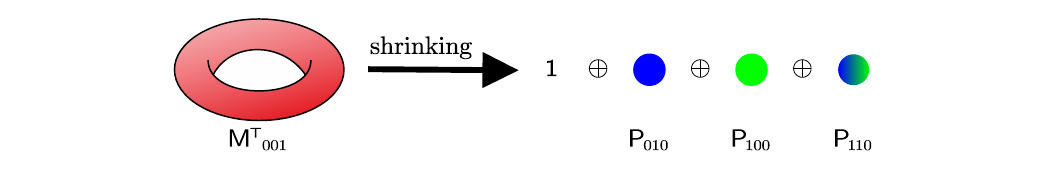}
	\caption{A typical example of shrinking rules without hierarchical structure for $T^2$ membranes in the $5$D twisted $BF$ theory with the $AAAdA$ twisted term, the TQFT action is given by eq.~(\ref{eqD10}). $\mathsf{M^T}_{001}$ is shrunk to a superposition of a trivial loop and $3$ nonequivalent particles in the first step. At the beginning of section ~\ref{ss22}, we have mentioned that a trivial loop is equivalent to a trivial particle, thus we conclude that $\mathsf{M^T}_{001}$ can be completely shrunk to particles in the first step.}
	\label{fig_hierarchy_trivial}
\end{figure}

In $5$D topological orders, spatially extended excitations include both loops and membranes, and the latter can have different shapes, e.g., $S^2$ and $T^2$. Here we only consider some simple shapes as shown in figure~\ref{fig_excitation}. Therefore, we expect the underlying shrinking rules will be significantly diversified, as shown in figures \ref{fig_shrinking}, \ref{fig_hierarchy_nontrivial},  and \ref{fig_hierarchy_trivial}. Intuitively,  an $S^2$ membrane can be directly shrunk to particles while a $T^2$ membrane can be shrunk to loops first, and then we can further shrink these loops to particles. As membranes exist in topological orders of $5$D and higher, this \textit{hierarchical structure} of shrinking rules can only appear in $5$D and higher. A typical example of hierarchical shrinking is shown in figure \ref{fig_hierarchy_nontrivial}. 

It seems that   two successive steps are required to shrink $T^2$ membranes completely into particles. But it is not always the case. As shown in figure \ref{fig_hierarchy_trivial}, some $T^2$ membranes in the twisted $BF$ theory with the $AAAdA$ twisted term, upon being shrunk in the first step, only lead to trivial loops and some particles. As trivial loops and trivial particles are equivalent to each other and all of them are identical to a vacuum, these $T^2$ membranes are completely shrunk to particles including vacua in the first step. If we need two steps to shrink a pure $T^2$ membrane to particles, we say this $T^2$ membrane has hierarchical shrinking rules.

To quantitatively calculate shrinking rules, we  define a shrinking operator $\mathcal{S}$ and the associated shrinking coefficient ${\mathrm{S}}_{\mathsf{b}}^{\mathsf{a}}$ as below:
\begin{align}
\mathcal{S}\left(\mathsf{a}\right)=\underset{X_{1}\rightarrow X_{2}}{\lim}\mathsf{a}=\oplus_{\mathsf{b}} {\mathrm{S}}_{\mathsf{b}}^{\mathsf{a}} \mathsf{b}\,,\label{eq_shrinking_coefficient}
\end{align}
where the shrinking coefficient ${\mathrm{S}}_{\mathsf{b}}^{\mathsf{a}}\in\Z$. $X_1$ and $X_2$ are respectively  spacetime trajectories of excitation $\mathsf{a}$ before and after shrinking, which respects $X_2\subset X_1$. We leave details of $X_1$ and $X_2$ to section~\ref{section_shrinking_nohier} and \ref{section_shrinking_hier} where concrete calculations are given.  The summation   $\oplus_{\mathsf{b}}$ exhausts all $29$ topologically distinct excitations.

In the path-integral representation, we may compute the following expectation value in the ground state:
\begin{align}
	\langle \mathcal{S} \left( \mathsf{a} \right) \rangle =&\langle \underset{X_1\rightarrow X_2}{\lim}\mathsf{a} \rangle =\underset{X_1\rightarrow X_2}{\lim}\frac{1}{\mathcal{Z}}\int{\mathcal{D} \left[ ABC \right] \exp \left( iS \right) \mathcal{O} _{\mathsf{a}}} \nonumber
	\\
	=&\sum_{\mathsf{b}}\frac{1}{\mathcal{Z}}\int{\mathcal{D} \left[ ABC \right] \exp \left( iS \right) \mathrm{S}_{\mathsf{b}}^{\mathsf{a}}\mathcal{O} _{\mathsf{b}}} =\langle \oplus_{\mathsf{b}}{\mathrm{S}}_{\mathsf{b}}^{\mathsf{a}}\mathsf{b} \rangle \,.
	\label{eq_shrinking_pathintegral}
\end{align}
 If the integer ${\mathrm{S}}_{\mathsf{b}}^{\mathsf{a}}=1$ for topological excitation $\mathsf{b}$ and vanishes for any other excitations,  the shrinking is said to be Abelian. If ${\mathrm{S}}_{\mathsf{b}}^{\mathsf{a}}$ is nonzero for more than one topological excitation or if there exists $\mathsf{b}$ such that  ${\mathrm{S}}_{\mathsf{b}}^{\mathsf{a}}\textgreater 1$,  the shrinking is said to be non-Abelian. We collect all shrinking rules in the  shrinking table~\ref{table_shrinking}.

\subsection{Shrinking rules without hierarchical structure\label{section_shrinking_nohier}}
Next, we present several typical examples in table~\ref{table_shrinking}.  In this subsection, we study shrinking rules without hierarchical structure. $X_1$ and $X_2$ meet the following conditions:
\begin{enumerate}
	\item If $\mathsf{a}$ is a particle ($\mathsf{1}$ and $\mathsf{P}_{001}$), then the world-line of the particle is unchanged, i.e., $X_1=X_2=\gamma$, which means that particles are unshrinkable.
	\item If $\mathsf{a}$ is a loop, then the world-sheet of the loop is shrunk into a world-line, i.e., $X_1=\sigma, X_2=\gamma$.
	\item If $\mathsf{a}$ is an $S^2$ membrane, then the world-volume  of the membrane is shrunk into a world-line and the manifold $\sigma$ is also shrunk into a world-line, i.e., $X_1=(\omega,\sigma), X_2=(\gamma,\gamma)$.
\end{enumerate}

\textbf{\textit{Shrinking a $\mathbb{Z}_{N_1}$ $B$-loop.}}
Shrinking a $\mathbb{Z}_{N_1}$ $B$-loop results in a collapse of  its world sheet to a world-line, which is $\sigma\rightarrow \gamma$. From the path integral, we obtain
\begin{align}
	\langle \mathcal{S} \left( \mathsf{L}_{100,} \right) \rangle &=\langle \underset{\sigma \rightarrow \gamma}{\lim}\mathsf{L}_{100,} \rangle =\underset{\sigma \rightarrow \gamma}{\lim}\frac{1}{\mathcal{Z}}\int{\mathcal{D} \left[ ABC \right] \exp \left( iS \right) \exp \left( i\int_{\sigma}{B^1} \right)}\nonumber
	\\
	&=\frac{1}{\mathcal{Z}}\int{\mathcal{D} \left[ ABC \right] \exp \left( iS \right) \exp \left( i0 \right)}=\langle \mathsf{1} \rangle\,. 
\end{align}
So we conclude that
 $	\mathcal{S} \left( \mathsf{L}_{100,} \right)=\mathsf{1}
$, which means that a $\mathbb{Z}_{N_1}$ $B$-loop denoted as $\mathsf{L}_{100,}$ is shrunk into a vacuum.

\textbf{\textit{Shrinking a $\mathbb{Z}_{N_1}$ $B$-loop decorated by $\mathbb{Z}_{N_3}$ particle.}}
Similarly, the shrinking rule for an $\mathsf{L}_{100,}^{001}$ loop can be obtained
\begin{align}
	\langle \mathcal{S} \left( \mathsf{L}_{100,}^{001} \right) \rangle &=\langle \underset{\sigma \rightarrow \gamma}{\lim}\mathsf{L}_{100,}^{001} \rangle =\underset{\sigma \rightarrow \gamma}{\lim}\frac{1}{\mathcal{Z}}\int{\mathcal{D} \left[ ABC\right] \exp \left( iS \right) \exp \left( i\int_{\sigma}{B^1}+i\int_{\gamma}{A^3} \right)}\nonumber
	\\
	&=\frac{1}{\mathcal{Z}}\int{\mathcal{D} \left[ ABC \right] \exp \left( iS \right) \exp \left( i0 \right) \exp \left( i\int_{\gamma}{A^3} \right)}=\langle \mathsf{P}_{001} \rangle \,.
\end{align}
Therefore, particle decoration is unaffected by the shrinking process. Also notice that 
\begin{align}
	\mathcal{S} \left( \mathsf{L}_{100,}^{001} \right)=\mathcal{S} \left( \mathsf{L}_{100,}\otimes \mathsf{P}_{001} \right)=\mathcal{S} \left( \mathsf{L}_{100,} \right)\otimes\mathcal{S} \left( \mathsf{P}_{001} \right)=\mathsf{P}_{001}\,,
\end{align}
implying that \textit{shrinking rules without hierarchical structure respect the fusion rules.} 

\textbf{\textit{Shrinking a $\mathbb{Z}_{N_1}$ $\tilde{B}$-loop.}}
From path integral we obtain 
\begin{align}
	&\langle \mathcal{S} \left( \mathsf{L}_{,100} \right) \rangle =\langle \underset{\sigma \rightarrow \gamma}{\lim}\mathsf{L}_{,100}\rangle \nonumber
	\\
	=&\underset{\sigma \rightarrow \gamma}{\lim}\frac{1}{\mathcal{Z}}\int{\mathcal{D} \left[ ABC \right] \exp \left( iS \right)}\times 2\exp \left[ i\int_{\sigma}{\tilde{B}^1-\frac{1}{2}\frac{2\pi q}{N_1}\left( d^{-1}A^3B^2+d^{-1}B^2A^3 \right)} \right]  \nonumber
	\\
	&\times \delta \left( \int_{\sigma}{B^2} \right) \delta \left( \int_{\gamma}{A^3} \right) \nonumber
	\\
	=&\frac{1}{\mathcal{Z}}\int{\mathcal{D} \left[ ABC \right] \exp \left( iS \right) 2\cdot \exp \left( i0 \right) \delta \left( 0 \right) \frac{1}{2}\left[ 1+\exp \left( i\int_{\gamma}{A^3} \right) \right]}=\langle \mathsf{1}\oplus\mathsf{P}_{001}\rangle \,.\label{eq_SL,100}
\end{align}
In other words, the shrinking process renders a superposition of a vacuum and a $\mathbb{Z}_{N_3}$ particle when shrinking a $\mathbb{Z}_{N_1}$ $\tilde{B}$-loop. This is a non-Abelian shrinking process since there are multiple shrinking channels. Notice that if we treat  $\mathsf{1}\oplus\mathsf{P}_{001}$ as a composite excitation we will find that it has a quantum dimension of $2$, which is equal to the quantum dimension of $\mathsf{L}_{,100}$. It may be seen as preserving the quantum dimension. As a matter of fact, by exhausting all shrinking processes, we can verify that \textit{all shrinking processes preserve quantum dimension:}
\begin{align}
	d_{\mathsf{a}}=\sum_{\mathsf{b}}{S_{\mathsf{b}}^{\mathsf{a}}d_{\mathsf{b}}}\,,
\end{align} 
where $d_{\mathsf{a}}$ and $d_{\mathsf{b}}$ are quantum dimensions of  ${\mathsf{a}}$ and ${\mathsf{b}}$ respectively, derivation of quantum dimension are shown in section~\ref{section_quantum_dimension}. The summation $\sum_{\mathsf{b}}$ exhausts all $29$ nonequivalent excitations in table~\ref{tab_excitation}.

\textbf{\textit{Shrinking a $\mathbb{Z}_{N_3}$ membrane in the shape of $S^2$.}}
From path integral we obtain 
\begin{align}
&	\langle \mathcal{S} \left( \mathsf{M^S}_{001} \right) \rangle 	=\langle \underset{\omega \rightarrow \gamma, \sigma \rightarrow \gamma}{\lim}\mathsf{M^S}_{001} \rangle \nonumber
	\\
	=&\underset{\omega \rightarrow \gamma, \sigma \rightarrow \gamma}{\lim}\frac{1}{\mathcal{Z}}\int{\mathcal{D} \left[ ABC \right] \exp \left( iS \right)}
	\times 2\exp \left[ i\int_{\omega}{C^3+\frac{1}{2}\frac{2\pi q}{N_3}\left( d^{-1}B^1B^2+d^{-1}B^2B^1 \right)} \right]  \nonumber
	\\
	&\times \delta \left( \int_{\sigma}{B^1} \right) \delta \left( \int_{\sigma }{B^2} \right) =\frac{1}{\mathcal{Z}}\int{\mathcal{D} \left[ ABC\right] \exp \left( iS \right) 2\cdot\exp \left( i0 \right) \delta \left( 0 \right) \delta \left( 0 \right)}=\langle 2\cdot\mathsf{1} \rangle \,.
\end{align}
This is also a non-Abelian shrinking process. The shrinking outcome is a superposition of $2$ vacua, which is reasonable because both $\mathsf{M^S}_{001}$ and $2\cdot\mathsf{1}$ have quantum dimension $=2$.

\subsection{Hierarchical shrinking process (i): shrinking a   $T^2$ membrane\label{section_shrinking_hier}}
Spatially, when shrinking a $T^2$ torus, we first get an $S^1$ closed line and then get a point. Thus for a $T^2$ membrane, we can first shrink it into loops and particles. If a $T^2$ membrane gives nontrivial loops after the first shrinking process, we can further shrink these loops into  particles. This is the aforementioned hierarchical shrinking. According to the discussion about loop decorated $T^2$ membranes in section~\ref{ss22}, we can treat all loop decorations on a contractible circle. Thus for a $T^2$ membrane with loop decorations, loops are simultaneously shrunk to particles in the first shrinking process. However we do not shrink $\sigma$ in delta functionals in the first shrinking process since $\delta \left( \int_{\sigma}{B^i} \right)$ enforces the $B$-loop to be trivial. When we shrink the world volume $\tilde{\omega}$ to world sheet $S^1\times S^1$, $\sigma$ is still a submanifold of this world sheet $S^1\times S^1$. Thus $\delta \left( \int_{\sigma}{B^i} \right)$ is still a valid constraint. $X_1$ and $X_2$ meet the following conditions:
\begin{enumerate}
	\item In the first step, $\mathcal{S}$ shrinks the world-volume of membrane and the world-sheet of loop decorations to a world-sheet and a world-line respectively. World-sheet in delta functionals is unaffected. That is, $X_1=(\tilde{\omega},\sigma_{c}), X_2=(\sigma,\gamma)$, where we use $\sigma_{c}$ to denote world sheet of loop decorations. 
	\item In the second step, $\mathcal{S}$ shrinks all world-sheets
	 to a world line, i.e., $X_1=\sigma$, $X_2=\gamma$.
\end{enumerate}

\textbf{\textit{Shrinking an $\mathsf{M^T}_{001}$ membrane.}}
From path integral we obtain
\begin{align}
	\langle \mathcal{S} \left( \mathsf{M^T}_{001} \right) \rangle =&\langle \underset{\tilde{\omega} \rightarrow \sigma}{\lim}\mathsf{M^T}_{001} \rangle \nonumber
	\\
	=&\underset{\tilde{\omega} \rightarrow \sigma}{\lim}\frac{1}{\mathcal{Z}}\int{\mathcal{D} \left[ ABC\right] \exp \left( iS \right)} \nonumber
	\\
	&\times 4\exp \left[ i\int_{\tilde{\omega}}{C^3+\frac{1}{2}\frac{2\pi q}{N_3}\left( d^{-1}B^1B^2+d^{-1}B^2B^1 \right)} \right]\delta \left( \int_{\sigma}{B^1} \right) \delta \left( \int_{\sigma}{B^2} \right) \nonumber
	\\
	=&\frac{1}{\mathcal{Z}}\int{\mathcal{D} \left[ ABC \right] \exp \left( iS \right)}\times 4\exp \left( i0 \right) \delta \left( \int_{\sigma}{B^1} \right) \delta \left( \int_{\sigma}{B^2} \right) \,.
\end{align}
By further expanding delta functionals, we have
\begin{align}
	\langle \mathcal{S} \left( \mathsf{M^T}_{001} \right) \rangle 
	=&\langle \mathsf{1}\oplus \mathsf{L}_{100,}\oplus \mathsf{L}_{010,}\oplus \mathsf{L}_{110,} \rangle \,.
\end{align}
If we regard $\mathsf{1}$ as a trivial loop, shrinking $\mathsf{M^T}_{001}$ once gives us a superposition of a vacuum and  all three Abelian pure loops in table~\ref{tab_excitation}. We also see that the normalization factor $4$ ensures that the shrinking coefficients are integers.

Since the above shrinking process leads to nontrivial loops, we can continue to shrink the loops to particles (the symbol $\mathcal{S}^2=\mathcal{S}\mathcal{S}$):
\begin{align}
	\langle \mathcal{S} ^2\left( \mathsf{M^T}_{001} \right) \rangle =&\langle \mathcal{S} \left( \mathsf{1}\oplus \mathsf{L}_{100,}\oplus \mathsf{L}_{010,}\oplus \mathsf{L}_{110,} \right) \rangle =\langle \underset{\sigma \rightarrow \gamma}{\lim}\left( \mathsf{1}\oplus \mathsf{L}_{100,}\oplus \mathsf{L}_{010,}\oplus \mathsf{L}_{110,} \right) \rangle \nonumber
	\\
	=&\underset{\sigma \rightarrow \gamma}{\lim}\frac{1}{\mathcal{Z}}\int{\mathcal{D} \left[ABC  \right] \exp \left( iS \right)}\nonumber
	\\
	&\times \left[ 1+\exp \left( i\int_{\sigma}{B^1} \right) +\exp \left( i\int_{\sigma}{B^2} \right) +\exp \left( i\int_{\sigma}{B^1+B^2} \right) \right] \nonumber
	\\
	=&\frac{1}{\mathcal{Z}}\int{\mathcal{D} \left[ ABC \right] \exp \left( iS \right) \times \left[ 1+\exp \left( i0 \right) +\exp \left( i0 \right) +\exp \left( i0 \right) \right]}\nonumber
	\\
	=&\frac{1}{\mathcal{Z}}\int{\mathcal{D} \left[ ABC \right] \exp \left( iS \right) \times 4\cdot \mathsf{1}}=\langle 4\cdot \mathsf{1} \rangle \,.
\end{align}
In summary,  after two steps of shrinking operation, we finally shrink a $\mathbb{Z}_{N_3}$ membrane in the shape of $T^2$ to $4$ vacua, which is    different from shrinking an $S^2$ membrane.

\textbf{\textit{Shrinking an $\mathsf{M^T}_{001}^{,100}$ membrane.}}
From path integral we obtain
\begin{align}
	&\langle \mathcal{S} \left( {\mathsf{M}^{\mathsf{T}}}_{001}^{,100} \right) \rangle =\langle \underset{\tilde{\omega}\rightarrow \sigma ,\sigma _c\rightarrow \gamma}{\lim}{\mathsf{M}^{\mathsf{T}}}_{001}^{,100}\rangle \nonumber
	\\
	=&\underset{\tilde{\omega}\rightarrow \sigma ,\sigma _c\rightarrow \gamma}{\lim}\frac{1}{\mathcal{Z}}\int{\mathcal{D} \left[ ABC \right] \exp \left( iS \right)}\times 8\exp \left[ i\int_{\tilde{\omega}}{C^3+\frac{1}{2}\frac{2\pi q}{N_3}\left( d^{-1}B^1B^2+d^{-1}B^2B^1 \right)} \right] \nonumber
	\\
	&\times \exp \left[ i\int_{\sigma _c}{\tilde{B}^1-\frac{1}{2}\frac{2\pi q}{N_1}\left( d^{-1}A^3B^2+d^{-1}B^2A^3 \right)} \right] \delta \left( \int_{\sigma}{B^1} \right) \delta \left( \int_{\sigma}{B^2} \right) \delta \left( \int_{\gamma}{A^3} \right) \nonumber
	\\
	=&\frac{1}{\mathcal{Z}}\int{\mathcal{D} \left[ ABC \right] \exp \left( iS \right)}\times 8\exp \left( i0 \right) \exp \left( i0 \right)  \delta \left( \int_{\sigma}{B^1} \right) \delta \left( \int_{\sigma}{B^2} \right) \delta \left( \int_{\gamma}{A^3} \right) \,.
\end{align}
By further expanding delta functionals, we have
\begin{align}
	\langle \mathcal{S} \left( {\mathsf{M}^{\mathsf{T}}}_{001}^{,100} \right) \rangle =\langle 1\oplus \mathsf{L}_{100,}\oplus \mathsf{L}_{010,}\oplus \mathsf{L}_{110,}\oplus \mathsf{P}_{001}\oplus \mathsf{L}_{100,}^{001}\oplus \mathsf{L}_{010,}^{001}\oplus \mathsf{L}_{110,}^{001}\rangle \,.
\end{align}
This is a superposition of all Abelian excitations. Continue to shrink the above result we obtain
\begin{align}
	\langle \mathcal{S} ^2\left( {\mathsf{M}^{\mathsf{T}}}_{001}^{,100} \right) \rangle =&\langle \mathcal{S} \left( 1\oplus \mathsf{L}_{100,}\oplus \mathsf{L}_{010,}\oplus \mathsf{L}_{110,}\oplus \mathsf{P}_{001}\oplus \mathsf{L}_{100,}^{001}\oplus \mathsf{L}_{010,}^{001}\oplus \mathsf{L}_{110,}^{001} \right) \rangle \nonumber
	\\
	=&\langle 4\cdot \left( 1\oplus \mathsf{P}_{001} \right) \rangle \,.
\end{align}
Notice that ${\mathsf{M}^{\mathsf{T}}}_{001}^{,100}={\mathsf{M}^{\mathsf{T}}}_{001}\otimes \mathsf{L}_{,100}$, we can verify that 
\begin{align}
	\mathcal{S} ^2\left( {\mathsf{M}^{\mathsf{T}}}_{001}^{,100} \right) =\mathcal{S} ^2\left( {\mathsf{M}^{\mathsf{T}}}_{001}\otimes \mathsf{L}_{,100} \right) =\mathcal{S} ^2\left( {\mathsf{M}^{\mathsf{T}}}_{001} \right) \otimes \mathcal{S} ^2\left( \mathsf{L}_{,100} \right) \,.
\end{align}
This result implies that \textit{hierarchical shrinking rules respect the fusion rules.}

\subsection{Hierarchical shrinking process (ii): shrinking an $\mathsf{M^{ST}}$-excitation}

For an $\mathsf{M^{ST}}$-excitation, we define that $X_1$ and $X_2$ meet the following conditions:
\begin{enumerate}
	\item In the first step, $\mathcal{S}$  shrink the world-volume $\omega$ and $\tilde{\omega}$ to a world-line $\gamma$ and a world-sheet $\sigma$ respectively. World-sheet $\sigma_{c}$ of loop decorations is also shrunk to a world-line $\gamma$. $X_1=(\omega,\tilde{\omega},\sigma_{c}), X_2=(\gamma,\sigma,\gamma)$.
	\item In the second step, $\mathcal{S}$ shrinks all world-sheets
	to a world line, i.e., $X_1=\sigma$, $X_2=\gamma$.
\end{enumerate}
Thus we obtain 
\begin{align}
	&\langle \mathcal{S} \left( \mathsf{M^{ST}} \right) \rangle =\langle \underset{\tilde{\omega}\rightarrow \sigma,\omega \rightarrow \gamma}{\lim}\mathsf{M^{ST}} \rangle \nonumber
	\\
	=&\underset{\tilde{\omega}\rightarrow \sigma,\omega \rightarrow \gamma}{\lim}\frac{1}{\mathcal{Z}}\int{\mathcal{D} \left[ ABC \right] \exp \left( iS \right)} \times 8\exp \left[ i\int_{\tilde{\omega}}{C^3+\frac{1}{2}\frac{2\pi q}{N_3}\left( d^{-1}B^1B^2+d^{-1}B^2B^1 \right)} \right] \nonumber
	\\
	&\times \exp \left[ i\int_{\omega}{C^3+\frac{1}{2}\frac{2\pi q}{N_3}\left( d^{-1}B^1B^2+d^{-1}B^2B^1 \right)} \right] \delta \left( \int_{\sigma}{B^1} \right) \delta \left( \int_{\sigma}{B^2} \right) \nonumber
	\\
	=&\frac{1}{\mathcal{Z}}\int{\mathcal{D} \left[ ABC\right] \exp \left( iS \right) 8\exp \left( i0 \right)}\exp \left( i0 \right) \times \delta \left( \int_{\sigma}{B^1} \right) \delta \left( \int_{\sigma}{B^2} \right) \nonumber
	\\
	=&\langle 2\cdot\left(\mathsf{1}\oplus \mathsf{L}_{100,}\oplus \mathsf{L}_{010,}\oplus \mathsf{L}_{110,}\right) \rangle \,.
\end{align}
We continue to shrink all loops to particles:  
\begin{align}
	\langle \mathcal{S} ^2\left( \mathsf{M^{ST}} \right) \rangle =\langle \mathcal{S} \left[ 2\cdot \left( \mathsf{1}\oplus \mathsf{L}_{100,}\oplus \mathsf{L}_{010,}\oplus \mathsf{L}_{110,} \right) \right]  \rangle =\langle 8\cdot \mathsf{1} \rangle \,.
\end{align}
Finally, we get $8$ vacua.

\subsection{Shrinking table\label{section_shrinking_table}}
All shrinking rules for the $BBA$ twisted term are shown in table~\ref{table_shrinking}. We can see that loops and $S^2$ membranes can be directly shrunk to particles in the first shrinking process, thus they do not have a hierarchical shrinking structure. $T^2$ membranes and $\mathsf{M^{ST}}$-excitations have hierarchical shrinking structure. We can also verify that \textit{quantum dimensions are preserved in all shrinking processes i.e.,} $d_{\mathsf{a}}=\sum_{\mathsf{b}}{S_{\mathsf{b}}^{\mathsf{a}}d_{\mathsf{b}}}$.

\begin{table}
	\caption{\label{table_shrinking}Shrinking table  for the $5$D twisted $BF$ theory with the $BBA$ twisted term.  Loops and $S^2$ membranes do not have hierarchical shrinking structures, thus $\mathcal{S}\mathcal{S}=\mathcal{S}$ for them. For excitations that have a hierarchical shrinking structure, we write $\mathcal{S}\mathcal{S}=\mathcal{S}^2$. The meanings of numbers in subscript and superscript are summarized at the end of section~\ref{ss23}. The Wilson operators for excitations are shown in table~\ref{tab_excitation}.}
	\centering
	\setlength{\extrarowheight}{2pt}
	\footnotesize
	\begin{tabular*}{\textwidth}{@{\extracolsep{\fill}}|c|c|c|c|}
		\hline
		\small excitation & \small $\mathcal{S} \left( \text{excitation}\right) $ &\small excitation & \small $\mathcal{S} \left( \text{excitation}\right) $\\
		\hline
		$\mathsf{L}_{100,}$&$\mathsf{1}$&$\mathsf{L}_{100,100}$&$\mathsf{1}\oplus \mathsf{P}_{001}$\\
		\hline
		$\mathsf{L}_{010,}$&$\mathsf{1}$&$\mathsf{L}_{010,010}$&$\mathsf{1}\oplus \mathsf{P}_{001}$\\
		\hline
		$\mathsf{L}_{110,}$&$\mathsf{1}$&$\mathsf{L}_{100,110}$&$\mathsf{1}\oplus \mathsf{P}_{001}$\\
		\hline
		$\mathsf{L}_{100,}^{001}$&$\mathsf{P}_{001}$&$\mathsf{M^S}_{001}$&$2\cdot\mathsf{1}$\\
		\hline
		$\mathsf{L}_{010,}^{001}$&$\mathsf{P}_{001}$&$\mathsf{M^S}_{001}^{001;}$&$2\cdot\mathsf{P_{100}}$\\
		\hline
		$\mathsf{L}_{110,}^{001}$&$\mathsf{P}_{001}$&$\mathsf{M^S}_{001}^{,100}$&$2\cdot\left(\mathsf{1}\oplus \mathsf{P}_{001}\right)$\\
		\hline
		$\mathsf{L}_{,100}$&$\mathsf{1}\oplus \mathsf{P}_{001}$&$\mathsf{M^S}_{001}^{,010}$&$2\cdot\left(\mathsf{1}\oplus \mathsf{P}_{001}\right)$\\
		\hline
		$\mathsf{L}_{,010}$&$\mathsf{1}\oplus \mathsf{P}_{001}$&$\mathsf{M^S}_{001}^{,110}$&$2\cdot\left(\mathsf{1}\oplus \mathsf{P}_{001}\right)$\\
		\hline
		$\mathsf{L}_{,110}$&$\mathsf{1}\oplus \mathsf{P}_{001}$&$-$&$-$\\
		\hline
		\hline
		\small excitation & \small $\mathcal{S} \left( \text{excitation}\right) $&\small $\mathcal{S} ^2\left( \text{excitation}\right) $&$-$\\
		\hline
		$\mathsf{M^T}_{001}$&$\mathsf{1}\oplus \mathsf{L}_{100,}\oplus \mathsf{L}_{010,}\oplus \mathsf{L}_{110,}$&$4\cdot\mathsf{1}$ &$-$\\
		\hline
		$\mathsf{M^T}_{001}^{001;}$&$\mathsf{P}_{001}\oplus \mathsf{L}_{100,}^{001}\oplus \mathsf{L}_{010,}^{001}\oplus \mathsf{L}_{110,}^{001}$&$4\cdot \mathsf{P}_{001}$&$-$\\
		\hline
		$\mathsf{M^{ST}}$&$2\cdot\left(\mathsf{1}\oplus \mathsf{L}_{100,}\oplus \mathsf{L}_{010,}\oplus \mathsf{L}_{110,}\right)$&$8\cdot\mathsf{1}$&$-$\\
		\hline
		$\mathsf{M^{ST}}^{001;}$&$2\cdot\left(\mathsf{P}_{001}\oplus \mathsf{L}_{100,}^{001}\oplus \mathsf{L}_{010,}^{001}\oplus \mathsf{L}_{110,}^{001}\right)$&$8\cdot \mathsf{P}_{001}$&$-$\\
		\hline

		$\mathsf{M^T}_{001}^{,100}$&$\mathsf{1}\oplus \mathsf{L}_{100,}\oplus \mathsf{L}_{010,}\oplus \mathsf{L}_{110,}\oplus\mathsf{P}_{001}\oplus \mathsf{L}_{100,}^{001}\oplus \mathsf{L}_{010,}^{001}\oplus \mathsf{L}_{110,}^{001}$&$4\cdot\left(\mathsf{1}\oplus\mathsf{P}_{001}\right)$&$-$\\
		\hline
		$\mathsf{M^T}_{001}^{,010}$&$\mathsf{1}\oplus \mathsf{L}_{100,}\oplus \mathsf{L}_{010,}\oplus \mathsf{L}_{110,}\oplus\mathsf{P}_{001}\oplus \mathsf{L}_{100,}^{001}\oplus \mathsf{L}_{010,}^{001}\oplus \mathsf{L}_{110,}^{001}$&$4\cdot\left(\mathsf{1}\oplus\mathsf{P}_{001}\right)$&$-$\\
		\hline
		$\mathsf{M^T}_{001}^{,110}$&$\mathsf{1}\oplus \mathsf{L}_{100,}\oplus \mathsf{L}_{010,}\oplus \mathsf{L}_{110,}\oplus\mathsf{P}_{001}\oplus \mathsf{L}_{100,}^{001}\oplus \mathsf{L}_{010,}^{001}\oplus \mathsf{L}_{110,}^{001}$&$4\cdot\left(\mathsf{1}\oplus\mathsf{P}_{001}\right)$&$-$\\
		\hline
		$\mathsf{M^{ST}}^{,100}$&$2\cdot\left(\mathsf{1}\oplus \mathsf{L}_{100,}\oplus \mathsf{L}_{010,}\oplus \mathsf{L}_{110,}\oplus\mathsf{P}_{001}\oplus \mathsf{L}_{100,}^{001}\oplus \mathsf{L}_{010,}^{001}\oplus \mathsf{L}_{110,}^{001}\right)$&$8\cdot\left(\mathsf{1}\oplus\mathsf{P}_{001}\right)$&$-$\\
		\hline
		$\mathsf{M^{ST}}^{,010}$&$2\cdot\left(\mathsf{1}\oplus \mathsf{L}_{100,}\oplus \mathsf{L}_{010,}\oplus \mathsf{L}_{110,}\oplus\mathsf{P}_{001}\oplus \mathsf{L}_{100,}^{001}\oplus \mathsf{L}_{010,}^{001}\oplus \mathsf{L}_{110,}^{001}\right)$&$8\cdot\left(\mathsf{1}\oplus\mathsf{P}_{001}\right)$&$-$\\
		\hline
		$\mathsf{M^{ST}}^{,110}$&$2\cdot\left(\mathsf{1}\oplus \mathsf{L}_{100,}\oplus \mathsf{L}_{010,}\oplus \mathsf{L}_{110,}\oplus\mathsf{P}_{001}\oplus \mathsf{L}_{100,}^{001}\oplus \mathsf{L}_{010,}^{001}\oplus \mathsf{L}_{110,}^{001}\right)$&$8\cdot\left(\mathsf{1}\oplus\mathsf{P}_{001}\right)$&$-$\\
		\hline
	\end{tabular*}
\end{table}

\section{Anomaly-free algebraic structure in fusion rules and shrinking rules}\label{section_algebraic}

It is fundamentally important to study the algebraic structure encoded in topological data that characterizes topological orders. Especially, all data, e.g., fusion coefficients and shrinking coefficients, should be consistent with each other such that they can together define an anomaly-free topological order. Otherwise, we expect that there should be a quantum anomaly of some kind that obstructs the existence of such a topological order. In section~\ref{section_quantum_dimension} and section~\ref{section_shrinking_table}, we have already concluded that: (i) \textit{quantum dimension equals to the normalization factor of the excitation}; (ii) \textit{quantum dimensions and fusion coefficients satisfy that} $	d_{\mathsf{a}}d_{\mathsf{b}}=\sum_{\mathsf{c}}{N_{\mathsf{c}}^{\mathsf{ab}}d_{\mathsf{c}}}$; (iii) \textit{shrinking process preserve quantum dimension, i.e.,} $d_{\mathsf{a}}=\sum_{\mathsf{b}}{S_{\mathsf{b}}^{\mathsf{a}}d_{\mathsf{b}}}$. In this section, we study the consistency relations between fusion and shrinking coefficients with the help of the results of fusion and shrinking rules obtained through the path-integral approach in the previous sections.

Using the fusion table and shrinking table, we can verify that if $\mathsf{a}$ and $ \mathsf{b}$ are excitations without hierarchical shrinking structure, then their shrinking rules respect fusion rules:
\begin{align}
	\mathcal{S} \left( \mathsf{a}\otimes \mathsf{b} \right) =\mathcal{S} \left( \mathsf{a} \right) \otimes \mathcal{S} \left( \mathsf{b} \right) \,. \label{eq_shrinking_and_fusion}
\end{align}
If at least one of $\mathsf{a}$ and $ \mathsf{b}$ have hierarchical shrinking structure, then their hierarchical shrinking rules respect fusion rules:
\begin{align}
	\mathcal{S}^{2} \left( \mathsf{a}\otimes \mathsf{b} \right) =\mathcal{S}^{2} \left( \mathsf{a} \right) \otimes \mathcal{S}^{2} \left( \mathsf{b} \right) \,.
	\label{eq_hieshrinking_and_fusion}
\end{align}

Eq.~(\ref{eq_shrinking_and_fusion}) and eq.~(\ref{eq_hieshrinking_and_fusion}) also imply that there are some algebraic relationships between the fusion coefficient and shrinking coefficient. Using eq.~(\ref{eq_general_fusion}) and eq.~(\ref{eq_shrinking_pathintegral}), we first calculate $\mathcal{S} \left( \mathsf{a}\otimes \mathsf{b} \right)$ for excitations without hierarchical shrinking structure:
\begin{align}
	\mathcal{S} \left( \mathsf{a}\otimes \mathsf{b} \right) =&\mathcal{S} \left( \oplus _{\mathsf{i}}N_{\mathsf{i}}^{\mathsf{ab}}\mathsf{i} \right) =\oplus _{\mathsf{i}}N_{\mathsf{i}}^{\mathsf{ab}}\mathcal{S} \left( \mathsf{i} \right) =\oplus _{\mathsf{i}}N_{\mathsf{i}}^{\mathsf{ab}}\left( \oplus _{\mathsf{c}}\mathrm{S}_{\mathsf{c}}^{\mathsf{i}}\mathsf{c} \right) 
	=\oplus _{\mathsf{c}}\left( \sum_{\mathsf{i}}{\mathrm{S}_{\mathsf{c}}^{\mathsf{i}}N_{\mathsf{i}}^{\mathsf{ab}}} \right) \mathsf{c}	\,.
	\label{eq15}
\end{align}
Now we calculate $\mathcal{S} \left( \mathsf{a} \right) \otimes \mathcal{S} \left( \mathsf{b} \right)$:
\begin{align}
	\mathcal{S} \left( \mathsf{a} \right) \otimes \mathcal{S} \left( \mathsf{b} \right) =&\left( \oplus _{\mathsf{i}}\mathrm{S}_{\mathsf{i}}^{\mathsf{a}}\mathsf{i} \right) \otimes \left( \oplus _{\mathsf{j}}\mathrm{S}_{\mathsf{j}}^{\mathsf{b}}\mathsf{j} \right) =\oplus _{\mathsf{i}}\oplus _{\mathsf{j}}\left[ \mathrm{S}_{\mathsf{i}}^{\mathsf{a}}\mathrm{S}_{\mathsf{j}}^{\mathsf{b}}\left( \mathsf{i}\otimes \mathsf{j} \right) \right] =\oplus _{\mathsf{i}}\oplus _{\mathsf{j}}\left[ \mathrm{S}_{\mathsf{i}}^{\mathsf{a}}\mathrm{S}_{\mathsf{j}}^{\mathsf{b}}\left( \oplus _{\mathsf{c}}N_{\mathsf{c}}^{\mathsf{ij}}\mathsf{c} \right) \right] 	\nonumber\\
	=&\oplus _{\mathsf{c}}\left( \sum_{\mathsf{ij}}{\mathrm{S}_{\mathsf{i}}^{\mathsf{a}}\mathrm{S}_{\mathsf{j}}^{\mathsf{b}}N_{\mathsf{c}}^{\mathsf{ij}}} \right) \mathsf{c}\,.
	\label{eq17}
\end{align}

For excitations that have hierarchical shrinking structure, we calculate $\mathcal{S} ^2\left( \mathsf{a}\otimes \mathsf{b} \right)$ as:
\begin{align}
	\mathcal{S} ^2\left( \mathsf{a}\otimes \mathsf{b} \right) =&\mathcal{S} \mathcal{S} \left( \mathsf{a}\otimes \mathsf{b} \right) =\mathcal{S} \mathcal{S} \left( \oplus _{\mathsf{i}}N_{\mathsf{i}}^{\mathsf{ab}}\mathsf{i} \right) =\mathcal{S} \left[ \oplus _{\mathsf{j}}\left( \sum_{\mathsf{i}}{{\mathrm{S}}_{\mathsf{j}}^{\mathsf{i}}N_{\mathsf{i}}^{\mathsf{ab}}} \right) \mathsf{j} \right] \nonumber
	\\
	=&\oplus _{\mathsf{j}}\left[ \left( \sum_{\mathsf{i}}{{\mathrm{S}}_{\mathsf{j}}^{\mathsf{i}}N_{\mathsf{i}}^{\mathsf{ab}}} \right) \mathcal{S} \left( \mathsf{j} \right) \right] =\oplus _{\mathsf{j}}\left[ \left( \sum_{\mathsf{i}}{{\mathrm{S}}_{\mathsf{j}}^{\mathsf{i}}N_{\mathsf{i}}^{\mathsf{ab}}} \right) \left( \oplus _{\mathsf{c}}\mathrm{S}_{\mathsf{c}}^{\mathsf{j}}\mathsf{c} \right) \right]\nonumber
	\\
	=&\oplus _{\mathsf{c}}\left( \sum_{\mathsf{ij}}{\mathrm{S}_{\mathsf{c}}^{\mathsf{j}}{\mathrm{S}}_{\mathsf{j}}^{\mathsf{i}}N_{\mathsf{i}}^{\mathsf{ab}}} \right) \mathsf{c}\,.
	\label{eq16}
\end{align}
Then we obtain $\mathcal{S} ^2\left( \mathsf{a} \right) \otimes \mathcal{S} ^2\left( \mathsf{b} \right)$:
\begin{align}
	&\mathcal{S} ^2\left( \mathsf{a} \right) \otimes \mathcal{S} ^2\left( \mathsf{b} \right) =\mathcal{S} \mathcal{S} \left( \mathsf{a} \right) \otimes \mathcal{S} \mathcal{S} \left( \mathsf{b} \right) =\mathcal{S} \left( \oplus _{\mathsf{i}}{\mathrm{S}}_{\mathsf{i}}^{\mathsf{a}}\mathsf{i} \right) \otimes \mathcal{S} \left( \oplus _{\mathsf{j}}{\mathrm{S}}_{\mathsf{j}}^{\mathsf{b}}\mathsf{j} \right) \nonumber
	\\
	=&\left[ \oplus _{\mathsf{k}}\left( \sum_{\mathsf{i}}{{\mathrm{S}}_{\mathsf{i}}^{\mathsf{a}}\mathrm{S}_{\mathsf{k}}^{\mathsf{i}}} \right) \mathsf{k} \right] \otimes \left[ \oplus _{\mathsf{l}}\left( \sum_{\mathsf{j}}{{\mathrm{S}}_{\mathsf{j}}^{\mathsf{b}}\mathrm{S}_{\mathsf{l}}^{\mathsf{j}}} \right) \mathsf{l} \right] =\oplus _{\mathsf{k}}\oplus _{\mathsf{l}}\left[ \left( \sum_{\mathsf{ij}}{{\mathrm{S}}_{\mathsf{i}}^{\mathsf{a}}\mathrm{S}_{\mathsf{k}}^{\mathsf{i}}{\mathrm{S}}_{\mathsf{j}}^{\mathsf{b}}\mathrm{S}_{\mathsf{l}}^{\mathsf{j}}} \right) \left( \mathsf{k}\otimes \mathsf{l} \right) \right] \nonumber
	\\
	=&\oplus _{\mathsf{k}}\oplus _{\mathsf{l}}\left[ \left( \sum_{\mathsf{ij}}{{\mathrm{S}}_{\mathsf{i}}^{\mathsf{a}}\mathrm{S}_{\mathsf{k}}^{\mathsf{i}}{\mathrm{S}}_{\mathsf{j}}^{\mathsf{b}}\mathrm{S}_{\mathsf{l}}^{\mathsf{j}}} \right) \left( \oplus _{\mathsf{c}}N_{\mathsf{c}}^{\mathsf{kl}}\mathsf{c} \right) \right] =\oplus _{\mathsf{c}}\left( \sum_{\mathsf{ijkl}}{{\mathrm{S}}_{\mathsf{i}}^{\mathsf{a}}{\mathrm{S}}_{\mathsf{j}}^{\mathsf{b}}\mathrm{S}_{\mathsf{k}}^{\mathsf{i}}\mathrm{S}_{\mathsf{l}}^{\mathsf{j}}}N_{\mathsf{c}}^{\mathsf{kl}} \right) \mathsf{c}\,.
	\label{eq18}
\end{align}

Comparing eq.~(\ref{eq15}) and eq.~(\ref{eq17}), eq.~(\ref{eq16}) and eq.~(\ref{eq18}), we obtain
\begin{gather}
	\sum_{\mathsf{i}}{\mathrm{S}_{\mathsf{c}}^{\mathsf{i}}N_{\mathsf{i}}^{\mathsf{ab}}}=\sum_{\mathsf{ij}}{\mathrm{S}_{\mathsf{i}}^{\mathsf{a}}\mathrm{S}_{\mathsf{j}}^{\mathsf{b}}N_{\mathsf{c}}^{\mathsf{ij}}}\,,\label{eq20-}
	\\ \sum_{\mathsf{ij}}{\mathrm{S}_{\mathsf{c}}^{\mathsf{j}}{\mathrm{S}}_{\mathsf{j}}^{\mathsf{i}}N_{\mathsf{i}}^{\mathsf{ab}}}=\sum_{\mathsf{ijkl}}{{\mathrm{S}}_{\mathsf{i}}^{\mathsf{a}}{\mathrm{S}}_{\mathsf{j}}^{\mathsf{b}}\mathrm{S}_{\mathsf{k}}^{\mathsf{i}}\mathrm{S}_{\mathsf{l}}^{\mathsf{j}}}N_{\mathsf{c}}^{\mathsf{kl}}\,.
	\label{eq20}
\end{gather}
Eq.~(\ref{eq20-}) only apply for excitations without hierarchical shrinking structure, i.e., $\mathsf{a}$ and $\mathsf{b}$ can only be particles, loops and $S^2$ membranes. The complete relation between all fusion coefficients and all shrinking coefficients is given by eq.~(\ref{eq20}), where $\mathsf{a}$ and $\mathsf{b}$ can be any excitations.

\section{General discussions on all twisted terms in 5D\label{section_general_discuss_twisted5D}}
In the above analysis, we have only considered the twisted $BF$ theory with the  $BBA$ twisted term, but there are many other legitimate twisted terms in $5$D topological orders, e.g., $AAAB$, $AAAAA$, $AAAdA$, $AAC$, $AdAdA$, $AdAB$, and $AAdB$. If gauge-invariant combinations among them are taken into account, there are even more twisted $BF$ theories as shown in ref.~\cite{Zhang:2021ycl}.  In principle, we can construct equivalence classes of  Wilson operators, and compute fusion and shrinking rules explicitly. For example, we find $AAAB$ also simultaneously supports non-Abelian fusion rules, non-Abelian shrinking rules and hierarchical shrinking rules (see appendix~\ref{ap4}). As shown in appendix~\ref{ap2}, we find that a simple method allows us to determine the general properties of fusion and shrinking rules from gauge transformations and some simple Wilson operators. We summarize these properties in table~\ref{table_comparison_5dtwisted} and discuss this table below. 
\begin{table}
	\caption{\label{table_comparison_5dtwisted}Comparison of fusion and shrinking rules in different twisted terms.}
	\centering
	\begin{tabular*}{\textwidth}{@{\extracolsep{\fill}}|c|c|c|c|}
		\hline
		twisted term & fusion rules & shrinking rules  & hierarchical shrinking rules \\
		\hline
		$BBA$ & non-Abelian & non-Abelian & yes \\
		\hline
		$AAAB$ & non-Abelian & non-Abelian & yes \\
		\hline
		$AAAAA$ & non-Abelian & non-Abelian & no \\
		\hline
		$AAAdA$ & non-Abelian & non-Abelian & no \\
		\hline
		$AAC$ & non-Abelian & non-Abelian & no \\
		\hline
		$AdAdA$ & Abelian & Abelian & no \\
		\hline
		$AdAB$ & Abelian & Abelian & no \\
		\hline
		$AAdB$ & Abelian & Abelian & no \\
		\hline
	\end{tabular*}
\end{table}
\begin{itemize}
\item Both $BBA$ and $AAAB$ twisted terms simultaneously support non-Abelian fusion rules, non-Abelian shrinking rules, and hierarchical shrinking rules. According to the discussion in appendix~\ref{ap2}, delta functionals carried by Wilson operators determine the type of fusion and shrinking rules. We find that in $BBA$ and $AAAB$, Wilson operators for membranes carry $\delta \left( \int_{\sigma}{B^i} \right) $, which  leads to hierarchical shrinking rules. The presence of hierarchical shrinking rules indicates that some fusion and shrinking rules are non-Abelian.

\item $AAAAA$, $AAAdA$, and $AAC$ support non-Abelian fusion rules and non-Abelian shrinking rules but they do not support hierarchical shrinking rules. As studied in ref.~\cite{Zhang:2021ycl}, the twisted $BF$ theories with these twisted terms contain only type-I $BF$ term, i.e., $CdA$. Since there are no $B^i$ fields, none of membranes carry $\delta \left( \int_{\sigma}{B^i} \right) $ and thus they do not have hierarchical shrinking. A pure $T^2$ membrane can only be shrunk to a superposition of  trivial loop (vacuum) and particles. Although $\delta \left( \int_{\sigma}{B^i} \right) $ is absent, some Wilson operators for membranes carry $\delta \left( \int_{\gamma}{A^i} \right) $, which can lead to non-Abelian fusion and shrinking rules.

\item $AdAdA$, $AdAB$, and $AAdB$ twisted terms only support Abelian fusion and Abelian shrinking rules. Some Wilson operators may contain nontrivial terms besides $A^i$ or $B^i$ or $C^i$ to compensate for the nontrivial shift in gauge transformation, but they do not carry delta functionals that lead to non-Abelian fusion rules and non-Abelian shrinking rules.
\end{itemize}

The presence of hierarchical shrinking leads to an interesting phenomenon: $S^2$ and $T^2$ membranes have different quantum dimensions. In a theory without hierarchical shrinking, they only differ by shape. Remember we request the fusion coefficients and the shrinking coefficients are all integers, this affects the normalization factors of Wilson operators (details are shown in appendix~\ref{ap_normalization_factors}). In the above analysis, we have seen that an $S^2$ membrane and a $T^2$ membrane carrying the same gauge charges have different normalization factors as long as the $T^2$ membrane can shrink to nontrivial loops. Different normalization factors lead to different fusion coefficients and thus lead to different fusion coefficient matrices. Since quantum dimension is defined as the greatest eigenvalue of the fusion coefficient matrix,  $S^2$ and $T^2$ membranes have different quantum dimensions. This is absent in the theory without hierarchical shrinking, since the $S^2$ and $T^2$ membranes carrying the same gauge charges have identical normalization factors. To be more specific, each gauge charge of $C^i$ field contributes an extra $2$ to the normalization factor of a $T^2$ membrane. This can be clearly viewed from $AAAB$ twisted term shown in appendix~\ref{ap4}. From eq.~(\ref{eq31}), we can see when the membrane carries $n$ units of gauge charges of $C^i$ field, we have $\tilde{N}=2^nN$. This result also applies to the $BBA$ twisted term since there are only $\mathbb{Z}_{N_3}$ membranes and we do find that $\mathsf{M^T}_{001}$ has normalization factor $4$ while $\mathsf{M^S}_{001}$ has $2$.

\section{Summary and outlook}\label{section_summary}

In this paper, as a series of work on $5$D topological orders, we   concentrated on the fusion and shrinking rules of $5$D topological orders by means of the path-integral formalism of TQFTs proposed in ref.~\cite{Zhang:2021ycl}. More concretely, we constructed Wilson operators for all topologically distinct excitations by carefully investigating equivalence classes among all possible gauge-invariant Wilson operators.  
Based on Wilson operators, we may derive fusion rules, quantum dimensions, and shrinking rules. We found exotic non-Abelian fusion and non-Abelian shrinking, and especially identified exotic hierarchical structure in shrinking processes.  The underlying algebraic structure of the fusion coefficients and shrinking coefficients was also found in terms of several exact algebraic equations about the coefficients, which sets an anomaly-free condition on $5$D topological orders.  Below we present several future directions motivated by this paper.  (i) Adopting the strategy in ref.~\cite{Ning2018prb}, we can explore symmetry-enrichment in $5$D topological orders. While the idea of ``mixed $3$-loop statistics'' plays an important role in characterizing and classifying SET phases in ref.~\cite{Ning2018prb}, it will be interesting to generalize such mixed type of braiding statistics to probe SET order in higher dimensions, especially when higher-form symmetry~\cite{mcgreevy2023} is considered and higher-form response gauge fields are involved.
(ii) In this work, we focused on simple-shaped topological excitations, including particles, unknotted loops, $S^2$ membranes, $T^2$ membranes, and combination of these membranes. However, it is important to explore more complicated spatially extended excitations, e.g., $T^2$ membranes with higher genus, to investigate if more exotic topological data that should be incorporated for a comprehensive understanding of $5$D topological orders. 
(iii) Finally, an interesting avenue for future research involves the exploration of fermionic topological orders by introducing TQFTs to a spacetime manifold with a spin structure. This could lead to novel insights into the behavior of fermionic systems in higher dimensions.

\acknowledgements
This work was supported by NSFC Grant No. 12074438, Guangdong Basic and Applied Basic Research Foundation under Grant No. 2020B1515120100, the Open Project of Guangdong Provincial Key Laboratory of Magnetoelectric Physics and Devices under Grant No. 2022B1212010008, and the Fundamental Research Funds for the Central Universities, Sun Yat-sen University (No. 23ptpy05).

\appendix
\section{Normalization factors of Wilson operators\label{ap_normalization_factors}}
In section~\ref{ss21}, we introduced a normalization factor for the Wilson operator corresponding to a membrane excitation in eq.~(\ref{eq_op_sphere_001}). However, it's important to note that all Wilson operators are associated with normalization factors. In this appendix, we will explain the necessity of these normalization factors and how they are derived. We will consider several principles related to the fusion and shrinking of topological excitations.
\begin{itemize}
	\item [1] Fundamental Excitations: We start by considering topological excitations with trivial shapes, i.e., particles in the shape of $S^0$ (a point), loops in the shape of $S^1$ (no knots or links), and membranes in the shape of $S^2$ (we exclude torus-shaped membranes for now). These excitations carry only one unit of one species of gauge charge. In the twisted $BF$ theory (eq.~(\ref{eq_action_BBA})) with the gauge group $G=\left(\mathbb{Z}_2\right)^3$, each excitation is its own anti-excitation due to the $\Z_2$ cyclicity. Fusion of two such excitations should yield \emph{one} vacuum, which is represented by $N^{\mathsf{a}\mathsf{a}}_{\mathsf{1}}=1$, where $\mathsf{a}$ is an excitation with a trivial shape and one unit of one species of gauge charge. This principle allows us to find the normalization factors for $\mathsf{P}_{001}$, $\mathsf{L}_{100,}$, and $\mathsf{L}_{010,}$ to be equal to $1$.
	
	\item [2] Multiple Species of Gauge Charges: We extend our consideration to excitations with trivial shapes but carrying multiple species of gauge charges. These excitations can be defined through the fusion of those carrying only one unit of one species of gauge charge. For example,  $\mathsf{L}_{110,}=\mathsf{L}_{100,} \otimes \mathsf{L}_{010,}=N\left(\mathsf{L}_{110,}\right)\cdot \exp\left(i\int_{\sigma} B^1 +i \int_{\sigma} B^2\right)$. By comparing the coefficients, we can determine the normalization factor for $\mathsf{L}_{110,}$ to be $1$. 
	
	\item [3] Integer Coefficients: Regardless of the shape of an excitation, the fusion and shrinking coefficients must be integers.
\end{itemize}
Guided by above principles, we can determine the normalization factors for Wilson operators of excitations with trivial topological shape (i.e., $S^0$, $S^1$, $S^2$; we will come to $T^2$ membrane shortly), no matter the excitation is Abelian or non-Abelian. More examples are as follows. Consider a $S^2$ membrane $\mathsf{M^S}_{001}$: 

%
\begin{gather}
	\mathsf{M^S}_{001}=N\exp \left[ i\int_{\omega}{C^3+\frac{1}{2}\frac{2\pi q}{N_3}\left( d^{-1}B^1B^2+d^{-1}B^2B^1 \right)} \right] \delta \left( \int_{\sigma}{B^1} \right) \delta \left( \int_{\sigma}{B^2} \right) \,,
\end{gather}
where $N$ is the normalization factors. Fusion of two copies of $\mathsf{M^S}_{001}$ is 
\begin{align}
	{\mathsf{M}^{\mathsf{S}}}_{001}\otimes {\mathsf{M}^{\mathsf{S}}}_{001}=\frac{N^2}{4} \left( \mathsf{1}\oplus \mathsf{L}_{100,}\oplus \mathsf{L}_{010,}\oplus \mathsf{L}_{110,} \right) \,.
\end{align}
The fusion coefficient of vacuum $\mathsf{1}$ should be one which implies that the normalization factor of $\mathsf{M^S}_{001}$ is $N=2$. 
If we consider a $S^{2}$ membrane decorated by a loop, e.g., 
\begin{align}
	\mathsf{M^{S}}_{001}^{,100}= & N\left(\mathsf{M^{S}}_{001}^{,100}\right)\nonumber \\
	& \times\exp\left[i\int_{\omega}C^{3}+\frac{1}{2}\frac{2\pi q}{N_{3}}\left(d^{-1}B^{1}B^{2}+d^{-1}B^{2}B^{1}\right)\right.\nonumber \\
	& \left.+i\int_{\sigma}\widetilde{B}^{1}-\frac{1}{2}\frac{2\pi q}{N_{1}}\left(d^{-1}A^{3}B^{2}+d^{-1}B^{2}A^{3}\right)\right]\nonumber \\
	& \times\delta\left(\int_{\sigma}B^{1}\right)\delta\left(\int_{\sigma}B^{2}\right)\delta\left(\int_{\gamma}A^{3}\right)\,,
\end{align}
such an excitation is defined by $\mathsf{M^{S}}_{001}^{,100}=\mathsf{M}_{001}^{\mathsf{S}}\otimes\mathsf{L}_{,100}$.
The Wilson operator for $\mathsf{L}_{,100}$ is $\mathsf{L}_{,100}=2\exp\left[i\int_{\sigma}\widetilde{B}^{1}-\frac{1}{2}\frac{2\pi q}{N_{1}}\left(d^{-1}A^{3}B^{2}+d^{-1}B^{2}A^{3}\right)\right]\delta\left(\int_{\sigma}B^{2}\right)\delta\left(\int_{\gamma}A^{3}\right)$
where the normalization factor $2$ is determined in a way similar
to that of $\mathsf{M}_{001}^{\mathsf{S}}$. By simply substituting
the Wilson operators in $\mathsf{M^{S}}_{001}^{,100}=\mathsf{M}_{001}^{\mathsf{S}}\otimes\mathsf{L}_{,100}$,
we obtain 
\begin{align}
	\mathsf{M^{S}}_{001}^{,100}= & 4\exp\left[i\int_{\omega}C^{3}+\frac{1}{2}\frac{2\pi q}{N_{3}}\left(d^{-1}B^{1}B^{2}+d^{-1}B^{2}B^{1}\right)\right.\nonumber \\
	& \left.+i\int_{\sigma}\widetilde{B}^{1}-\frac{1}{2}\frac{2\pi q}{N_{1}}\left(d^{-1}A^{3}B^{2}+d^{-1}B^{2}A^{3}\right)\right]\nonumber \\
	& \times\delta\left(\int_{\sigma}B^{1}\right)\delta\left(\int_{\sigma}B^{2}\right)\delta\left(\int_{\gamma}A^{3}\right)\,,
\end{align}
i.e., the normalization of $\mathsf{M^{S}}_{001}^{,100}$ is $4$. 

Following the procedures described above, we can determine normalization factors of all excitations as long as its shape has \emph{trivial topology}. Once the normalization factors are fixed, we obtain all fusion and shrinking rules involving these excitations.

However, when dealing with membrane excitations with nontrivial topological shapes, such as a $T^2$ membrane, we need to approach the normalization factor with extra care. These membranes, having nontrivial shapes, may introduce internal degrees of freedom that the first principle, which worked for trivial shapes, no longer fully addresses. For example, consider the following $T^2$ membrane:
 \begin{align}
	\mathsf{M^T}_{001}=\tilde{N}\exp \left[ i\int_{\tilde{\omega}}{C^3+\frac{1}{2}\frac{2\pi q}{N_3}\left( d^{-1}B^1B^2+d^{-1}B^2B^1 \right)} \right] \delta \left( \int_{\sigma}{B^1} \right) \delta \left( \int_{\sigma}{B^2} \right)\,,
\end{align}
and fusion of two copies of $\mathsf{M^T}_{001}$ as well as its shrinking:
\begin{align}
	{\mathsf{M}^{\mathsf{T}}}_{001}\otimes {\mathsf{M}^{\mathsf{T}}}_{001}= & \frac{\tilde{N}^2}{4}\cdot \left( \mathsf{1}\oplus \mathsf{L}_{100,}\oplus \mathsf{L}_{010,}\oplus \mathsf{L}_{110,} \right) \,,\label{eq_fusion_two_torus}\\
	\mathcal{S} \left( {\mathsf{M}^{\mathsf{T}}}_{001} \right) =&\frac{\tilde{N}}{4}\cdot \left( \mathsf{1}\oplus \mathsf{L}_{100,}\oplus \mathsf{L}_{010,}\oplus \mathsf{L}_{110,} \right) \label{eq_shrink_torus}\,.
\end{align}
Our third principle requires that the fusion and shrinking coefficients are integers. A natural and economic choice is $\tilde{N}=4$ which indicates 
\begin{align}
	{\mathsf{M}^{\mathsf{T}}}_{001}\otimes {\mathsf{M}^{\mathsf{T}}}_{001}=&4\cdot \left( 1\oplus \mathsf{L}_{100,}\oplus \mathsf{L}_{010,}\oplus \mathsf{L}_{110,} \right)   \,,\label{eq_fusion_two_torus_factor_set}\\
	\mathcal{S} \left( {\mathsf{M}^{\mathsf{T}}}_{001} \right) =&  \mathsf{1}\oplus \mathsf{L}_{100,}\oplus \mathsf{L}_{010,}\oplus \mathsf{L}_{110,}  \label{eq_shrink_torus_factor_set}\,.
\end{align}
Here we make some explanations to justify the normalization factor of ${\mathsf{M}^{\mathsf{T}}}_{001}$. Geometrically, shrinking a torus results in a loop. The original gauge charge of $3$-form gauge field $C^3$ cannot be carried by a loop since the dimension reduces. Therefore, a trivial loop that is equivalent to a vacuum is expected. This is why we set the normalization factor $\tilde{N}=4$ such that one vacuum (trivial loop) is obtained after shrinking a $T^2$ membrane. As for other outputs of shrinking like $\mathsf{L}_{100,}\oplus \mathsf{L}_{010,}\oplus \mathsf{L}_{110,}$, they can be traced back to the fact that some loops can be decorated on a $T^2$ membrane without changing the correlation functions (formally indicated by the delta functionals). Notice the hierarchical shrinking property of $T^2$ membrane, $\mathcal{S}^2 \left({\mathsf{M}^{\mathsf{T}}}_{001}\right)=4\cdot \mathsf{1}$. After this process, a $T^2$ membrane is shrunk to a point. The $4$ vacua can be understood as the results of shrinking $4$ loops (one trivial loop and three nontrivial loops) leads to $4$ trivial particles (i.e., vacua). 

On the other hand, eq.~(\ref{eq_fusion_two_torus_factor_set}) indicates that fusion of a $T^2$ membrane ${\mathsf{M}^{\mathsf{T}}}_{001} $ and its anti-membrane (which is just itself due to $\mathbb{Z}_{N_3}=\Z_2$ ) results in a superposition of 4 vacua. The multiplicity of vacua implies the existence of some internal degrees of freedom, which may be introduced by the nontrivial topology of torus. There is an analog in $\left(3+1\right)$D topological order: in ref.~\cite{ypdw}, the authors studied loop excitations in shape of nontrivial knots and links and found that they have larger quantum dimensions than those in shape of trivial $S^1$. This shows that the shape of topological excitation may bring in new degrees of freedom which definitely deserves further exploration. 

In summary, we follow from three principles based on physical intuition to determine normalization factors for Wilson operators, as presented in table~\ref{tab_excitation}. With these normalization factors in place, we can derive the complete fusion and shrinking algebra, allowing us to compute the quantum dimensions of excitations.

\section{Nontrivial terms in Wilson operators\label{ap1}}
When we construct the Wilson operators, we have to make sure they are gauge invariant. If the gauge field transforms with nontrivial shifts, we will have to add some nontrivial terms to compensate for these nontrivial shifts in Wilson operators. For example in $BBA$ twisted term, $C^3$ transforms with nontrivial shifts
\begin{align}
	C^3\rightarrow C^3+dT^3+\frac{2\pi q}{N_3}\left( -V^1B^2-B^1V^2-\frac{1}{2}V^1dV^2-\frac{1}{2}V^2dV^1 \right) \,,
\end{align}
thus we construct the Wilson operator for $\mathsf{M^S}_{001}$ as 
\begin{align}
	\mathsf{M^S}_{001}=2\exp \left[ i\int_{\omega}{C^3+\frac{1}{2}\frac{2\pi q}{N_3}\left( d^{-1}B^1B^2+d^{-1}B^2B^1 \right)} \right] \delta \left( \int_{\sigma}{B^1} \right) \delta \left( \int_{\sigma}{B^2} \right) \,,
\end{align}
where $\frac{1}{2}\frac{2\pi q}{N_3}\left( d^{-1}B^1B^2+d^{-1}B^2B^1 \right)$ is introduced to compensate for the nontrivial shifts.

When we consider $\langle \mathsf{M^S}_{001}\otimes \mathsf{M^S}_{001} \rangle $, we encounter 
\begin{align}
	&\exp \left[ i2\int_{\omega}{\frac{1}{2}\frac{2\pi}{N_3}\frac{pN_1N_2N_3}{\left( 2\pi \right) ^2N_{123}}\left( d^{-1}\hat{B}^1\hat{B}^2+d^{-1}\hat{B}^2\hat{B}^1 \right)} \right] \nonumber
	\\
	=&\exp \left[ i2\times\frac{1}{2}\frac{2\pi}{N_3}\frac{pN_1N_2N_3}{\left( 2\pi \right) ^2N_{123}}\left( \frac{2\pi k_1}{N_1}\frac{2\pi m_2}{N_2}+\frac{2\pi m_1}{N_1}\frac{2\pi k_2}{N_2} \right) \right] \nonumber
	\\
	=&\exp \left[ i\pi p\left( k_1m_2+m_1k_2 \right) \right] \,.
\end{align}
At the first sight, $\exp \left[ i\pi p\left( k_1m_2+m_1k_2 \right) \right]$ can be $+1$ or $-1$ since $k_1,k_2,m_1,m_2$ are all integers. However we can find that $k_1m_2$ is actually equal to $m_1k_2$, because $d^{-1}B^1B^2$ is equal to $d^{-1}B^2B^1$ up to a boundary term. Notice that the delta functionals enforce $\int_{\sigma}{B^1}$ and $\int_{\sigma}{B^2}$ are $0$ mod $2\pi$, thus $B^1$ and $B^2$ are exact on $\omega$. We can find two gauge fields $F_1$ and $F_2$ satisfy that $dF_1=B^1$ and $dF_2=B^2$, thus we have
\begin{align}
	\int{d^{-1}B^2B^1}=\int{F_2dF_1}=\int{dF_2F_1-d\left( F_2F_1 \right)}=\int{F_1dF_2}=\int{d^{-1}B^1B^2}\,.
\end{align}
We drop the boundary term because it does not contribute to the integral. Now we can see that $\frac{2\pi k_1}{N_1}\frac{2\pi m_2}{N_2}=\frac{2\pi m_1}{N_1}\frac{2\pi k_2}{N_2}$ and thus $k_1m_2=m_1k_2$. Finally, we obtain 
\begin{align}
	\exp \left[ i\pi p\left( k_1m_2+m_1k_2 \right) \right] =1\,.
\end{align}

Similarly, in $AAAB$ twisted term, we construct the gauge invariant Wilson operator for $\mathsf{M^S}_{1000}$ as 
\begin{align}
	\mathsf{M^S}_{1000}=&4\exp \left[ i\int_{\omega}{C^1-\frac{1}{3}\frac{2\pi q}{N_1}\left( B^4A^2d^{-1}A^3-B^4A^3d^{-1}A^2-A^2A^3d^{-1}B^4 \right)} \right] \nonumber
	\\
	&\times \delta \left( \int_{\gamma}{A^2} \right) \delta \left( \int_{\gamma}{A^3} \right) \delta \left( \int_{\sigma}{B^4} \right) \,.
\end{align}
Due to delta functionals, we can see that $B^4$, $A^2$ and $A^3$ are all exact on $\omega$ and we can find $dF_4=B^4$, $dF_2=A^2$ and $dF_3=A^3$. Thus
\begin{align}
	\int{B^4A^2d^{-1}A^3}=&\int{dF_4dF_2F_3}=\int{d\left( F_4dF_2 \right) F_3}=\int{d\left( F_4dF_2F_3 \right) -F_4dF_2dF_3} \nonumber
	\\
	=&\int{-dF_2dF_3F_4}=\int{-A^2A^3d^{-1}B^4}\,,\nonumber
	\\
	\int{B^4A^3d^{-1}A^2}=&\int{dF_4dF_3F_2}=\int{d\left( F_4dF_3 \right) F_2}=\int{d\left( F_4dF_3F_2 \right) -F_4dF_3dF_2}\nonumber
	\\
	=&\int{dF_2dF_3F_4}=\int{A^2A^3d^{-1}B^4}\,.
\end{align}
We obtain that $B^4A^2d^{-1}A^3=-B^4A^3d^{-1}A^2=-A^2A^3d^{-1}B^4$ after dropping some boundary terms. When we consider fusing two $\mathsf{M^S}_{1000}$, we encounter 
\begin{align}
	&\exp \left[ i2\int_{\omega}{\frac{1}{3}\frac{2\pi}{N_1}\frac{pN_1N_2N_3N_4}{\left( 2\pi \right) ^3N_{1234}}\left( \hat{B}^4\hat{A}^2d^{-1}\hat{A}^3-\hat{B}^4\hat{A}^3d^{-1}\hat{A}^2-\hat{A}^2\hat{A}^3d^{-1}\hat{B}^4 \right)} \right] \nonumber
	\\
	=&\exp \left[ i2\times \frac{1}{3}\frac{2\pi}{N_1}\frac{pN_1N_2N_3N_4}{\left( 2\pi \right) ^3N_{1234}} \right. \nonumber
	\\
	&\left. \times \left( \frac{2\pi m_1}{N_4}\frac{2\pi m_2}{N_2}\frac{2\pi k_3}{N_3}-\frac{2\pi m_4}{N_4}\frac{2\pi m_3}{N_3}\frac{2\pi k_2}{N_2}-\frac{2\pi m_2}{N_2}\frac{2\pi m_3}{N_3}\frac{2\pi k_4}{N_4} \right) \right] \nonumber
	\\
	=&\exp \left[ \frac{i2\pi p}{3}\left( m_1m_2k_3-m_4m_3k_2-m_2m_3k_4 \right) \right] =\exp \left[ \frac{i2\pi p}{3}\left( 3m_1m_2k_3\right) \right]=1\,.
\end{align}
For other excitations, we have similar calculations. 

In short, when calculating fusion rules, we have $\langle \exp \left[ i2\int_{X_{\mathsf{i}}}{F^{\mathsf{i}}+\left( \cdots \right)} \right] \rangle =1$ in our theory since $G=\left( \mathbb{Z} _2 \right) ^3$, where $X_{\mathsf{i}}$ is the world-line (sheet or volume) and $F^{\mathsf{i}}$ is the gauge field. 
\section{Fusion and shrinking rules for the $AAAB$ twisted term\label{ap4}}
We have already seen that $BBA$ twisted term can present non-Abelian fusion and shrinking rules and hierarchical shrinking rules. $AAAB$ twisted term can also present these properties. In this appendix, we study fusion and shrinking rules for the $AAAB$ term. We can clearly see that for $T^2$ membranes, each gauge charge of $C^i$ field contributes an extra $2$ factor to the quantum dimension compared to $S^2$ membranes.

When the gauge group is $G={\left( \mathbb{Z} _{2} \right)}^4$, we find that there are $165$ nonequivalent excitations. We will present  some typical fusion rules and give several formulas to describe all shrinking rules. Since we deal with the $AAAB$ term in a similar way to the $BBA$ term, we will only show our results. 
\subsection{TQFT action and gauge transformations for $AAAB$ twisted term}
For the $AAAB$ twisted term, the index of all $A$ and $B$ fields must be different. Thus the simplest case is take the gauge group $G={\left( \mathbb{Z} _{2} \right)}^4$. Then we can write down the TQFT action:
\begin{align}
	S=\int{\sum_{i=1}^3{\frac{N_i}{2\pi}C^idA^i}+}\frac{N_4}{2\pi}\tilde{B}^4dB^4+qA^1A^2A^3B^4\,,
\end{align}
where $q=\frac{pN_1N_2N_3N_4}{\left( 2\pi \right) ^3N_{1234}}$, $p\in \mathbb{Z} _{N_{1234}}$, $N_{1,2,3,4}=N_{1234}=2$. The gauge transformations are:
\begin{align}
	&A^i\rightarrow A^i+d\chi ^i,B^4\rightarrow B^4+dV^4\,,
	\label{eq21}
	\\
	&C^i\rightarrow C^i+dT^i+\frac{2\pi q}{N_i}\sum_{j,k}{\epsilon ^{ijk}}\left[ \left( A^j\chi ^k-\frac{1}{2}\chi ^jd\chi ^k \right) \left( B^4+dV^4 \right) -\frac{1}{2}A^jA^kV^4 \right] \,,
	\\
	&\tilde{B}^4\rightarrow \tilde{B}^4+d\tilde{V}^4+\frac{2\pi q}{N_4}\sum_{i,j,k}{\epsilon ^{4ijk}}\left( -\frac{1}{2}\chi ^iA^jA^k+\frac{1}{2}A^i\chi ^jd\chi ^k-\frac{1}{6}\chi ^id\chi ^jd\chi ^k \right) \,,
\end{align}
where $\epsilon$ is the anti-symmetric tensor and $i,j,k=1,2,3$.
\subsection{Nonequivalent Wilson operators for excitations}
Now we can construct Wilson operators for different excitations. We will illustrate it in $4$ different cases: particles, loops, membranes, and $\mathsf{M^{ST}}$-excitations.

\textbf{\textit{Particles.}}
There are $8$ different particles and  they are all Abelian excitations because in eq.~(\ref{eq21}), $A^i$ transform normally (without nontrivial shifts). Thus we can easily write down their Wilson operators in a general way:
\begin{align}
	\mathsf{P}_{n_{1}n_{2}n_{3}0}=\exp \left( in_{1}\int_{\gamma}{A^1}+in_{2}\int_{\gamma}{A^2}+in_{3}\int_{\gamma}{A^3} \right)\,,
	\label{AAABparticle}
\end{align}
where $n_{1},n_{2},n_{3}=0,1$ and $n_{4}=0$.

\textbf{\textit{Loops.}}
All $B$-loops are Abelian excitations and their Wilson operators are:
\begin{align}
	\mathsf{L}_{000n_4,}^{n_1n_2n_30}=\exp \left( in_4\int_{\sigma}{B^4}+in_1\int_{\gamma}{A^1}+in_2\int_{\gamma}{A^2}+in_3\int_{\gamma}{A^3} \right) \,.
	\label{AAABloop1}
\end{align}
Other loops are non-Abelian excitations:
\begin{align}
	&\mathsf{L}_{000n_4,0001}\nonumber
	\\
	=&8\exp \left[ in_4\int_{\sigma}{B^4}+i\int_{\sigma}{\tilde{B}^4}-\frac{1}{3}\frac{2\pi q}{N_4}\left( A^1A^2d^{-1}A^3-A^1A^3d^{-1}A^2+A^2A^3d^{-1}A^1 \right) \right] \nonumber 
	\\
	&\times \delta \left( \int_{\gamma}{A^1} \right) \delta \left( \int_{\gamma}{A^2} \right) \delta \left( \int_{\gamma}{A^3} \right) \,.
	\label{AAABloop2}
\end{align}
Where $n_{1},n_{2},n_{3},n_{4}=0,1$. There are $8$ Abelian loops and $2$ non-Abelian loops. For non-Abelian loops, all particle decorations are trivial (i.e., particle decorations only lead to equivalent excitations).

\textbf{\textit{$S^2$ and $T^2$ membranes.}}
All membranes are non-Abelian excitations. We give a general formula for all nonequivalent $S^2$ membranes first, they can be written as 
\begin{align}
	{\mathsf{M}^{\mathsf{S}}}_{n_1n_2n_30}^{m_1m_2m_30;,000\tilde{l}_4}=&N_{n_1n_2n_3\tilde{l}_4}\exp \left[ im_1\int_{\gamma}{A^1}+im_2\int_{\gamma}{A^2}+im_3\int_{\gamma}{A^3} \right. \nonumber
	\\
	&+in_1\int_{\omega}{C^1-\frac{1}{3}\frac{2\pi q}{N_1}\left( B^4A^2d^{-1}A^3-B^4A^3d^{-1}A^2-A^2A^3d^{-1}B^4 \right)} \nonumber
	\\
	&+in_2\int_{\omega}{C^2+\frac{1}{3}\frac{2\pi q}{N_2}\left( B^4A^1d^{-1}A^3-B^4A^3d^{-1}A^1-A^1A^3d^{-1}B^4 \right)}\nonumber
	\\
	&+in_3\int_{\omega}{C^3-\frac{1}{3}\frac{2\pi q}{N_3}\left( B^4A^1d^{-1}A^2-B^4A^2d^{-1}A^1-A^1A^2d^{-1}B^4 \right)} \nonumber
	\\
	&\left.+i\tilde{l}_4\int_{\sigma}{\tilde{B}^4}-\frac{1}{3}\frac{2\pi q}{N_4}\left( A^1A^2d^{-1}A^3-A^1A^3d^{-1}A^2+A^2A^3d^{-1}A^1 \right) \right] \nonumber
	\\
	&\times \delta \left[ \int_{\gamma}{n_1A^2-n_2A^1} \right] \delta \left[ \int_{\gamma}{n_1A^3-n_3A^1} \right] \delta \left[ \int_{\gamma}{n_2A^3-n_3A^2} \right] \nonumber
	\\
	&\times \delta \left( \int_{\gamma}{\tilde{l}_4A^1} \right) \delta \left( \int_{\gamma}{\tilde{l}_4A^2} \right) \delta \left( \int_{\gamma}{\tilde{l}_4A^3} \right) \delta \left( \int_{\sigma}{B^4} \right) \,.
	\label{eq8}
\end{align}
Although eq.~(\ref{eq8}) only describes $S^2$ membranes, $T^2$ membranes have a similar expression. We just need to replace all $\mathsf{M^S}$, $\omega$ and $N_{n_1n_2n_3\tilde{l}_4}$ with $\mathsf{M^T}$, $\tilde{\omega}$ and $\tilde{N}_{n_1n_2n_3\tilde{l}_4}$ respectively in eq.~(\ref{eq8}), then we get the correct formula for $T^2$ membranes. In eq.~(\ref{eq8}), $N_{n_1n_2n_3\tilde{l}_4}$ is normalization factor, $n_1,n_2,n_3,\tilde{l}_4=0,1$ and at least one of $n_1,n_2,n_3$ is nonzero. So there are $7$ possible combinations when $\tilde{l}_4=0$:
\begin{gather}
	\left\{ \left( n_1n_2n_3 \right) \right\} =\left\{ \left( 100 \right) ,\left( 010 \right) ,\left( 001 \right) ,\left( 110 \right) ,\left( 101 \right) ,\left( 011 \right) ,\left( 111 \right) \right\}\,,
	\label{eq31-}  
	\\
	N_{n_1n_2n_30}=\left\{4,4,4,8,8,8,16\right\}\,,\nonumber
	\\
	\tilde{N}_{n_1n_2n_30}=\left\{8,8,8,32,32,32,128\right\}\,.
	\label{eq31}
\end{gather}
Normalization factors are determined to ensure all fusion coefficients and shrinking coefficients are integers. When $\tilde{l}_4=0$, for a specific $\left( n_1n_2n_3 \right)$, we need to choose the corresponding normalization factor shown in eq.~(\ref{eq31}). Also, given a $\left( n_1n_2n_3 \right)$, the possible values of $\left( m_1m_2m_3 \right)$ are also restricted. When $\left( n_1n_2n_3 \right)$ takes value in the order of eq.~(\ref{eq31-}), correspondingly we can choose
\begin{align}
	\left\{ \left( m_1m_2m_3 \right) \right\} =\left\{ \left( m_100 \right) ,\left( 0m_20 \right) ,\left( 00m_3 \right) ,\left( m_100 \right) ,\left( m_100 \right) ,\left( 0m_20 \right) ,\left( m_100 \right) \right\}  \,,
	\label{eq9}
\end{align}
where $m_1,m_2,m_3=0,1$. We can see that when $\tilde{l}_4=0$, once given a specific $\left( n_1n_2n_3 \right)$, two of $m_1,m_2,m_3$ are forced to zero and hence there are only two possible $\left( m_1m_2m_3 \right)$. For example, when we choose $\left( n_1n_2n_3 \right)=\left( 011 \right)$, then $\left( m_1m_2m_3 \right)=\left( 010 \right)$ or $\left( 000 \right)$. Eq.~(\ref{eq9}) is not the only choice, other choices can also correctly represent these nonequivalent excitations, but these choices are equivalent to eq.~(\ref{eq9}). In the following discussion, we fix our choice in eq.~(\ref{eq9}).

When $\tilde{l}_4=1$, we have $m_1,m_2,m_3=0$ and:
\begin{align}
	N_{n_1n_2n_31}=&8N_{n_1n_2n_30}=\left\{32,32,32,64,64,64,128\right\} \,,\nonumber
	\\
	\tilde{N}_{n_1n_2n_31}=&8\tilde{N}_{n_1n_2n_30}=\left\{64,64,64,256,256,256,1024\right\}\,.
\end{align}
Thus ${\mathsf{M}^{\mathsf{S}}}_{n_1n_2n_30}^{m_1m_2m_30;,0001}$ can be written as ${\mathsf{M}^{\mathsf{S}}}_{n_1n_2n_30}^{,0001}$.

Eq.~(\ref{eq8}) contains $21$ nonequivalent $S^2$ membranes since there are $14$ different $\left( n_1n_2n_3\tilde{l}_4 \right)$ and each $\left( n_1n_2n_30 \right)$ correspond to $2$ nonequivalent $\left( m_1m_2m_3 \right)$. For $T^2$ membranes we have a similar discussion, just replace all $\mathsf{M^S}$, $\omega$ and $N_{n_1n_2n_3\tilde{l}}$ with $\mathsf{M^T}$, $\tilde{\omega}$ and $\tilde{N}_{n_1n_2n_3\tilde{l}}$ respectively in eq.~(\ref{eq8}), we obtain $21$ nonequivalent $T^2$ membranes.

\textbf{\textit{$\mathsf{M^{ST}}$-excitations.}}
Similar to $BBA$ twisted term, $\mathsf{M^{ST}}$-excitations in $AAAB$ twisted term are also created by fusing $S^2$ membranes $\mathsf{M^S}_{n_1n_2n_30}^{m_1m_2m_30;,000\tilde{l}_4}$ and $T^2$ membranes $\mathsf{M^T}_{{n_1}^{\prime}{n_2}^{\prime}{n_3}^{\prime}0}^{{m_1}^{\prime}{m_2}^{\prime}{m_3}^{\prime}0;,000{\tilde{l}_4}^{\prime}}$. When $\tilde{l}_4$ and ${\tilde{l}_4}^{\prime}$ are not both $1$, we label the result as ${\mathsf{M}^{\mathsf{ST}}}_{n_1n_2n_30,{n_1}^{\prime}{n_2}^{\prime}{n_3}^{\prime}0}^{\left( m_1+{m_1}^{\prime} \right) \left( m_2+{m_2}^{\prime} \right) \left( m_3+{m_3}^{\prime} \right) 0;,000\left( \tilde{l}_4+{\tilde{l}_4}^{\prime} \right)}$. Any addition that appears in subscript and superscript is addition mod $2$, which means $1+1=0$ and $0+1=1$. Notice that ${\mathsf{M}^{\mathsf{S}}}_{n_1n_2n_30}^{m_1m_2m_30;,000\tilde{l}_4}\otimes {\mathsf{M}^{\mathsf{T}}}_{{n_1}^{\prime}{n_2}^{\prime}{n_3}^{\prime}0}^{{m_1}^{\prime}{m_2}^{\prime}{m_3}^{\prime}0;,000{\tilde{l}_4}^{\prime}}$ is not equal to ${\mathsf{M}^{\mathsf{T}}}_{n_1n_2n_30}^{m_1m_2m_30;,000\tilde{l}_4}\otimes {\mathsf{M}^{\mathsf{S}}}_{{n_1}^{\prime}{n_2}^{\prime}{n_3}^{\prime}0}^{{m_1}^{\prime}{m_2}^{\prime}{m_3}^{\prime}0;,000{\tilde{l}_4}^{\prime}}$ generally. Also, we find that ${\mathsf{M}^{\mathsf{ST}}}_{n_1n_2n_30,{n_1}^{\prime}{n_2}^{\prime}{n_3}^{\prime}0}^{\left( m_1+{m_1}^{\prime} \right) \left( m_2+{m_2}^{\prime} \right) \left( m_3+{m_3}^{\prime} \right) 0;,000\left( \tilde{l}_4+{\tilde{l}_4}^{\prime} \right)}$ with different subscripts and superscripts can be equivalent in our calculation. These $\mathsf{M^{ST}}$-excitations can be classified into $105$ equivalence classes and we are going to give two formulas to describe all nonequivalent $\mathsf{M^{ST}}$-excitations. When $\tilde{l}_4\times{\tilde{l}_4}^{\prime}=0$ and $\left(n_1n_2n_3\right)$ is equal to $\left({n_1}^{\prime}{n_2}^{\prime}{n_3}^{\prime}\right)$, i.e., $n_1={n_1}^{\prime},n_2={n_2}^{\prime},n_3={n_3}^{\prime}$, an $\mathsf{M^{ST}}$-excitation can be represented by:
\begin{align}
	&{\mathsf{M}^{\mathsf{ST}}}_{n_1n_2n_30,n_1n_2n_30}^{{m_1}^{\prime\prime}{m_2}^{\prime\prime}{m_3}^{\prime\prime}0;,000{\tilde{l}_4}^{\prime\prime}}={\mathsf{M}^{\mathsf{S}}}_{n_1n_2n_30}^{m_1m_2m_30;,000\tilde{l}_4}\otimes {\mathsf{M}^{\mathsf{T}}}_{n_1n_2n_30}^{{m_1}^{\prime}{m_2}^{\prime}{m_3}^{\prime}0;,000{\tilde{l}_4}^{\prime}}\nonumber
	\\
	=&N_{n_1n_2n_3{\tilde{l}_4}^{\prime\prime}}\tilde{N}_{n_1n_2n_30}\exp \left[ {im_1}^{\prime\prime}\int_{\gamma}{A^1}+{im_2}^{\prime\prime}\int_{\gamma}{A^2}+{im_3}^{\prime\prime}\int_{\gamma}{A^3} \right. \nonumber
	\\
	&+in_1\int_{\omega}{C^1-\frac{1}{3}\frac{2\pi q}{N_1}\left( B^4A^2d^{-1}A^3-B^4A^3d^{-1}A^2-A^2A^3d^{-1}B^4 \right)}\nonumber
	\\
	&+in_2\int_{\omega}{C^2+\frac{1}{3}\frac{2\pi q}{N_2}\left( B^4A^1d^{-1}A^3-B^4A^3d^{-1}A^1-A^1A^3d^{-1}B^4 \right)}\nonumber
	\\
	&+in_3\int_{\omega}{C^3-\frac{1}{3}\frac{2\pi q}{N_3}\left( B^4A^1d^{-1}A^2-B^4A^2d^{-1}A^1-A^1A^2d^{-1}B^4 \right)}\nonumber
	\\
	&+in_1\int_{\tilde{\omega}}{C^1-\frac{1}{3}\frac{2\pi q}{N_1}\left( B^4A^2d^{-1}A^3-B^4A^3d^{-1}A^2-A^2A^3d^{-1}B^4 \right)}\nonumber
	\\
	&+in_2\int_{\tilde{\omega}}{C^2+\frac{1}{3}\frac{2\pi q}{N_2}\left( B^4A^1d^{-1}A^3-B^4A^3d^{-1}A^1-A^1A^3d^{-1}B^4 \right)}\nonumber
	\\
	&+in_3\int_{\tilde{\omega}}{C^3-\frac{1}{3}\frac{2\pi q}{N_3}\left( B^4A^1d^{-1}A^2-B^4A^2d^{-1}A^1-A^1A^2d^{-1}B^4 \right)} \nonumber
	\\
	&\left. +{i\tilde{l}_4}^{\prime\prime}\int_{\sigma}{\tilde{B}^4}-\frac{1}{3}\frac{2\pi q}{N_4}\left( A^1A^2d^{-1}A^3-A^1A^3d^{-1}A^2+A^2A^3d^{-1}A^1 \right) \right] \nonumber
	\\
	&\times \delta \left[ \int_{\gamma}{n_1A^2-n_2A^1} \right] \delta \left[ \int_{\gamma}{n_1A^3-n_3A^1} \right] \delta \left[ \int_{\gamma}{n_2A^3-n_3A^2} \right] \nonumber
	\\
	&\times \delta \left( \int_{\gamma}{{\tilde{l}_4}^{\prime\prime}A^1} \right) \delta \left( \int_{\gamma}{{\tilde{l}_4}^{\prime\prime}A^2} \right) \delta \left( \int_{\gamma}{{\tilde{l}_4}^{\prime\prime}A^3} \right) \delta \left( \int_{\sigma}{B^4} \right) \,.
	\label{eq_MST1}
\end{align}
Where ${\tilde{l}_4}^{\prime\prime}=\tilde{l}_4+{\tilde{l}_4}^{\prime}$. Possible values of $\left( {m_1}^{\prime\prime}{m_2}^{\prime\prime}{m_3}^{\prime\prime} \right)$ still follows eq.~(\ref{eq9}) when ${\tilde{l}_4}^{\prime\prime}=0$. Thus two of ${m_1}^{\prime\prime}$, ${m_2}^{\prime\prime}$ and ${m_3}^{\prime\prime}$ are forced to zero once $\left( n_1n_2n_3 \right)$ is given, the last remaining one is determined by 
\begin{align}
	{m_1}^{\prime\prime}&=m_1+{m_1}^{\prime} \quad \text{mod} \quad 2  \nonumber 
	\\
	\text{or}\quad {m_2}^{\prime\prime}&=m_2+{m_2}^{\prime} \quad \text{mod} \quad 2  \nonumber 
	\\
	\text{or}\quad {m_3}^{\prime\prime}&=m_3+{m_3}^{\prime} \quad \text{mod} \quad 2 \,.
\end{align}
When ${\tilde{l}_4}^{\prime\prime}=1$, then $\left( {m_1}^{\prime\prime}{m_2}^{\prime\prime}{m_3}^{\prime\prime} \right)$ are all forced to zero and ${\mathsf{M}^{\mathsf{ST}}}_{n_1n_2n_30,n_1n_2n_30}^{{m_1}^{\prime\prime}{m_2}^{\prime\prime}{m_3}^{\prime\prime}0;,0001}$ can be written as ${\mathsf{M}^{\mathsf{ST}}}_{n_1n_2n_30,n_1n_2n_30}^{,0001}$.

When $\tilde{l}_4\times{\tilde{l}_4}^{\prime}=0$ and $\left(n_1n_2n_3\right)$ is not equal to $\left({n_1}^{\prime}{n_2}^{\prime}{n_3}^{\prime}\right)$, we obtain
\begin{align}
	&{\mathsf{M}^{\mathsf{ST}}}_{n_1n_2n_30,{n_1}^{\prime}{n_2}^{\prime}{n_3}^{\prime}0}^{,000{\tilde{l}_4}^{\prime\prime}}={\mathsf{M}^{\mathsf{S}}}_{n_1n_2n_30}^{m_1m_2m_30;,000\tilde{l}_4}\otimes {\mathsf{M}^{\mathsf{T}}}_{{n_1}^{\prime}{n_2}^{\prime}{n_3}^{\prime}0}^{{m_1}^{\prime}{m_2}^{\prime}{m_3}^{\prime}0;,000{\tilde{l}_4}^{\prime}}\nonumber
	\\
	=&N_{n_1n_2n_3{\tilde{l}_4}^{\prime\prime}}\tilde{N}_{{n_1}^{\prime}{n_2}^{\prime}{n_3}^{\prime}0}\nonumber
	\\
	&\times\exp \left[ in_1\int_{\omega}{C^1-\frac{1}{3}\frac{2\pi q}{N_1}\left( B^4A^2d^{-1}A^3-B^4A^3d^{-1}A^2-A^2A^3d^{-1}B^4 \right)} \right. \nonumber
	\\
	&+in_2\int_{\omega}{C^2+\frac{1}{3}\frac{2\pi q}{N_2}\left( B^4A^1d^{-1}A^3-B^4A^3d^{-1}A^1-A^1A^3d^{-1}B^4 \right)}\nonumber
	\\
	&+in_3\int_{\omega}{C^3-\frac{1}{3}\frac{2\pi q}{N_3}\left( B^4A^1d^{-1}A^2-B^4A^2d^{-1}A^1-A^1A^2d^{-1}B^4 \right)}\nonumber
	\\
	&+{in_1}^{\prime}\int_{\tilde{\omega}}{C^1-\frac{1}{3}\frac{2\pi q}{N_1}\left( B^4A^2d^{-1}A^3-B^4A^3d^{-1}A^2-A^2A^3d^{-1}B^4 \right)}\nonumber
	\\
	&+{in_2}^{\prime}\int_{\tilde{\omega}}{C^2+\frac{1}{3}\frac{2\pi q}{N_2}\left( B^4A^1d^{-1}A^3-B^4A^3d^{-1}A^1-A^1A^3d^{-1}B^4 \right)}\nonumber
	\\
	&+{in_3}^{\prime}\int_{\tilde{\omega}}{C^3-\frac{1}{3}\frac{2\pi q}{N_3}\left( B^4A^1d^{-1}A^2-B^4A^2d^{-1}A^1-A^1A^2d^{-1}B^4 \right)}\nonumber
	\\
	&\left. +{i\tilde{l}_4}^{\prime\prime}\int_{\sigma}{\tilde{B}^4}-\frac{1}{3}\frac{2\pi q}{N_4}\left( A^1A^2d^{-1}A^3-A^1A^3d^{-1}A^2+A^2A^3d^{-1}A^1 \right) \right] \nonumber
	\\
	&\times \delta \left( \int_{\gamma}{A^1} \right) \delta \left( \int_{\gamma}{A^2} \right) \delta \left( \int_{\gamma}{A^3} \right) \delta \left( \int_{\sigma}{B^4} \right) \,.
	\label{eq_MST2}
\end{align}
All particle decorations become trivial in this case and this can be seen from the delta functionals. 

We conclude that for $\tilde{l}_4\times{\tilde{l}_4}^{\prime}=0$, ${\mathsf{M}^{\mathsf{ST}}}_{n_1n_2n_30,{n_1}^{\prime}{n_2}^{\prime}{n_3}^{\prime}0}^{\left( m_1+{m_1}^{\prime} \right) \left( m_2+{m_2}^{\prime} \right) \left( m_3+{m_3}^{\prime} \right) 0;,000\left( \tilde{l}_4+{\tilde{l}_4}^{\prime} \right)}$ is equivalent to ${\mathsf{M}^{\mathsf{ST}}}_{n_1n_2n_30,n_1n_2n_30}^{{m_1}^{\prime\prime}{m_2}^{\prime\prime}{m_3}^{\prime\prime}0;,000{\tilde{l}_4}^{\prime\prime}}$ when $\left(n_1n_2n_3\right)=\left({n_1}^{\prime}{n_2}^{\prime}{n_3}^{\prime}\right)$ and ${\mathsf{M}^{\mathsf{ST}}}_{n_1n_2n_30,{n_1}^{\prime}{n_2}^{\prime}{n_3}^{\prime}0}^{\left( m_1+{m_1}^{\prime} \right) \left( m_2+{m_2}^{\prime} \right) \left( m_3+{m_3}^{\prime} \right) 0;,000\left( \tilde{l}_4+{\tilde{l}_4}^{\prime} \right)}$ is equivalent to ${\mathsf{M}^{\mathsf{ST}}}_{n_1n_2n_30,{n_1}^{\prime}{n_2}^{\prime}{n_3}^{\prime}0}^{,000{\tilde{l}_4}^{\prime\prime}}$ when $\left(n_1n_2n_3\right)\ne\left({n_1}^{\prime}{n_2}^{\prime}{n_3}^{\prime}\right)$. For $\tilde{l}_4={\tilde{l}_4}^{\prime}=1$, we conclude that
\begin{gather}
	{\mathsf{M}^{\mathsf{S}}}_{n_1n_2n_30}^{,0001}\otimes {\mathsf{M}^{\mathsf{T}}}_{n_1n_2n_30}^{,0001}	=32\cdot {\mathsf{M}^{\mathsf{ST}}}_{n_1n_2n_30,n_1n_2n_30}\oplus 32\cdot {\mathsf{M}^{\mathsf{ST}}}_{n_1n_2n_30,n_1n_2n_30}^{\left( 1 \right) \left( 1 \right) \left( 1 \right) 0;} \,,\label{eqb15}
	\\
	{\mathsf{M}^{\mathsf{S}}}_{n_1n_2n_30}^{m_1m_2m_30;,0001}\otimes {\mathsf{M}^{\mathsf{T}}}_{{n_1}^{\prime}{n_2}^{\prime}{n_3}^{\prime}0}^{{m_1}^{\prime}{m_2}^{\prime}{m_3}^{\prime}0;,0001}=64\cdot {\mathsf{M}^{\mathsf{ST}}}_{n_1n_2n_30,{n_1}^{\prime}{n_2}^{\prime}{n_3}^{\prime}0}\,.
\end{gather}
Where ``$\left( 1 \right) \left( 1 \right) \left( 1 \right)$'' in eq.~(\ref{eqb15}) means that $2$ of them are forced to zero depending on the value of $n_1n_2n_3$. 

Eq.~(\ref{eq_MST1}) and eq.~(\ref{eq_MST2}) contain $105$ nonequivalent non-Abelian excitations. Now we obtain all equivalence classes. There are $16$ Abelian excitations and $149$ non-Abelian excitations.

\subsection{Fusion rules for the $AAAB$ twisted term}
Since it is hard to list all fusion rules, we only consider some typical fusion rules here.  

\textbf{\textit{Fusing $2$ Abelian excitations.}}
All $16$ Abelian excitations can be represented by $\mathsf{L}_{000n_4,}^{n_1n_2n_30}$, thus the fusion rules between them are 
\begin{align}
	\mathsf{L}_{000n_4,}^{n_1n_2n_30}\otimes \mathsf{L}_{000{n_4}^{\prime},}^{{n_1}^{\prime}{n_2}^{\prime}{n_3}^{\prime}0}=\mathsf{L}_{000\left( n_4+{n_4}^{\prime} \right),}^{\left( n_1+{n_1}^{\prime} \right) \left( n_2+{n_2}^{\prime} \right) \left( n_3+{n_3}^{\prime} \right) 0}\,.
\end{align}

\textbf{\textit{Fusing $2$ $\tilde{B}$-loop.}}
Fusing $2$ $\tilde{B}$-loop are
\begin{align}
	\mathsf{L}_{,0001}\otimes \mathsf{L}_{,0001}=\oplus _{n_1n_2n_3}8\cdot \mathsf{P}_{n_1n_2n_30}\,.
\end{align}

\textbf{\textit{Fusing a loop and an $S^2$ membrane.}}
If the $S^2$ membrane is not decorated by a $\tilde{B}$-loop, then
\begin{gather}
	\mathsf{L}_{000n_4,}^{m_1m_2m_30}\otimes {\mathsf{M}^{\mathsf{S}}}_{n_1n_2n_30}^{{m_1}^{\prime}{m_2}^{\prime}{m_3}^{\prime}0;}={\mathsf{M}^{\mathsf{S}}}_{n_1n_2n_30}^{\left( m_1+{m_1}^{\prime} \right) \left( m_2+{m_2}^{\prime} \right) \left( m_3+{m_3}^{\prime} \right) 0;}\,,
	\\
	\mathsf{L}_{000n_4,0001}\otimes {\mathsf{M}^{\mathsf{S}}}_{n_1n_2n_30}^{m_1m_2m_30;}={\mathsf{M}^{\mathsf{S}}}_{n_1n_2n_30}^{,0001}\,.
\end{gather}
Here only one of $\left( m_1+{m_1}^{\prime} \right)$, $\left( m_2+{m_2}^{\prime} \right)$ and $\left( m_3+{m_3}^{\prime} \right)$ can be nonzero, see  eq.~(\ref{eq9}).

If the $S^2$ membrane is decorated by a $\tilde{B}$-loop, then all particle decorations become trivial and we have:
\begin{gather}
	\mathsf{L}_{000n_4,}^{m_1m_2m_30}\otimes {\mathsf{M}^{\mathsf{S}}}_{n_1n_2n_30}^{,0001}={\mathsf{M}^{\mathsf{S}}}_{n_1n_2n_30}^{,0001}\,,
	\\
	\mathsf{L}_{000n_4,0001}\otimes {\mathsf{M}^{\mathsf{S}}}_{n_1n_2n_30}^{,0001}=32\cdot {\mathsf{M}^{\mathsf{S}}}_{n_1n_2n_30}^{\left( 1 \right) \left( 1 \right) \left( 1 \right) 0;}\oplus 32\cdot {\mathsf{M}^{\mathsf{S}}}_{n_1n_2n_30}\,.
\end{gather}

\textbf{\textit{Fusing $2$ membranes without $\tilde{B}$-loop decoration.}}
Fusion rules for membranes are much more complicated and it is hard to find a general formula. We have to calculate these fusion rules carefully and finally, we derive a fusion table for two $S^2$ membranes shown in table~\ref{table6}. 
\begin{sidewaystable*}
	\caption{\label{table6}Fusion rules for $\mathsf{M^S}_{n_1n_2n_30}^{m_1m_2m_30;}\otimes \mathsf{M^S}_{{n_1}^{\prime}{n_2}^{\prime}{n_3}^{\prime}0}^{{m_1}^{\prime}{m_2}^{\prime}{m_3}^{\prime}0;}$. This table is also symmetric, so we only have to write down the upper triangular part of this table. Repeated indices imply summation. $\oplus _{\left( m_1m_2m_3 \right) m_4}$ means that $m_1,m_2,m_3$ are not independent to each other, we constrain $\left\{\left( m_1m_2m_3 \right)\right\}=\left\{\left( 000 \right),\left( 110 \right),\left( 101 \right),\left( 011 \right)\right\}$. To be more specifically, $32\cdot\oplus _{\left( m_1m_2m_3 \right) m_4}\mathsf{L}_{000m_4,}^{\left(m_1+n_1+{n_1}^{\prime}\right)m_2m_30}=32\cdot \mathsf{P}_{\left(n_1+{n_1}^{\prime}\right)000}\oplus 32\cdot \mathsf{L}_{0001,}^{\left(n_1+{n_1}^{\prime}\right)000}\oplus 32\cdot \mathsf{L}_{0001,}^{\left(1+n_1+{n_1}^{\prime}\right)100}\oplus 32\cdot \mathsf{L}_{0001,}^{\left(1+n_1+{n_1}^{\prime}\right)010}\oplus 32\cdot \mathsf{L}_{0001,}^{\left(n_1+{n_1}^{\prime}\right)110}\oplus 32\cdot \mathsf{P}_{\left(1+n_1+{n_1}^{\prime}\right)100}\oplus 32\cdot \mathsf{P}_{\left(1+n_1+{n_1}^{\prime}\right)010}\oplus 32\cdot \mathsf{P}_{\left(n_1+{n_1}^{\prime}\right)110} $.}
	\centering
	\begin{tabular*}{\textwidth}{@{\extracolsep{\fill}}|c|c|c|c|c|c|c|c|}
		\hline
		& $\mathsf{M^S}_{1000}^{{n_1}^{\prime}000;}$ & $\mathsf{M^S}_{0100}^{0{n_2}^{\prime}00;}$ & $\mathsf{M^S}_{0010}^{00{n_3}^{\prime}0;}$ & $\mathsf{M^S}_{1100}^{{n_1}^{\prime}000;}$ & $\mathsf{M^S}_{1010}^{{n_1}^{\prime}000;}$ & $\mathsf{M^S}_{0110}^{0{n_2}^{\prime}00;}$ & $\mathsf{M^S}_{1110}^{{n_1}^{\prime}000;}$ \\
		\hline
		$\mathsf{M^S}_{1000}^{{n_1}000;}$ & \makecell{$2\cdot\oplus _{m_2m_3m_4}$\\$\mathsf{L}_{000m_4,}^{\left(n_1+{n_1}^{\prime}\right)m_2m_30}$} & \makecell{$\mathsf{M^S}_{1100}^{1000;}$\\$\oplus \mathsf{M^S}_{1100}$} & \makecell{$\mathsf{M^S}_{1010}^{1000;}$\\$\oplus \mathsf{M^S}_{1010}$} & \makecell{$4\cdot \mathsf{M^S}_{0100}^{0100;}$\\$\oplus4\cdot \mathsf{M^S}_{0100}$} & \makecell{$4\cdot \mathsf{M^S}_{0010}^{0010;}$\\$\oplus4\cdot \mathsf{M^S}_{0010}$} & \makecell{$\mathsf{M^S}_{1110}^{1000;}$\\$\oplus \mathsf{M^S}_{1110}$} & \makecell{$4\cdot \mathsf{M^S}_{0110}^{0100;}$\\$\oplus4\cdot \mathsf{M^S}_{0110}$} \\
		\hline
		$\mathsf{M^S}_{0100}^{0{n_2}00;}$ &  & \makecell{$2\cdot\oplus _{m_1m_3m_4}$\\  $\mathsf{L}_{000m_4,}^{m_1\left(n_2+{n_2}^{\prime}\right)m_30}$} & \makecell{$\mathsf{M^S}_{0110}^{0100;}$\\$\oplus \mathsf{M^S}_{0110}$} & \makecell{$4\cdot \mathsf{M^S}_{1000}^{1000;}$\\$\oplus4\cdot \mathsf{M^S}_{1000}$} & \makecell{$\mathsf{M^S}_{1110}^{1000;}$\\$\oplus \mathsf{M^S}_{1110}$} & \makecell{$4\cdot \mathsf{M^S}_{0010}^{0010;}$\\$\oplus4\cdot \mathsf{M^S}_{0010}$} & \makecell{$4\cdot \mathsf{M^S}_{1010}^{1000;}$\\$\oplus4\cdot \mathsf{M^S}_{1010}$} \\
		\hline
		$\mathsf{M^S}_{0010}^{00{n_3}0;}$ &  &  & \makecell{$2\cdot\oplus _{m_1m_2m_4}$\\$\mathsf{L}_{000m_4,}^{m_1m_2\left(n_3+{n_3}^{\prime}\right)0}$} & \makecell{$\mathsf{M^S}_{1110}^{1000;}$\\$\oplus \mathsf{M^S}_{1110}$} & \makecell{$4\cdot \mathsf{M^S}_{1000}^{1000;}$\\$\oplus4\cdot \mathsf{M^S}_{1000}$} & \makecell{$4\cdot \mathsf{M^S}_{0100}^{0100;}$\\$\oplus4\cdot \mathsf{M^S}_{0100}$} & \makecell{$4\cdot \mathsf{M^S}_{1100}^{1000;}$\\$\oplus4\cdot \mathsf{M^S}_{1100}$} \\
		\hline
		$\mathsf{M^S}_{1100}^{{n_1}000;}$ &  &  &  & \makecell{$8\cdot\oplus _{m_1m_3m_4}$\\$\mathsf{L}_{000m_4,}^{\left(m_1+n_1+{n_1}^{\prime}\right)m_1m_30}$} & \makecell{$4\cdot \mathsf{M^S}_{0110}^{0100;}$\\$\oplus4\cdot \mathsf{M^S}_{0110}$} & \makecell{$4\cdot \mathsf{M^S}_{1010}^{1000;}$\\$\oplus4\cdot \mathsf{M^S}_{1010}$} & \makecell{$16\cdot \mathsf{M^S}_{0010}^{0010;}$\\$\oplus16\cdot \mathsf{M^S}_{0010}$} \\
		\hline
		$\mathsf{M^S}_{1010}^{{n_1}000;}$ &  &  &  &  & \makecell{$8\cdot\oplus _{m_1m_2m_4}$\\$\mathsf{L}_{000m_4,}^{\left(m_1+n_1+{n_1}^{\prime}\right)m_2m_10}$} & \makecell{$4\cdot \mathsf{M^S}_{1100}^{1000;}$\\$\oplus4\cdot \mathsf{M^S}_{1100}$} & \makecell{$16\cdot \mathsf{M^S}_{0100}^{0100;}$\\$\oplus16\cdot \mathsf{M^S}_{0100}$} \\
		\hline
		$\mathsf{M^S}_{0110}^{0{n_2}00;}$ &  &  &  &  &  & \makecell{$8\cdot\oplus _{m_1m_2m_4}$\\$\mathsf{L}_{000m_4,}^{m_1\left(m_2+n_2+{n_2}^{\prime}\right)m_20}$} & \makecell{$16\cdot \mathsf{M^S}_{1000}^{1000;}$\\$\oplus16\cdot \mathsf{M^S}_{1000}$} \\
		\hline
		$\mathsf{M^S}_{1110}^{{n_1}000;}$ &  &  &  &  &  &  & \makecell{$32\cdot\oplus _{\left( m_1m_2m_3 \right) m_4}$\\$\mathsf{L}_{000m_4,}^{\left(m_1+n_1+{n_1}^{\prime}\right)m_2m_30}$} \\
		\hline
	\end{tabular*}
\end{sidewaystable*}

For two $T^2$ membranes, we have a similar table. First, we replace all $\mathsf{M^S}$ with $\mathsf{M^T}$ in table~\ref{table6}. Notice that fusion coefficients larger than $1$ in table~\ref{table6} can be $2,4,8,16,32$, we replace them with $8,16,128,256,2048$ respectively, then we have a fusion table for $\mathsf{M^T}_{n_1n_2n_30}^{m_1m_2m_30;}\otimes \mathsf{M^T}_{{n_1}^{\prime}{n_2}^{\prime}{n_3}^{\prime}0}^{{m_1}^{\prime}{m_2}^{\prime}{m_3}^{\prime}0;}$. We do this replacement because the coefficients of Wilson operators for $T^2$ membranes are larger than $S^2$ membranes so that all shrinking coefficients are integers. For example, we have 
\begin{gather}
	\mathsf{M^S}_{1100}^{n_1000;}\otimes \mathsf{M^S}_{1100}^{{n_1}^{\prime}000;}=8\cdot \oplus _{m_1m_3m_4}\mathsf{L}_{000m_4,}^{\left( m_1+n_1+{n_1}^{\prime} \right) m_1m_30}\,,
	\\
	\mathsf{M^S}_{1110}^{n_1000;}\otimes \mathsf{M^S}_{0110}^{0{n_2}^{\prime}00;}=16\cdot \mathsf{M^S}_{1000}^{1000;}\oplus 16\cdot \mathsf{M^S}_{1000} \,,
\end{gather}
in table~\ref{table6}. Do the replacement we obtain 
\begin{gather}
	\mathsf{M^T}_{1100}^{n_1000;}\otimes \mathsf{M^T}_{1100}^{{n_1}^{\prime}000;}=128\cdot \oplus _{m_1m_3m_4}\mathsf{L}_{000m_4,}^{\left( m_1+n_1+{n_1}^{\prime} \right) m_1m_30}\,,
	\\
	\mathsf{M^T}_{1110}^{n_1000;}\otimes \mathsf{M^T}_{0110}^{0{n_2}^{\prime}00;}=256\cdot \mathsf{M^T}_{1000}^{1000;}\oplus 256\cdot \mathsf{M^T}_{1000} \,,
\end{gather}

As for fusing an $S^2$ membrane and a $T^2$ membrane, we have already discussed it when constructing Wilson operators for $\mathsf{M^{ST}}$-excitations.  

\textbf{\textit{Fusion rules for $\mathsf{M^{ST}}$-excitations.}}
At the beginning of section~\ref{s3}, we illustrate that fusion rules in our theory should satisfy commutative and associative law, i.e., fusion rules satisfy eq.~(\ref{eq_fusion_algebra_rule}). We can use this powerful condition to calculate fusion rules for $\mathsf{M^{ST}}$-excitations. 

For example ${\mathsf{M}^{\mathsf{ST}}}_{n_1n_2n_30,n_1n_2n_30}^{m_1m_2m_30;}$ is equivalent to:
\begin{align}
	{\mathsf{M}^{\mathsf{ST}}}_{n_1n_2n_30,n_1n_2n_30}^{m_1m_2m_30;}={\mathsf{M}^{\mathsf{S}}}_{n_1n_2n_30}^{m_1m_2m_30;}\otimes {\mathsf{M}^{\mathsf{T}}}_{n_1n_2n_30} \,.
\end{align}
Thus when we consider ${\mathsf{M}^{\mathsf{ST}}}_{n_1n_2n_30,n_1n_2n_30}^{m_1m_2m_30;}\otimes {\mathsf{M}^{\mathsf{S}}}_{{n_1}^{\prime}{n_2}^{\prime}{n_3}^{\prime}0}^{{m_1}^{\prime}{m_2}^{\prime}{m_3}^{\prime}0;}$, we have 
\begin{align}
	{\mathsf{M}^{\mathsf{ST}}}_{n_1n_2n_30,n_1n_2n_30}^{m_1m_2m_30;}\otimes {\mathsf{M}^{\mathsf{S}}}_{{n_1}^{\prime}{n_2}^{\prime}{n_3}^{\prime}0}^{{m_1}^{\prime}{m_2}^{\prime}{m_3}^{\prime}0;}=&{\mathsf{M}^{\mathsf{S}}}_{n_1n_2n_30}^{m_1m_2m_30;}\otimes {\mathsf{M}^{\mathsf{T}}}_{n_1n_2n_30}\otimes {\mathsf{M}^{\mathsf{S}}}_{{n_1}^{\prime}{n_2}^{\prime}{n_3}^{\prime}0}^{{m_1}^{\prime}{m_2}^{\prime}{m_3}^{\prime}0;}\nonumber
	\\
	=&{\mathsf{M}^{\mathsf{S}}}_{n_1n_2n_30}^{m_1m_2m_30;}\otimes {\mathsf{M}^{\mathsf{S}}}_{{n_1}^{\prime}{n_2}^{\prime}{n_3}^{\prime}0}^{{m_1}^{\prime}{m_2}^{\prime}{m_3}^{\prime}0;}\otimes {\mathsf{M}^{\mathsf{T}}}_{n_1n_2n_30}	\,.
\end{align}
Now we can use table~\ref{table6} to derive the fusion rules instead of directly calculate it from path integral. We can consider $\mathsf{M^{ST}}_{1100,1100}^{1000;}\otimes \mathsf{M^S}_{0110}^{0100;}$ as an example, following the method in this section we have 
\begin{align}
	&\mathsf{M^{ST}}_{1100,1100}^{1000;}\otimes \mathsf{M^S}_{0110}^{0100;}\nonumber
	\\
	=&\mathsf{M^S}_{1100}^{1000;}\otimes \mathsf{M^T}_{1100}\otimes \mathsf{M^S}_{0110}^{0100;}=\mathsf{M^S}_{1100}^{1000;}\otimes \mathsf{M^S}_{0110}^{0100;}\otimes \mathsf{M^T}_{1100} \nonumber
	\\
	=&4\cdot \left( \mathsf{M^S}_{1010}^{1000;}\oplus \mathsf{M^S}_{1010} \right) \otimes \mathsf{M^T}_{1100}=4\cdot \left( \mathsf{M^{ST}}_{1010,1100}\oplus \mathsf{M^{ST}}_{1010,1100} \right) \nonumber
	\\
	=&8\cdot \mathsf{M^{ST}}_{1010,1100}\,.
\end{align}

\subsection{Shrinking rules for the $AAAB$ twisted term}
We have already defined shrinking operator $\mathcal{S}$ for different excitations in section~\ref{s4}, we continue to use this definition. 

\textbf{\textit{Shrinking loops.}}
For Abelian loops, shrinking rules are
\begin{align}
	\mathcal{S} \left( \mathsf{L}_{000n_4,}^{n_1n_2n_30} \right) =\mathsf{P}_{n_1n_2n_30}\,.
\end{align}
For non-Abelian loops, shrinking rules are
\begin{align}
	\mathcal{S} \left( \mathsf{L}_{000n_4,0001} \right) =&\mathsf{1}\oplus \mathsf{P}_{1000}\oplus \mathsf{P}_{0100}\oplus \mathsf{P}_{0010}\oplus \mathsf{P}_{1100}\oplus \mathsf{P}_{1010}\oplus \mathsf{P}_{0110}\oplus \mathsf{P}_{1110}\nonumber
	\\
	=&\oplus _{n_1n_2n_3}\mathsf{P}_{n_1n_2n_30}\,.
\end{align}

\textbf{\textit{Shrinking $S^2$ membranes.}}
Shrinking rules for $S^2$ membranes can be written as 
\begin{align}
	\mathcal{S} \left( {\mathsf{M}^{\mathsf{S}}}_{n_1n_2n_30}^{m_1m_2m_30;} \right) =&\frac{N_{n_1n_2n_30}}{8}\left\{ m_1\mathsf{P}_{m_1000}\otimes m_2\mathsf{P}_{0m_200}\otimes m_3\mathsf{P}_{00m_30}\otimes \right. \nonumber 
	\\
	&\otimes \left[ 1\oplus \left( n_1\mathsf{P}_{0100}\otimes n_2\mathsf{P}_{1000} \right) \right] \left[ 1\oplus \left( n_1\mathsf{P}_{0010}\otimes n_3\mathsf{P}_{1000} \right) \right] \nonumber
	\\
	&\left.  \otimes \left[ 1\oplus \left( n_2\mathsf{P}_{0010}\otimes n_3\mathsf{P}_{0100} \right) \right] \right\} \,,
	\\
	\mathcal{S} \left( {\mathsf{M}^{\mathsf{S}}}_{n_1n_2n_30}^{,0001} \right) =&\frac{N_{n_1n_2n_31}}{8}\oplus _{m_1m_2m_3}\mathsf{P}_{n_1n_2n_30}\,.
\end{align}
In the above equation, if the coefficient of the excitation is zero, we treat it as vacuum $\mathsf{1}$. For example terms like $\otimes \left[ \mathsf{1}\oplus \left( 1\mathsf{P}_{0010}\otimes 0\mathsf{P}_{1000} \right) \right] $ are treated as $\otimes \left[ \mathsf{1}\oplus \left( \mathsf{P}_{0010}\otimes \mathsf{1} \right) \right]=\otimes \left( \mathsf{1}\oplus \mathsf{P}_{0010} \right)$, $\otimes \left[ \mathsf{1}\oplus \left( 0\mathsf{P}_{0010}\otimes 0\mathsf{P}_{1000} \right) \right] $ are treated as $\otimes \left[ \mathsf{1}\oplus \left( \mathsf{1}\otimes \mathsf{1} \right) \right]=\otimes 2\cdot \mathsf{1}$. If we consider shrinking an $\mathsf{M^S}_{1010}^{1000;}$, we have 
\begin{align}
	\mathcal{S} \left( {\mathsf{M}^{\mathsf{S}}}_{1010}^{1000;} \right) =&\frac{8}{8}\left\{ \mathsf{P}_{1000}\otimes \left( \mathsf{1}\oplus \mathsf{P}_{0100} \right) \otimes \left[ \mathsf{1}\oplus \left( \mathsf{P}_{0010}\otimes \mathsf{P}_{1000} \right) \right] \otimes \left( \mathsf{1}\oplus \mathsf{P}_{0100} \right)  \right\} \nonumber
	\\
	=&2\cdot \left( \mathsf{P}_{1000}\oplus \mathsf{P}_{0010}\oplus \mathsf{P}_{1100}\oplus \mathsf{P}_{0110} \right) \,.
\end{align}

\textbf{\textit{Shrinking $T^2$ membranes.}}
Shrinking rules for $T^2$ membranes without $\tilde{B}$-loop decorations can be written as 
\begin{align}
	\mathcal{S} \left( {\mathsf{M}^{\mathsf{T}}}_{n_1n_2n_30}^{m_1m_2m_30;} \right) =&\frac{\tilde{N}_{n_1n_2n_30}}{16}\left\{ m_1\mathsf{P}_{m_1000}\otimes m_2\mathsf{P}_{0m_200}\otimes m_3\mathsf{P}_{00m_30}\otimes \right. \nonumber
	\\
	&\otimes \left[ 1\oplus \left( n_1\mathsf{P}_{0100}\otimes n_2\mathsf{P}_{1000} \right) \right] \left[ 1\oplus \left( n_1\mathsf{P}_{0010}\otimes n_3\mathsf{P}_{1000} \right) \right] \nonumber
	\\
	&\left. \otimes \left[ 1\oplus \left( n_2\mathsf{P}_{0010}\otimes n_3\mathsf{P}_{0100} \right) \right] \otimes \left( 1\oplus \mathsf{L}_{0001,} \right) \right\} \,,
	\\
	\mathcal{S} ^2\left( {\mathsf{M}^{\mathsf{T}}}_{n_1n_2n_30}^{m_1m_2m_30;} \right) =&\frac{\tilde{N}_{n_1n_2n_30}}{N_{n_1n_2n_30}}\mathcal{S} \left( {\mathsf{M}^{\mathsf{S}}}_{n_1n_2n_30}^{m_1m_2m_30;} \right) \,.
\end{align} 

If $T^2$ membranes are decorated by $\tilde{B}$-loops, we obtain:
\begin{align}
	\mathcal{S}\left( {\mathsf{M}^{\mathsf{T}}}_{n_1n_2n_30}^{,0001} \right) =&\frac{\tilde{N}_{n_1n_2n_31}}{16}\oplus _{m_1m_2m_3m_4}\mathsf{L}_{000m_4,}^{m_1m_2m_30} \,,
	\\
	\mathcal{S} ^2\left( {\mathsf{M}^{\mathsf{T}}}_{n_1n_2n_30}^{,0001} \right) =&\frac{\tilde{N}_{n_1n_2n_31}}{8}\oplus _{m_1m_2m_3}\mathsf{P}_{m_1m_2m_30}\nonumber
	\\
	=&\frac{\tilde{N}_{n_1n_2n_30}}{N_{n_1n_2n_30}}\mathcal{S} \left( {\mathsf{M}^{\mathsf{S}}}_{n_1n_2n_30}^{,0001} \right) \,.
\end{align}
Where $\mathcal{S}^2=\mathcal{S} \mathcal{S}$. If a $T^2$ and an $S^2$ membrane carry the same gauge charges (of 3-form fields) and decorations, they can eventually be shrunk to the same superposition of particles up to a global coefficient $\frac{\tilde{N}_{n_1n_2n_30}}{N_{n_1n_2n_30}}$. When they only carry one unit of gauge charge (of 3-form fields), this global coefficient is $2^1=2$, carry two units is $2^2=4$, and carry three units is $2^3=8$. Every unit of gauge charge contributes an extra $2$ to the normalization factor $\tilde{N}_{n_1n_2n_30}$ in the front of Wilson operators for $T^2$ membranes.

\textbf{\textit{Shrinking an $\mathsf{M^{ST}}$-excitation.}}
When $\left(n_1n_2n_3\right)=\left({n_1}^{\prime}{n_2}^{\prime}{n_3}^{\prime}\right)$, shrinking rules for $\mathsf{M^{ST}}$-excitations can be written as 
\begin{align}
	\mathcal{S} \left( {\mathsf{M}^{\mathsf{ST}}}_{n_1n_2n_30,n_1n_2n_30}^{m_1m_2m_30;,} \right) =&N_{n_1n_2n_30}\mathcal{S} \left( {\mathsf{M}^{\mathsf{T}}}_{n_1n_2n_30}^{m_1m_2m_30;} \right) \,,
	\\
	\mathcal{S} ^2\left( {\mathsf{M}^{\mathsf{ST}}}_{n_1n_2n_30,n_1n_2n_30}^{m_1m_2m_30;,} \right) =&\tilde{N}_{n_1n_2n_30}\mathcal{S} \left( {\mathsf{M}^{\mathsf{S}}}_{n_1n_2n_30}^{m_1m_2m_30;} \right) \,,
\end{align}
and
\begin{align}
	\mathcal{S} \left( {\mathsf{M}^{\mathsf{ST}}}_{n_1n_2n_30,n_1n_2n_30}^{,0001} \right) =&N_{n_1n_2n_30}\mathcal{S} \left( {\mathsf{M}^{\mathsf{T}}}_{n_1n_2n_30}^{,0001} \right) \,,
	\\
	\mathcal{S} ^2\left( {\mathsf{M}^{\mathsf{ST}}}_{n_1n_2n_30,n_1n_2n_30}^{;0001} \right) =&\tilde{N}_{n_1n_2n_30}\mathcal{S} \left( {\mathsf{M}^{\mathsf{S}}}_{n_1n_2n_30}^{,0001} \right) \,.
\end{align}

When $\left(n_1n_2n_3\right)\ne\left({n_1}^{\prime}{n_2}^{\prime}{n_3}^{\prime}\right)$, shrinking rules for $\mathsf{M^{ST}}$-excitations can be written as
\begin{align}
	\mathcal{S} \left( {\mathsf{M}^{\mathsf{ST}}}_{n_1n_2n_30,{n_1}^{\prime}{n_2}^{\prime}{n_3}^{\prime}0} \right) =&\frac{N_{n_1n_2n_30}}{8}\mathcal{S} _1\left( {\mathsf{M}^{\mathsf{T}}}_{{n_1}^{\prime}{n_2}^{\prime}{n_3}^{\prime}0}^{,0001} \right) \,,
	\\
	\mathcal{S} ^2\left( {\mathsf{M}^{\mathsf{ST}}}_{n_1n_2n_30,{n_1}^{\prime}{n_2}^{\prime}{n_3}^{\prime}0} \right) =&\frac{\tilde{N}_{n_1n_2n_30}}{8}\mathcal{S} \left( {\mathsf{M}^{\mathsf{S}}}_{{n_1}^{\prime}{n_2}^{\prime}{n_3}^{\prime}0}^{,0001} \right) \,,
\end{align}
and
\begin{align}
	\mathcal{S}\left( {\mathsf{M}^{\mathsf{ST}}}_{n_1n_2n_30,{n_1}^{\prime}{n_2}^{\prime}{n_3}^{\prime}0}^{,0001} \right) =&N_{n_1n_2n_30}\mathcal{S}\left( {\mathsf{M}^{\mathsf{T}}}_{{n_1}^{\prime}{n_2}^{\prime}{n_3}^{\prime}0}^{,0001} \right) \,,
	\\
	\mathcal{S} ^2\left( {\mathsf{M}^{\mathsf{ST}}}_{n_1n_2n_30,{n_1}^{\prime}{n_2}^{\prime}{n_3}^{\prime}0}^{;0001} \right) =&\tilde{N}_{n_1n_2n_30}\mathcal{S} \left( {\mathsf{M}^{\mathsf{S}}}_{{n_1}^{\prime}{n_2}^{\prime}{n_3}^{\prime}0}^{,0001} \right) \,.
\end{align}

Now we have obtained all shrinking rules for $AAAB$ twisted term. One can verify that shrinking rules and fusion rules satisfy eq.~(\ref{eq_hieshrinking_and_fusion}):
\begin{align}
	\mathcal{S}^2 \left( \mathsf{a}\otimes \mathsf{b} \right) =\mathcal{S}^2 \left( \mathsf{a} \right) \otimes \mathcal{S}^2 \left( \mathsf{b} \right)\,.
\end{align}
If $\mathsf{a}$ and $\mathsf{b}$ do not have hierarchical shrinking structure, then we obtain eq.~(\ref{eq_shrinking_and_fusion}):
\begin{align}
	\mathcal{S} \left( \mathsf{a}\otimes \mathsf{b} \right) =\mathcal{S} \left( \mathsf{a} \right) \otimes \mathcal{S} \left( \mathsf{b} \right)\,.
\end{align}

\section{General method to determine the type of fusion and shrinking rules\label{ap2}}
Given a specific twisted term, we do not have to derive all fusion and shrinking rules explicitly to find out whether the fusion and shrinking rules are Abelian or non-Abelian. We just need to derive the gauge transformations and a few Wilson operators, then we can determine the type of fusion and shrinking rules.

Generally, Wilson operators for topological excitations in $BF$ theory can be written in these two forms 
\begin{align}
	\mathsf{E_a}=&N_{\mathsf{a}}\prod_{\mathsf{i}}{\exp \left( i\int_{X _{\mathsf{i}}}{F^{\mathsf{i}}+f^{\mathsf{i}}} \right)}\,, \label{eq1ap2}
	\\
	\mathsf{E_b}=&N_{\mathsf{b}}\left[ \prod_{\mathsf{j}}{\exp \left( i\int_{X _{\mathsf{j}}}{F^{\mathsf{j}}+f^{\mathsf{j}}} \right)} \right] \left[ \prod_{\mathsf{k}}{\delta \left( \int_{X _{\mathsf{k}}}{F^{\mathsf{k}}} \right)} \right] \,, \label{eq2ap2}
\end{align}
where $N_{\mathsf{a}}$ and $N_{\mathsf{b}}$ are normalization factors, $X _{\mathsf{i}}$ and $X _{\mathsf{j}}$ are spacetime trajectories of excitations, $F^{\mathsf{i}}$ and $F^{\mathsf{j}}$ are gauge fields, $f^{\mathsf{i}}$ and $f^{\mathsf{j}}$ are nontrivial terms. If $F^{\mathsf{i}}$ transforms without nontrivial shifts, $f^{\mathsf{i}}$ vanishes. 

We conclude that all excitations in the form of eq.~(\ref{eq1ap2}) are Abelian excitation because we can always find that fusing two $\mathsf{E_a}$-form excitations gives another excitation that still has $\mathsf{E_a}$-form, fusing a $\mathsf{E_a}$-form excitations and a $\mathsf{E_b}$-form excitations gives another excitation that has $\mathsf{E_b}$-form. Since all fusion channels for $\mathsf{E_a}$ are single, this is an Abelian excitation.

For all excitations in the form of eq.~(\ref{eq2ap2}), we conclude that they are non-Abelian. Notice that $\delta \left( \int_{X _{\mathsf{k}}}{F^{\mathsf{k}}} \right) $ can be expanded as $\frac{1}{2}\left[ 1+\exp \left( \frac{i2\pi m_{\mathsf{k}}}{2} \right) \right] $ when the corresponding gauge subgroup of $F^{\mathsf{k}}$ is a $\mathbb{Z}_2$ group, thus 
\begin{align}
	\mathsf{E_b}\otimes \mathsf{E_b}=N\left[ \otimes _{\mathsf{k}}\left( \mathsf{1}\oplus \mathsf{E}_{\mathsf{k}} \right) \right] \,.
\end{align}
This is sufficient to show that $\mathsf{E_b}$ in the form of eq.~(\ref{eq2ap2}) is non-Abelian. When $F^{\mathsf{k}}$ corresponds to a $\mathbb{Z}_N$ gauge subgroup, $\delta \left( \int_{X _{\mathsf{k}}}{F^{\mathsf{k}}} \right) $ can be expanded as $\delta \left( \int_{X_{\mathsf{k}}}{F^{\mathsf{k}}} \right) =\frac{1}{N}\sum_{n=0}^{N-1}{\exp \left( \frac{i2\pi m_{\mathsf{k}}}{N}\cdot n \right)}
$. Thus we still can see that $\mathsf{E_b}$ in the form of eq.~(\ref{eq2ap2}) is non-Abelian.

Now we consider shrinking excitations. When we apply the shrinking operator to the excitation, $\prod_{\mathsf{j}}{\exp \left( i\int_{X _{\mathsf{j}}}{F^{\mathsf{j}}+f^{\mathsf{j}}} \right)}$ vanishes, only some delta functionals may survive and these delta functionals can be expanded, causing other nontrivial excitations. Thus, only excitations that carry delta functionals may have non-Abelian shrinking rules. If we want to have hierarchical shrinking rules, which can only happen in $T^2$ membranes, there must be delta functionals in the form of $\delta \left( \int_{\sigma}{B^i} \right) $. When we shrink the $T^2$ membranes, $\delta \left( \int_{\sigma}{B^i} \right) $ will not vanish and we can expand it as $\frac{1}{2}\left[ 1+\exp \left( \frac{i2\pi m_i}{2} \right) \right] $, causing $\left( \mathsf{1}\oplus \mathsf{L}_i \right)$ to appear in the shrinking results.

From the above discussion, we can see that delta functionals determine whether there exist non-Abelian fusion rules and non-Abelian shrinking rules. We can list the basic Wilson operators whose corresponding gauge field transformations with nontrivial shifts to see if there are delta functionals.

For $AAC$ twisted term, TQFT action is 
\begin{align}
	S=\int{\sum_{i=1}^3{\frac{N_i}{2\pi}C^idA^i+qA^1A^2C^3}}\,.
\end{align} 
Gauge transformations are 
\begin{align}
	&A^{1,2}\rightarrow A^{1,2}+d\chi ^{1,2},\qquad C^3\rightarrow C^3+dT^3\,, \nonumber
	\\
	&C^1\rightarrow C^1+dT^1-\frac{2\pi q}{N_1}\left( \chi ^2C^3-A^2T^3+\chi ^2dT^3 \right) \,,\nonumber
	\\
	&C^2\rightarrow C^2+dT^2+\frac{2\pi q}{N_2}\left( \chi ^1C^3-A^1T^3+\chi ^1dT^3 \right) \,,\nonumber
	\\
	&A^3\rightarrow A^3+d\chi ^3-\frac{2\pi q}{N_3}\left[ \left( \chi ^1A^2+\frac{1}{2}\chi ^1d\chi ^2 \right) - \left( \chi ^2A^1+\frac{1}{2}\chi ^2d\chi ^1 \right)\right]  \,.
\end{align}
Basic nontrivial Wilson operators are
\begin{align}
	&W_1=N_1\exp \left[ i\int_{\omega}{C^1+\frac{1}{2}\frac{2\pi q}{N_1}\left( d^{-1}A^2C^3-d^{-1}C^3A^2 \right)} \right] \delta \left( \int_{\gamma}{A^2} \right) \delta \left( \int_{\omega}{C^3} \right) \,,\nonumber
	\\
	&W_2=N_2\exp \left[ i\int_{\omega}{C^2+\frac{1}{2}\frac{2\pi q}{N_2}\left( d^{-1}C^3A^1-d^{-1}A^1C^3 \right)} \right] \delta \left( \int_{\gamma}{A^1} \right) \delta \left( \int_{\omega}{C^3} \right) \,,\nonumber
	\\
	&W_3=N_3\exp \left[ i\int_{\omega}{C^3+\frac{1}{2}\frac{2\pi q}{N_3}\left( d^{-1}A^1A^2-d^{-1}A^2A^1 \right)} \right] \delta \left( \int_{\gamma}{A^1} \right) \delta \left( \int_{\gamma}{A^2} \right) \,.
\end{align}
Other Wilson operators that have nontrivial terms and delta functionals can always be constructed from the above three basic nontrivial Wilson operators. We can see that there exist $\delta \left( \int_{\gamma}{A^1}\right)$, $\delta \left( \int_{\gamma}{A^2}\right)$ and $\delta \left( \int_{\omega}{C^3} \right)$, but $\delta \left( \int_{\sigma}{B^i} \right)$ is absents. Thus we can conclude that $AAC$ twisted terms have non-Abelian fusion rules and non-Abelian shrinking rules, but do not have hierarchical shrinking rules.

Similarly, for $AAAAA$ twisted term, TQFT action is
\begin{align}
	S=\int{\sum_{i=1}^5{\frac{N_i}{2\pi}C^idA^i+qA^1A^2A^3A^4A^5}}\,.
\end{align}
Gauge transformations are 
\begin{align}
	A^i\rightarrow& A^i+d\chi ^i \,,\nonumber
	\\
	C^i\rightarrow& C^i+dT^i-\frac{2\pi q}{N_i}\sum_{k<l<m}{\epsilon ^{ijklm}}\chi ^jA^kA^lA^m-\frac{2\pi q}{N_i}\sum_{j<k,l<m}{\epsilon ^{ijklm}}\chi ^jd\chi ^kA^lA^m \nonumber
	\\
	&-\frac{2\pi q}{N_i}\sum_{j<k<l}{\sum_{m=1}^5{\epsilon ^{ijklm}}}\chi ^jd\chi ^kd\chi ^lA^m-\frac{2\pi q}{N_i}\sum_{j<k<l<m}{\epsilon ^{ijklm}}\chi ^jd\chi ^kd\chi ^ld\chi ^m \,.
\end{align}
Basic nontrivial Wilson operators are
\begin{align}
	W_i=N_i\exp \left[ i\int_{\omega}{C^1+\frac{1}{4}\frac{2\pi q}{N_i}\sum_{k<l<m}{\sum_{j=1}^5{\epsilon ^{ijklm}}}d^{-1}A^jA^kA^lA^m} \right] \prod_{n=1}^5{\delta \left( \int_{\gamma}{A^n} \right)}\,.
\end{align}
Thus $AAAAA$ twisted terms have non-Abelian fusion rules and non-Abelian shrinking rules, but do not have hierarchical shrinking rules.

For $AAAdA$ twisted term, TQFT action is
\begin{align}
	S=\int{\sum_{i=1}^3{\frac{N_i}{2\pi}}C^idA^i}+qA^1A^2A^3dA^1\,.\label{eqD10}
\end{align}
Gauge transformations are
\begin{align}
	&A^1\rightarrow A^1+d\chi ^1,C^1\rightarrow C^1+dT^1-\frac{2\pi q}{N_1}\left( A^1d\chi ^2A^3+A^1A^2d\chi ^3+A^1d\chi ^2d\chi ^3 \right) \,,\nonumber
	\\
	&A^2\rightarrow A^2+d\chi ^2,C^2\rightarrow C^2+dT^2+\frac{2\pi q}{N_2}\left( d\chi ^1A^3A^1+d\chi ^1d\chi ^3A^1 \right) \,,\nonumber
	\\
	&A^3\rightarrow A^3+d\chi ^3,C^3\rightarrow C^3+dT^3-\frac{2\pi q}{N_3}\left( d\chi ^1A^2A^1+d\chi ^1d\chi ^2A^1 \right) \,.
\end{align}
Basic nontrivial Wilson operators are
\begin{align}
	W_1=&N_1\exp \left[ i\int_{\omega}{C^1+\frac{1}{2}\frac{2\pi q}{N_1}\left( d^{-1}A^2A^3dA^1-d^{-1}A^3A^2dA^1 \right)} \right] \delta \left( \int_{\gamma}{A^2} \right) \delta \left( \int_{\gamma}{A^3} \right) \,,\nonumber
	\\
	W_2=&N_2\exp \left[ i\int_{\omega}{C^2+\frac{1}{2}\frac{2\pi q}{N_2}\left( d^{-1}A^1A^3dA^1-d^{-1}A^3A^1dA^1 \right)} \right] \delta \left( \int_{\gamma}{A^1} \right) \delta \left( \int_{\gamma}{A^3} \right) \,,\nonumber
	\\
	W_3=&N_3\exp \left[ i\int_{\omega}{C^3+\frac{1}{2}\frac{2\pi q}{N_3}\left( d^{-1}A^1A^2dA^1-d^{-1}A^2A^1dA^1 \right)} \right] \delta \left( \int_{\gamma}{A^1} \right) \delta \left( \int_{\gamma}{A^2} \right) \,.
\end{align}
Thus $AAAdA$ twisted terms have non-Abelian fusion rules and non-Abelian shrinking rules, but do not have hierarchical shrinking rules.

For $AdAdA$ twisted term, TQFT action is
\begin{align}
	S=\int{\sum_{i=1}^3{\frac{N_i}{2\pi}C^idA^i+qA^1dA^2dA^3}}\,.
\end{align}
Gauge transformations are 
\begin{align}
	A^i\rightarrow A^i+d\chi ^i,\qquad C^i\rightarrow C^i+dT^i\,.
\end{align}
All fields transform without nontrivial shifts and thus all Wilson operators do not have nontrivial terms and delta functionals. Thus $AdAdA$ twisted term only has Abelian fusion rules and Abelian shrinking rules.

For $AdAB$ twisted term, TQFT action is
\begin{align}
	S=\int{\frac{N_1}{2\pi}C^1dA^1+\frac{N_2}{2\pi}C^2dA^2+\frac{N_2}{2\pi}\tilde{B}^3dB^3+qA^1dA^2B^3}\,.
\end{align}
Gauge transformations are 
\begin{align}
	&A^1\rightarrow A^1+d\chi ^1,\qquad C^1\rightarrow C^1+dT^1-\frac{2\pi q}{N_1}dV^3A^2 \,,\nonumber
	\\
	&A^2\rightarrow A^2+d\chi ^2,\qquad C^2\rightarrow C^2+dT^2 \,,\nonumber
	\\
	&B^3\rightarrow B^3+dV^3,\qquad \tilde{B}^3\rightarrow \tilde{B}^3+d\tilde{V}^3-\frac{2\pi q}{N_3}d\chi ^1A^2\,.
\end{align}
Basic nontrivial Wilson operators are
\begin{align}
	W_1=&N_1\exp \left( i\int_{\omega}{C^1+\frac{2\pi q}{N_i}}\int_{\Xi}{B^3dA^2} \right) \,,\nonumber
	\\
	W_3=&N_3\exp \left( i\int_{\sigma}{\tilde{B}^3-\frac{2\pi q}{N_3}}\int_{\Omega}{A^1dA^2} \right) \,.
\end{align}
Although some Wilson operators have nontrivial terms, delta functionals are still absent. These nontrivial terms cannot change fusion and shrinking rules. Thus $AdAB$ twisted term only has Abelian fusion rules and Abelian shrinking rules.

For $AAdB$ twisted term, notice that 
\begin{align}
	d\left( A^1A^2B^3 \right) =A^1A^2dB^3+A^2dA^1B^3-A^1dA^2B^3\,.
\end{align}
This means $AAdB$ twisted term can always be written as two $AdAB$ twisted terms up to a boundary term. Thus $AAdB$ twisted term only has Abelian fusion rules and Abelian shrinking rules.

\bibliographystyle{JHEP}

\end{document}